\documentclass[aps,prd,twocolumn,showpacs,superscriptaddress,groupedaddress,preprintnumbers,noshowpacs]{revtex4}  
\usepackage{graphicx}  
\usepackage{dcolumn}   
\usepackage{bm}        
\usepackage{amssymb}   
\usepackage{amsmath}
\usepackage{slashed}
\usepackage{natbib}
\usepackage{stackrel}

\usepackage{multirow}
\usepackage{array}

\allowdisplaybreaks

\newcommand{\beq}{\begin{equation}}
\newcommand{\eeq}{\end{equation}}
\newcommand{\bqa}{\begin{eqnarray}}
\newcommand{\eqa}{\end{eqnarray}}

\newcommand{\nn}{\pagebreak[0] \nonumber \\}

\def\Formcalc{{{\sc FormCalc}}}

\def\C++{{{\sc c++ }}}
\def\Feynarts{{{\sc FeynArts}}}

\newcommand{\cp}{\mbox{c.p.}}
\newcommand{\QQ}{Q}
\newcommand{\GG}{G}
\newcommand{\GH}{\hat G}
\newcommand{\GA}{\Gamma}

\def\spa#1.#2{\langle#1\,#2\rangle}
\def\spb#1.#2{[#1\,#2]}
\def\spab#1.#2.#3{\langle\mskip-1mu{#1}
                  | #2 | {#3}]}
\def\spba#1.#2.#3{[\mskip-1mu{#1}
                  | #2 | {#3}\rangle}
\def\spbb#1.#2.#3.#4{[\mskip-1mu{#1}
                     | {#2} \ {#3} | {#4}]}
\def\spaa#1.#2.#3.#4{\langle\mskip-1mu{#1}
                     | {#2} \ {#3} | {#4}\rangle}

\newcommand{\bea}{\begin{eqnarray}}
\newcommand{\eea}{\end{eqnarray}}
\newcommand{\bean}{\begin{eqnarray*}}
\newcommand{\eean}{\end{eqnarray*}}

\newcommand{\mpi}{Max-Planck-Institut f\"ur Physik, F\"ohringer Ring 6, 80805 M\"unchen, Germany}
\newcommand{\padova}{Dipartimento di Fisica e Astronomia, Universit\`a di Padova, and INFN  \\   Sezione di Padova, via Marzolo 8, 35131 Padova, Italy}                       
\newcommand{\bogota}{Departamento de F\'isica, Universidad Nacional de Colombia,  Ciudad Universitaria, Bogot\`a, D.C. Colombia}

\begin{document}

\preprint{MPP-2014-110}

\title{On the Four-Dimensional Formulation 
of Dimensionally Regulated Amplitudes}
\author{Raffaele A. Fazio}
\affiliation{\bogota}
\author{Pierpaolo Mastrolia}
\affiliation{\mpi}
\affiliation{\padova}
\author{Edoardo Mirabella}
\affiliation{\mpi}
\author{William J. Torres Bobadilla}
\affiliation{\bogota}
\affiliation{\padova}
\date{\today}

\begin{abstract}
We propose a pure four-dimensional formulation (FDF) of the $d$-dimensional regularization of one-loop scattering amplitudes.
In our formulation particles  propagating inside the loop  are represented by  massive internal states regulating the divergences.
The latter obey Feynman rules containing multiplicative selection rules which automatically account for the effects of the extra-dimensional regulating terms of the amplitude.
The equivalence between the FDF  and the Four Dimensional Helicity scheme is discussed.
We present explicit representations of the polarization and helicity states of the four-dimensional particles propagating in the loop. They allow for a complete,
four-dimensional, unitarity-based construction of $d$-dimensional amplitudes.
Generalized unitarity within the FDF does not require any higher-dimensional extension of the Clifford and the spinor algebra.
Finally we show how the FDF allows for the recursive construction of $d$-dimensional one-loop integrands, generalizing the four-dimensional open-loop approach. 
\end{abstract}

\maketitle

\input{feynarts.sty}
\section{Introduction}
\label{sec:intro}

The recent development of novel methods for computing one-loop scattering amplitudes
has been highly stimulated by a deeper understanding 
of their multi-channel factorization properties in special
kinematic conditions enforced by on-shellness \cite{Cachazo:2004kj,Britto:2004ap} and  
generalized unitarity \cite{Bern:1994zx,Britto:2004nc}, strengthened by the complementary classification of the mathematical 
structures present in the residues at the singular points \cite{Ossola:2006us,Zhang:2012ce,Mastrolia:2012an}. 

The unitarity-based methods, reviewed 
in \cite{Alday:2008yw,Britto:2010xq,Henn:2011xk,Bern:2011qt,Carrasco:2011hw,Dixon:2011xs,Ellis:2011cr,Ita:2011hi},
use two general properties of scattering amplitudes such as analyticity and unitarity. 
The former grants  that the amplitudes can be reconstructed from 
 their singularity-structure while  the latter grants that 
the residues at the singular points factorize into products of simpler amplitudes.

Integrand-reduction methods~\cite{Ossola:2006us,Ossola:2007bb}, instead,
allow one to decompose the  integrands of  scattering amplitudes are  into multi-particle poles,
and the multi-particle residues are expressed in terms of 
{\it irreducible scalar products}  formed by the loop momenta and 
either external momenta or polarization vectors constructed out of them.
The polynomial structure of the multi-particle residues is a {\it qualitative} information
that turns into a {\it quantitative} algorithm for decomposing arbitrary 
amplitudes in terms of Master Integrals (MIs) by polynomial fitting 
at the integrand level. In this context the on-shell conditions have 
been used  as a computational tool reducing the complexity of the algorithm.
A more intimate connection among the idea of  reduction under the integral sign and
analyticity and unitarity has been pointed out recently. Using basic
principles of algebraic geometry, Refs.~\cite{Mastrolia:2011pr,Badger:2012dp,Zhang:2012ce,Mastrolia:2012an,Mastrolia:2013kca}
have shown that  the structure of the multi-particle poles is determined by the zeros of the denominators
involved in the corresponding multiple cut.
This new  approach to integrand reduction methods allows for their
systematization and for their all-loop extension.  \\
Moreover, the proper understanding of  the integrands of the amplitudes  paved the way to the  recent proposal of a four-dimensional renormalization scheme, 
which allows one the recognize and subtract UV-divergent contributions already at the integrand level~\cite{Pittau:2012zd,Donati:2013iya,Donati:2013voa}.

Dimensionally-regulated amplitudes are constituted by terms containing  (poly)logarithms, also called cut-constructible terms, and rational terms.   
The former may be obtained by the discontinuity structure of integrals  over the four-dimensional  loop momentum. The latter ones, instead, 
 escape any  four-dimensional detectability and require  to cope with 
integrations including also the  $(d-4)$ components of the loop momentum.

Within generalized-unitarity methods both terms can be in principle 
 obtained by performing $d$-dimensional generalized cuts~\cite{Mahlon:1993fe,Bern:1995db,Anastasiou:2006jv,Anastasiou:2006gt,Giele:2008ve,Ellis:2008ir}. 
 In this context, the issue of addressing factorization in conjunction with regularization clearly emerges,
 since  $d$-dimensional unitarity requires  to  work with tree-level amplitudes
involving external particles in arbitrary, non-integer dimensions.
Polarization states, dimensionality of  the on-shell  momenta, 
and the completeness relations  for the particles wavefunctions 
have to be consistently  handled since the number of spin eigenstates
depends on the space-time dimension. 
Therefore, in many cases generalized unitarity in arbitrary non-integer dimensions is avoided
and cut-constructible and rational terms are obtained 
in separate steps.  The former are computed by performing four-dimensional 
 generalized cuts in the  un-regularized amplitudes. If possible the 
 rational terms are obtained  by using special properties of the amplitude
 under consideration, like the supersymmetric decomposition~\cite{Bern:1993mq, Bern:1994cg}.

Within integrand reduction methods,  different approaches  are available,
according to the strategies  adopted for the determination of cut-constructible 
and rational terms.\\
In some algorithms, the  computation of  the two ingredients proceeds in two
steps: the cut-constructible part is obtained by reducing the un-regularized integrand while the 
rational one is computed by introducing new  counterterm-like diagrams
which depend on the model under consideration~\cite{Ossola:2008xq,Garzelli:2010qm}. \\
Other methods, instead, aim at the combined determination of the two ingredients  
by reducing the dimensionally regulated integrand. Therefore the numerator
of the integrand has to be generated and manipulated in $d$ dimensions 
and acquires a dependence on $(d-4)$ and on the square of the
$(d-4)$-dimensional components of the loop momentum, $\mu^2$~\cite{Giele:2008ve,Ellis:2008ir,Melnikov:2010iu}.  
The multi-particle residues are finally determined by performing 
generalized cuts by setting $d$-dimensional massive particles on shell. 
This is equivalent to have on-shell four-dimensional states whose squared mass is shifted by $\mu^2$.  

If the integrand at a generic multiple cut is  obtained as a product of  tree-level amplitudes, the issues
related to factorization in presence of dimensional regularization 
have to be addressed. An interesting  approach~\cite{Giele:2008ve} uses the linear dependence of the amplitude
on the space-time dimensionality to compute  the $d$-dimensional amplitude. In particular
the latter is obtained by interpolating the values of the one-loop amplitude
in correspondence to two  different integer values of the space-time.    When
 fermions are involved, the space-time  dimensions have to admit  an explicit  representation
 of the Clifford algebra~\cite{Ellis:2008ir}. 
More recently, this idea has been combined with
 the six-dimensional helicity formalism~\cite{Davies:2011vt} for the analytic reconstruction 
of one-loop scattering amplitudes in QCD via generalized unitarity. \\

In this article, we propose a four dimensional formulation (FDF) of the
$d$-dimensional regularization scheme which, at one loop, turns out to
be equivalent to the four-dimensional helicity (FDH) scheme~\cite{Bern:1991aq,Bern:1995db,Bern:2002zk}, and which
allows for a purely four-dimensional regularization of the amplitudes.
Within FDF, the states in the loop are described as 
four dimensional massive particles. The four-dimensional degrees
of freedom of the gauge bosons are carried by  {\it massive vector bosons}
of mass $\mu$ and their 
 $(d-4)$-dimensional ones by {\it real scalar particles} obeying 
a simple set of four-dimensional Feynman rules. A $d$-dimensional fermion 
of mass $m$ is instead traded for a {\it tachyonic  Dirac field} with
mass $m +i \mu \gamma^5$.   The $d$ dimensional algebraic manipulations are 
replaced by four-dimensional ones complemented by    
a set of multiplicative selection rules. The latter are treated as an algebra
describing internal symmetries.  

Within integrand reduction methods, our regularization scheme 
allows for the simultaneous  computation of both the cut-constructible and the rational terms
by employing a purely four-dimensional formulation of the integrands. 
As a consequence,  an explicit four-dimensional representation of
generalized states propagating around the loop can be
formulated. Therefore, a straightforward implementation of $d$-dimensional generalized unitarity 
within  exactly four space-time dimensions can be realized, avoiding any higher-dimensional 
extension  of  either the  Dirac~\cite{Giele:2008ve,Ellis:2008ir}  or the spinor algebra \cite{Cheung:2009dc}.

Another interesting consequence of our framework is the possibility to extend 
to $d$ dimensions the recursive generation of the integrand from
off-shell currents and open loops, 
now limited to four dimensions~\cite{Cascioli:2011va,Hirschi:2011pa,Actis:2012qn}.

The  paper is organized  as follows.  Section~\ref{sec:4D} is devoted to the description
of our regularization method, while  Section~\ref{sec:onshell} describes how 
generalized  unitarity method can be  applied in presence of a  FDF 
of one-loop amplitudes. Section~\ref{sec:4P} shows 
the decomposition in terms of MIs 
of certain classes of $2 \to 2$ one-loop amplitudes. It is preliminary to 
Sections~\ref{sec:gggg}, \ref{sec:ggqq} and~\ref{sec:gggH},  which collect the applications of 
generalized unitarity methods within the FDF.  In particular they present results
 for representative helicity amplitudes of  $gg
\to gg$, $q{\bar q} \to gg$ with massless quarks, and $gg \to Hg$ in
the heavy-top limit.  Section~\ref{sec:ol}
describes how the integrand of the FDF of one-loop amplitudes
can be generated recursively within the open-loop approach.

\section{Four-dimensional Feynman rules}
\label{sec:4D}

The FDH scheme~\cite{Bern:1991aq,Bern:1995db,Bern:2002zk}   defines a $d$-dimensional vector space embedded in a larger  $d_s$-dimensional 
space,  $d_s  \equiv(4-2\epsilon) > d > 4$.  The scheme is 
determined by the following rules
\begin{itemize}
\item The loop momenta are considered to be $d$-dimensional. All  observed external states are considered as four-dimensional.
All unobserved  internal states, {\it i.e.} virtual states in loops and intermediate states in trees,  are treated as  $d_s$-dimensional.
\item Since $d_s > d > 4$, the scalar product of any  $d$- or $d_s$-dimensional vector with a four-dimensional vector is a
four-dimensional scalar product.  Moreover any dot product between a $d_s$-dimensional tensor and a $d$-dimensional one
is a $d$-dimensional dot product. 
\item  The Lorentz and the Clifford algebra are performed in $d_s$ dimensions, which has to be kept distinct from $d$. The matrix $\gamma^5$
is treated using the  't Hooft-Veltman prescription, {\it i.e.} $\gamma^5$ commutes with the Dirac matrices carrying $-2\epsilon$
indices.
\item After the $\gamma$-matrix algebra has been performed,  the limit $d_s \to 4$ has to be performed, keeping $d$ fixed. The limit $d\to 4$
is taken at the very end. 
\end{itemize}
In the following $d_s$-dimensional quantities are denoted by a bar. One can split the $d_s$-dimensional metric tensor as follows 
\begin{align}
\bar g^{\mu \nu} = g^{\mu \nu} + \tilde g^{\mu \nu}  \, ,
\end{align}
in terms of  a four-dimensional tensor  $g$ and a $-2 \epsilon$-dimensional one, $ \tilde g$,
such that
\begin{align}
 \tilde g^{\mu \rho} \, g_{\rho \nu} =0 \, , \qquad 
 \tilde g^{\mu}_{\phantom{\mu}  \mu}  =  -2 \epsilon \stackrel[d_s\to4]{}{\longrightarrow} 0 \, , \qquad 
 g^{\mu}_{\phantom{\mu} \mu}  = 4 \, , 
 \label{Eq:OrthoGs}
\end{align}  
The  tensors $g$ and $\tilde g$  project a $d_s$-dimensional vector $\bar q$ into the four-dimensional  and the  
$-2 \epsilon$-dimensional subspaces respectively, 
\begin{align}
q^\mu &\equiv  g^{\mu}_{\phantom{\mu} \nu} \,  \bar q^\nu\, ,  &  \tilde q^\mu &\equiv \tilde g^{\mu}_{\phantom{\mu} \nu} \, \bar q^\nu\, .
\label{Eq:Projection}
\end{align}
At one loop the only $d$-dimensional object is the loop momentum $\bar \ell$. The square of its  $-2\epsilon$
dimensional component is defined as:
\begin{align}
\tilde \ell^2 = \tilde g^{\mu \nu} \,  \bar \ell_\mu\,  \bar \ell_\nu  \equiv -\mu^2  \, .
\end{align}
The properties of the matrices $\tilde \gamma^\mu = \tilde g^{\mu}_{\phantom{\mu} \nu} \, \bar \gamma^\nu$ can be obtained from Eq.~(\ref{Eq:OrthoGs})
\begin{subequations}
\begin{align}
[ \tilde \gamma^{\alpha}, \gamma^{5}   ] &= 0 \, , & 
\{\tilde \gamma^{\alpha}, \gamma^{\mu}  \} &=0 \ , \label{Eq:Gamma01} \\
\{\tilde \gamma^{\alpha}, \tilde \gamma^{\beta}  \} &= 2 \,  \tilde g^{\alpha \beta} \, . \label{Eq:Gamma02}
\end{align}
\end{subequations}

We remark that the $-2\epsilon$  tensors can not  have a four-dimensional representation. Indeed
the metric tensor $\tilde g$ is a tripotent matrix
\begin{align}
\tilde g^{\mu \rho} \tilde g_{\rho \nu} \tilde g^{\nu\sigma} =  \tilde g^{\mu \sigma} \, , \quad
\label{Eq:GG2}
\end{align} 
and its square is  traceless 
\begin{align}
\tilde g^{\mu \rho} \tilde g_{\rho \mu} =   \tilde  g^{\mu}_{\phantom{\mu} \mu}   \stackrel[d_s\to4]{}{\longrightarrow} 0 \, ,
\label{Eq:GG3}
\end{align}
but in any  integer-dimension space the  square of any non-null tripotent matrix has an integer,
positive trace~\cite{bookA3}.   Moreover the component $\tilde \ell$ of the loop momentum vanishes when contracted with the metric tensor $g$, 
\begin{align}
\tilde \ell^\mu \; g_{\mu \nu}  = \bar \ell_\rho \,  \tilde g^{\rho \mu} \,  g_{\mu \nu} = 0\, ,
\label{Eq:QQ1}
\end{align}
and in four dimensions the only four vector fulfilling~(\ref{Eq:QQ1}) is the null one.
Finally  in four dimensions the only non-null  matrices  fulfilling the 
conditions~(\ref{Eq:Gamma01}) are proportional to $\gamma^5$, hence $\tilde \gamma \sim \gamma^5$.
However the  matrices $\tilde \gamma$
fulfill the Clifford algebra~(\ref{Eq:Gamma02}), thus
\begin{align} 
\tilde \gamma^\mu \, \tilde \gamma_\mu \stackrel[d_s\to4]{}{\longrightarrow} 0\, ,  \quad \mbox{ while }  \quad   \gamma^5 \gamma^5 = \mathbb{I} \,.
\label{Eq:GA1}
\end{align}

These arguments exclude   any  four-dimensional representation
of the $-2\epsilon$ subspace. It is possible, however, to  find such a
representation by introducing additional rules, called in the following
$-2 \epsilon$ {\it selection rules}, 
($-2\epsilon$)-SRs.   Indeed, as shown in Appendix~\ref{App:Equiv}, 
the Clifford algebra~(\ref{Eq:Gamma02}) is equivalent to
\begin{align}
  \cdots \tilde \gamma^\alpha  \cdots  \cdots \tilde \gamma_\alpha  \cdots =  0, \, \qquad    
\tilde{ \slashed \ell}\tilde{ \slashed \ell} =  -\mu^2\, .
\label{Eq:Gamma03}
\end{align}
Therefore any regularization scheme which is equivalent of FDH has to fulfill the conditions~(\ref{Eq:OrthoGs}) -- (\ref{Eq:Gamma01}),  and~(\ref{Eq:Gamma03}).
The orthogonality conditions  (\ref{Eq:OrthoGs}) and (\ref{Eq:Projection}) are fulfilled  by splitting a $d_s$-dimensional gluon
onto  a four-dimensional one and a colored scalar, $s_g$, while the other  conditions  are fulfilled by performing the substitutions:
\begin{align}
\tilde g^{\alpha \beta} \to   G^{AB}, \qquad  \tilde \ell^{\alpha} \to i \, \mu \, \QQ^A \; , \qquad  \tilde \gamma^\alpha \to \gamma^5 \, \GA^A\, .
\label{Eq:SubF}
\end{align} 
The $-2\epsilon$-dimensional vectorial
 indices  are thus  traded for  ($-2\epsilon$)-SRs  such that
\begin{align}
\GG^{AB}\GG^{BC} &= \GG^{AC},    & \GG^{AA}&=0,  &   \GG^{AB}&=\GG^{BA},  \nn
 \GA^A \GG^{AB} &= \GA^B,              &   \GA^A \GA^{A} &=0,  & \QQ^A \GA^{A}  &=1, \nn
\QQ^A \GG^{AB} &= \QQ^B,              & \QQ^A \QQ^{A} &=1.  
\label{Eq:2epsA}
\end{align}
The exclusion  of the terms containing odd powers of $\mu$ completely
defines the FDF, and allows one to build integrands which, upon
integration, yield to the same result as in the FDH scheme.

The rules~(\ref{Eq:2epsA}) constitute an abstract algebra which  is similar to the algebras implementing internal symmetries. For instance, 
in a Feynman diagrammatic approach the ($-2\epsilon$)-SRs can be handled as the color 
algebra and  performed for each diagram once and for all. In each diagram, the indices  of the ($-2\epsilon$)-SRs
are fully contracted and the outcome of their manipulation is  either  $0$ or $\pm1$.  
It is worth to  remark that  the  replacement of   $\tilde \gamma^\alpha$  with  $\gamma^5$  takes care of the $d_s$-dimensional Clifford algebra automatically, thus
we do not need to introduce any additional scalar particle for each fermion flavor. 
These particles and their interactions  have been instead introduced in Ref.~\cite{Pittau:2011qp},
where a method for the  reconstruction of  the $\mu^2$-dependent part of the numerator
has been proposed.

To summarize, the QCD  $d$-dimensional Feynman rules  in the 't Hooft-Feynman gauge, collected in Ref.~\cite{Ellis:1991qj},  may have the following four-dimensional formulation:
\begin{subequations}
\vspace{-0.4cm}
\begin{align}
\parbox{20mm}{\unitlength=0.20bp%
\begin{feynartspicture}(300,300)(1,1)
\FADiagram{}
\FAProp(4.,10.)(16.,10.)(0.,){/Cycles}{0}
\FALabel(5.5,8.93)[t]{\tiny $a, \alpha$}
\FALabel(14.5,8.93)[t]{\tiny $b, \beta$}
\FALabel(10.,12.5)[]{\tiny $k$}
\FAVert(4.,10.){0}
\FAVert(16.,10.){0}
\end{feynartspicture}} &= -i \, \delta^{ab} \,\frac{ g^{\alpha \beta} }{k^2 -\mu^2 +i 0} \,  \quad (\mbox{gluon}), \label{Eq:FRglu}\\[-3.0ex]
\parbox{20mm}{\unitlength=0.20bp%
\begin{feynartspicture}(300,300)(1,1)
\FADiagram{}
\FAProp(4.,10.)(16.,10.)(0.,){/GhostDash}{1}
\FALabel(5.5,8.93)[t]{\tiny $a$}
\FALabel(14.5,8.93)[t]{\tiny $b$}
\FALabel(10.,12.5)[]{\tiny $k$}
\FAVert(4.,10.){0}
\FAVert(16.,10.){0}
\end{feynartspicture}} &= i \, \delta^{ab} \,\frac{ 1 }{k^2 -\mu^2 + i0}  \,   \quad (\mbox{ghost}), \\[-3.0ex]
\parbox{20mm}{\unitlength=0.20bp%
\begin{feynartspicture}(300,300)(1,1)
\FADiagram{}
\FAProp(4.,10.)(16.,10.)(0.,){/ScalarDash}{0}
\FALabel(5.5,8.93)[t]{\tiny $a, A$}
\FALabel(14.5,8.93)[t]{\tiny $b, B$}
\FALabel(10.,12.5)[]{\tiny $k$}
\FAVert(4.,10.){0}
\FAVert(16.,10.){0}
\end{feynartspicture}} &= -i \, \delta^{ab} \,\frac{ G^{AB} }{k^2 -\mu^2+ i0} \, , \quad (\mbox{scalar}),  \\[-3.0ex]
\parbox{20mm}{\unitlength=0.20bp%
\begin{feynartspicture}(300,300)(1,1)
\FADiagram{}
\FAProp(4.,10.)(16.,10.)(0.,){/Straight}{1}
\FALabel(5.5,8.93)[t]{\tiny $i$}
\FALabel(14.5,8.93)[t]{\tiny $j$}
\FALabel(10.,12.5)[]{\tiny $k$}
\FAVert(4.,10.){0}
\FAVert(16.,10.){0}
\end{feynartspicture}} &= i \, \delta^{ij} \,\frac{ \slashed k + i \mu \gamma^5 +m }{k^2 -m^2 -\mu^2+i0}  \, ,  \nonumber \\[-3.5ex] 
&\qquad \qquad  (\mbox{fermion}), 
 \label{Eq:FRfer} \\[-0.0ex]
 \parbox{20mm}{\unitlength=0.20bp%
\begin{feynartspicture}(300,300)(1,1)
\FADiagram{}
\FAProp(3.,10.)(10.,10.)(0.,){/Cycles}{0}
\FALabel(5.3,8.93)[t]{\tiny $1, a, \alpha$}
\FAProp(16.,15.)(10.,10.)(0.,){/Cycles}{0}
\FALabel(12.2273,13.5749)[br]{\tiny $2, b, \beta$}
\FAProp(16.,5.)(10.,10.)(0.,){/Cycles}{0}
\FALabel(12.8873,5.81315)[tr]{\tiny $3, c, \gamma$}
\FAVert(10.,10.){0}
\end{feynartspicture}} &=  -g \, f^{abc} \, \big [  (k_1-k_2)^\gamma g^{\alpha\beta}  \nonumber \\[-5.5ex]
 &\qquad \qquad +(k_2-k_3)^\alpha g^{\beta\gamma}  \nn
 &\qquad \qquad + (k_3-k_1)^\beta g^{\gamma\alpha} \big ]   \, ,   \\[-0.ex]
  \parbox{20mm}{\unitlength=0.20bp%
\begin{feynartspicture}(300,300)(1,1)
\FADiagram{}
\FAProp(3.,10.)(10.,10.)(0.,){/Cycles}{0}
\FALabel(5.3,8.93)[t]{\tiny $1, a, \alpha$}
\FAProp(16.,15.)(10.,10.)(0.,){/GhostDash}{-1}
\FALabel(12.2273,13.5749)[br]{\tiny $2, b$}
\FAProp(16.,5.)(10.,10.)(0.,){/GhostDash}{1}
\FALabel(12.8873,5.81315)[tr]{\tiny $3, c$}
\FAVert(10.,10.){0}
\end{feynartspicture}} &= - g\, f^{abc} \, k_2^\alpha  \, ,  \\[-3.0ex]
 \parbox{20mm}{\unitlength=0.20bp%
\begin{feynartspicture}(300,300)(1,1)
\FADiagram{}
\FAProp(3.,10.)(10.,10.)(0.,){/Cycles}{0}
\FALabel(5.3,8.93)[t]{\tiny $1,a, \alpha$}
\FAProp(16.,15.)(10.,10.)(0.,){/ScalarDash}{0}
\FALabel(12.2273,13.5749)[br]{\tiny $2, b, B$}
\FAProp(16.,5.)(10.,10.)(0.,){/ScalarDash}{0}
\FALabel(12.8873,5.81315)[tr]{\tiny $3,c, C$}
\FAVert(10.,10.){0}
\end{feynartspicture}} &=  -g \, f^{abc} \, (k_2-k_3)^\alpha  \, \GG^{BC}  \, ,  \\[-3.0ex]
%
 \parbox{20mm}{\unitlength=0.20bp%
\begin{feynartspicture}(300,300)(1,1)
\FADiagram{}
\FAProp(3.,10.)(10.,10.)(0.,){/Cycles}{0}
\FALabel(5.3,8.93)[t]{\tiny $1, a, \alpha$}
\FAProp(16.,15.)(10.,10.)(0.,){/ScalarDash}{0}
\FALabel(12.2273,13.5749)[br]{\tiny $2, b, B$}
\FAProp(16.,5.)(10.,10.)(0.,){/Cycles}{0}
\FALabel(12.8873,5.81315)[tr]{\tiny $3, c, \gamma$}
\FAVert(10.,10.){0}
\end{feynartspicture}} &= \mp g \, f^{abc} \,   ( i \mu) \, g^{\gamma\alpha}\, \QQ^B  \, , \nn[-5.5ex]
 & \qquad    (\tilde k_1 = 0, \quad  \tilde k_3 =\pm \tilde \ell)   \label{Eq:FRggs} \\[-0.0ex]
\parbox{20mm}{\unitlength=0.20bp%
\begin{feynartspicture}(300,300)(1,1)
\FADiagram{}
\FAProp(4.,15.)(10.,10.)(0.,){/Cycles}{0}
\FALabel(3.0,12.)[]{\tiny $1,a,\alpha$}
\FAProp(10.,10.)(4.,5.)(0.,){/Cycles}{0}
\FALabel(3.0,8.)[]{\tiny $4,d,\delta$}
\FAProp(16.,15.)(10.,10.)(0.,){/Cycles}{0}
\FALabel(11.,15.)[]{\tiny $2,b,\beta$}
\FAProp(10.,10.)(16.,5.)(0.,){/Cycles}{0}
\FALabel(11,5.)[]{\tiny $3,c,\gamma$}
\FAVert(10.,10.){0}
\end{feynartspicture}} &= - i g^2 \big [ \;   \nonumber  \\[-5.5ex]
   &\; +        f^{xad} \, f^{xbc} \, \left ( g^{\alpha \beta} g^{\delta \gamma} - g^{\alpha\gamma} g^{\beta \delta}\right )  \nn
   &\; +        f^{xac} \, f^{xbd} \, \left ( g^{\alpha \beta} g^{\delta \gamma} - g^{\alpha\delta} g^{\beta \gamma}\right )  \nn 
   &\; +        f^{xab} \, f^{xdc} \, \left ( g^{\alpha \delta} g^{\beta \gamma} - g^{\alpha\gamma} g^{\beta \delta}\right )  \big ]  \, ,  \\[-3.0ex]
\parbox{20mm}{\unitlength=0.20bp%
\begin{feynartspicture}(300,300)(1,1)
\FADiagram{}
\FAProp(4.,15.)(10.,10.)(0.,){/Cycles}{0}
\FALabel(3.0,12.)[]{\tiny $1,a,\alpha$}
\FAProp(10.,10.)(4.,5.)(0.,){/Cycles}{0}
\FALabel(3.0,8.)[]{\tiny $4,d,\delta$}
\FAProp(16.,15.)(10.,10.)(0.,){/ScalarDash}{0}
\FALabel(11.,15.)[]{\tiny $2,b,B$}
\FAProp(10.,10.)(16.,5.)(0.,){/ScalarDash}{0}
\FALabel(11,5.)[]{\tiny $3,c,C$}
\FAVert(10.,10.){0}
\end{feynartspicture}} &= 2 i g^2 \, g^{\alpha \delta} \; \big (  f^{xab} \, f^{xcd} \nonumber \\[-5.5ex]
 &\qquad \qquad + f^{xac} \, f^{xbd} \big ) \; \GG^{BC}  \, ,   \\[-0.ex]
 \parbox{20mm}{\unitlength=0.20bp%
\begin{feynartspicture}(300,300)(1,1)
\FADiagram{}
\FAProp(3.,10.)(10.,10.)(0.,){/Straight}{1}
\FALabel(5.3,8.93)[t]{\tiny $1, i$}
\FAProp(16.,15.)(10.,10.)(0.,){/Cycles}{0}
\FALabel(12.2273,13.5749)[br]{\tiny $2, b, \beta$}
\FAProp(16.,5.)(10.,10.)(0.,){/Straight}{-1}
\FALabel(12.8873,5.81315)[tr]{\tiny $3, j$}
\FAVert(10.,10.){0}
\end{feynartspicture}} &= - i g \, \left ( t^{b}\right)_{ji} \,\gamma^\beta \, ,  \\[-3.0ex]
\parbox{20mm}{\unitlength=0.20bp%
\begin{feynartspicture}(300,300)(1,1)
\FADiagram{}
\FAProp(3.,10.)(10.,10.)(0.,){/Straight}{1}
\FALabel(5.3,8.93)[t]{\tiny $1, i$}
\FAProp(16.,15.)(10.,10.)(0.,){/ScalarDash}{0}
\FALabel(12.2273,13.5749)[br]{\tiny $2, b, B$}
\FAProp(16.,5.)(10.,10.)(0.,){/Straight}{-1}
\FALabel(12.8873,5.81315)[tr]{\tiny $3, j$}
\FAVert(10.,10.){0}
\end{feynartspicture}} &= - i g \, \left ( t^{b}\right)_{ji} \,\gamma^5 \,  \GA^B  \, .
\end{align}
\label{Eq:FR4}
\end{subequations}
In the Feynman rules~(\ref{Eq:FR4})  all the momenta are incoming and   the scalar particle  $s_g$ can circulate in  the loop only. 
The terms  $\mu^2$ appearing in the the propagators~(\ref{Eq:FRglu})--(\ref{Eq:FRfer}) enter only  if the corresponding momentum 
$k$ is $d$-dimensional,  {\it i.e.} only if  the corresponding particle  circulates in  the loop.  In the vertex~(\ref{Eq:FRggs}) the momentum 
$k_1$ is four-dimensional  while the other two are $d$-dimensional.  The possible combinations of the $-2 \epsilon$ components 
of the momenta involved are
\begin{align}
\{ \tilde k_1\, ,  \tilde k_2\,, \tilde k_3\, \}  = \{ 0 \, , \mp \tilde \ell  \, , \pm \tilde \ell  \, \} \, .
\label{Eq:Combo}
\end{align}
The overall sign of the Feynman rule~(\ref{Eq:FRggs}) depends on which of the combinations~(\ref{Eq:Combo}) is present in the vertex.
\medskip

The ($-2\epsilon$)-SRs~(\ref{Eq:2epsA})  and the Feynman rules~(\ref{Eq:FR4}) have been implemented in \Feynarts~\cite{Hahn:2000kx} and \Formcalc~\cite{Hahn:1998yk,Agrawal:2011tm,Nejad:2013ina} and have been
used to generate the numerators of the one-loop  integrands of the processes
\begin{align}
    q\, \bar q \, &\to \, t \, \bar t   \, , 
& g\,  g \,&\to\, t \, \bar t    \, , 
&   t \, \bar t \,    &\to \,  t \,  \bar t   \, , 
\nn
     g \, g \, &\to\, g \,g  \, ,  
& q \, \bar q \,&\to\, t\, \bar t\, g \, , 
& g \, g\,  &\to\,  t\,  \bar t\, g   \, ,
\nn
   q\,  \bar   q \,&\to \,t \,\bar t\, q' \,\bar q '  \, .
\end{align}
We have analytically checked that the numerators of the integrands
obtained using FDF are equivalent 
to the corresponding ones obtained using the  FDH scheme. In
particular, we have verified that their difference is 
spurious, i.e. it vanishes upon integration over the loop momentum.
\medskip

Our prescriptions, Eq.~(\ref{Eq:SubF}), can be related to a five-dimensional theory characterized by $g^{55}=-1$, $\ell^5 =  \mu$
and a $4\times 4$ representation of the Clifford algebra, $\{\gamma^0, \ldots, \gamma^3, i \gamma^5 \}$. Regularization
methods in five dimensions have been proposed as an alternative formulation of the Pauli-Villars regularization~\cite{PTP.5.14}
or as regulators of massless pure Yang Mills theories at one loop~\cite{tHooft:1971fh}.  Our method distinguishes itself
by the presence of the  ($- 2\epsilon$)-SRs, a crucial ingredient  for the correct reconstruction of dimensionally-regularized  amplitudes. 

\section{Generalized Unitarity}
\label{sec:onshell}

Generalized-unitarity methods in $d$ dimensions require an explicit representation of the polarization
vectors and the spinors of $d$-dimensional particles.  The latter ones are essential ingredients 
for the construction of the tree-level amplitudes that are sewn along the generalized cuts.  In this
respect, the FDF scheme  is suitable for the four-dimensional formulation of $d$-dimensional 
generalized unitarity. The main advantage of the FDF is that the  four-dimensional 
expression of the propagators of the particles  in the loop admits an  explicit 
representation  in  terms of generalized spinors and polarization expressions,
whose expression is collected below. 
\smallskip

In the following discussion we will decompose a   $d$-dimensional  momentum $\bar \ell$ as follows
\begin{align}
\bar \ell = \ell + \tilde \ell \, , \qquad \bar \ell^2  = \ell^2 -\mu^2 = m^2  \,  ,
\label{Eq:Dec0}
\end{align}
while its four-dimensional component $\ell$ will be expressed as 
\begin{align}
\ell = \ell^\flat + \hat q_\ell \, , \qquad \hat q_\ell \equiv \frac{m^2+\mu^2 }{2\, \ell \cdot q_\ell} q_\ell  \,  ,
\label{Eq:Dec}
\end{align}
in terms of the two massless   momenta $\ell^\flat$ and $q_\ell$.
\medskip

\paragraph*{Spinors --}
The spinors of a $d$-dimensional fermion have to fulfill a completeness relation which
reconstructs the numerator of the cut  propagator,
\begin{align}
\sum_{\lambda=1}^{2^{(d_s-2)/2}}  u_{\lambda, \, (d)}\left(\bar \ell \right)\bar{u}_{\lambda, \, (d)}\left(\bar \ell \right) & =\bar{ \slashed \ell}+m \, , \nn
\sum_{\lambda=1}^{2^{(d_s-2)/2}}  v_{\lambda, \, (d)}\left(\bar \ell \right)\bar{v}_{\lambda, \, (d)}\left(\bar \ell \right) & =\bar{ \slashed \ell}-m \, .
\label{Eq:CompFD}
\end{align}
The  substitutions~(\ref{Eq:SubF}) allow one to express  Eq.~(\ref{Eq:CompFD}) as follows:
\begin{align}
\sum_{\lambda=\pm}u_{\lambda}\left(\ell \right)\bar{u}_{\lambda}\left(\ell \right) & = \slashed \ell + i \mu \gamma^5 + m \, , \nn
\sum_{\lambda=\pm}v_{\lambda}\left(\ell  \right)\bar{v}_{\lambda}\left(\ell \right)  & = \slashed \ell + i \mu \gamma^5  - m \, .
\label{Eq:CompF4}
\end{align}
As shown in the Appendix~\ref{App:Completeness}, the generalized massive spinors
\begin{subequations}
\begin{align}
u_{+}\left(\ell \right) & =\left| \ell^{\flat}\right\rangle +\frac{\left(m - i\mu\right)}{\left[ \ell^{\flat}\,q_\ell \right]}\left|q_\ell \right],\nn
u_{-}\left(\ell \right) & =\left| \ell^{\flat}\right]+\frac{\left(m  +  i\mu\right)}{\left\langle \ell^{\flat}\, q_\ell \right\rangle }\left|q_\ell \right\rangle , \nn
v_{-}\left(\ell \right) & =\left| \ell^{\flat}\right\rangle -\frac{\left(m  -  i\mu\right)}{\left[ \ell^{\flat}\, q_\ell \right]}\left|q_\ell \right],\nn
\label{Eq:SpinorG}
v_{+}\left(\ell \right) & =\left| \ell^{\flat}\right]-\frac{\left(m +  i\mu\right)}{\left\langle \ell^{\flat}\, q_\ell \right\rangle }\left|q_\ell \right\rangle , \\[3ex]
\bar{u}_{+}\left(\ell \right) & =\left[\ell^{\flat}\right|+\frac{\left(m + i\mu\right)}{\left\langle q_\ell\, \ell^{\flat}\right\rangle }\left\langle q_\ell\right|\, , \nn
\bar{u}_{-}\left(\ell \right) & =\left\langle \ell^{\flat}\right|+\frac{\left(m -  i\mu\right)}{\left[q_\ell\,  \ell^{\flat}\right]}\left[q_\ell\right| \, ,\nn
\bar{v}_{-}\left(\ell \right) & =\left[\ell^{\flat}\right|-\frac{\left(m  +  i\mu\right)}{\left\langle q_\ell  \, \ell^{\flat}\right\rangle }\left\langle q_\ell \right| \, ,  \nn
\bar{v}_{+}\left(\ell \right) & =\left\langle \ell^{\flat}\right|-\frac{\left(m - i\mu\right)}{\left[q_\ell\,  \ell^{\flat}\right]}\left[q_\ell \right| \, ,
\end{align}
\label{Eq:SpinorF}
\end{subequations}
fulfill the completeness relation~(\ref{Eq:CompF4}).  The spinors~(\ref{Eq:SpinorG}) are solutions of the  tachyonic Dirac equations~\cite{PTP.5.14,Leiter,Trzetrzelewski:2011vr,Jentschura:2012vp}
\begin{align}
\left(\slashed \ell + i\mu\gamma^{5} + m \right)\, u_\lambda\left(\ell \right)  &= 0 \, , \nn
\left(\slashed \ell + i\mu\gamma^{5}  - m \right)\, v_\lambda \left(\ell \right) &=0 \, ,
\end{align}
which leads to  a Hermitian Hamiltonian. It is worth to notice that  the spinors~(\ref{Eq:SpinorF})  fulfill the Gordon's identities
\begin{equation}
\frac{\bar u_\lambda\left  (\ell \right) \;  \gamma^\nu  \;  u_\lambda\left  (\ell \right) }{2}  = 
\frac{\bar v_\lambda\left  (\ell \right) \;  \gamma^\nu  \;  v_\lambda\left  (\ell \right) }{2}   = \ell^\nu \, .
\label{Eq:SpinMom}
\end{equation}

\paragraph*{Polarization vectors --}
The $d$-dimensional polarization vectors of a spin-1 particle 
fulfill the following relation
\begin{equation}
\sum_{i=1}^{d -2} \, \varepsilon_{i\, (d)}^\mu\left (\bar \ell , \bar \eta \right )\varepsilon_{i\, (d)}^{\ast \nu}\left (\bar \ell , \bar \eta \right ) = 
- \bar g^{\mu \nu} +\frac{\bar \ell^\mu \, \bar \eta^\nu + \bar \ell^\nu \,\bar \eta^\mu}{\bar \ell \cdot \bar \eta} \,  , 
\label{Eq:CompGD}
\end{equation}
where $\bar \eta$ is an arbitrary $d$-dimensional massless momentum such that $\bar \ell \cdot \bar \eta \neq 0$.  Gauge invariance
in $d$ dimensions guarantees that the cut is independent of  $\bar \eta$. In particular the choice
\begin{equation}
\bar \eta^\mu = \ell^\mu - \tilde \ell^\mu \, ,
\end{equation}
with $\ell$, $\tilde \ell$ defined in Eq.~(\ref{Eq:Dec0}), 
allows one to disentangle the four-dimensional contribution form the $d$-dimensional one:  
\begin{eqnarray}
\sum_{i=1}^{d -2} \, \varepsilon_{i\, (d)}^\mu\left (\bar \ell , \bar \eta \right )\varepsilon_{i\, (d)}^{\ast \nu}\left (\bar \ell , \bar \eta \right ) &=&\left (   - g^{\mu \nu}  +\frac{ \ell^\mu \ell^\nu}{\mu^2} \right) \nonumber \\
&-&\left (  \tilde g^{\mu \nu}  +\frac{ \tilde \ell^\mu \tilde \ell^\nu}{\mu^2} \right ) \, .
\label{Eq:CompGD2}
\end{eqnarray}
The first term is  related to the cut propagator of a massive gluon  and can be expressed as follows
\begin{align}
 -g^{\mu\nu}+\frac{\ell^{\mu}\ell^{\nu}}{\mu^{2}} &=  \sum_{\lambda=\pm,0}\varepsilon_{\lambda}^{\mu}(\ell) \, \varepsilon_{\lambda}^{*\nu}(\ell)  \, ,  \label{flat}
 \end{align}
in terms of the polarization vectors of a vector boson of mass $\mu$~\cite{Mahlon:1998jd},  
\begin{align}
\varepsilon_{+}^{\mu}\left(\ell \right)    &= -\frac{\left[\ell^{\flat}\left|\gamma^{\mu}\right|  \hat q_\ell \right\rangle }{\sqrt{2}\mu} \, ,  &  &
\varepsilon_{-}^{\mu}\left(\ell \right)    = - \frac{\left\langle \ell^{\flat}\left|\gamma^{\mu}\right| \hat q_\ell \right]}{\sqrt{2}\mu}\, ,    \nn      
\varepsilon_{0}^{\mu}\left(\ell   \right)  &=  \phantom{-} \frac{\ell^{\flat\mu}-\hat q_\ell^{\mu}}{\mu} \, .
\label{emass1}
\end{align}
The latter  fulfill the  well-known relations
 \begin{align}
 \varepsilon^2_{\pm}(\ell) & =\phantom{-} 0\, ,  & \varepsilon_{\pm}(\ell)\cdot\varepsilon_{\mp}(\ell)&=-1\, , \nn
 \varepsilon_{0}^2(\ell)& =-1\, , &  \varepsilon_{\pm}(\ell)\cdot\varepsilon_{0}(\ell)   &=\phantom{-} 0\, ,  \nn
  \varepsilon_{\lambda}(\ell) \cdot \ell &=\phantom{-} 0 \, .
  \label{Eq:propEps}
 \end{align}
 The second term of the r.h.s. of Eq.~(\ref{Eq:CompGD2})  is related to the numerator of cut propagator of the scalar $s_g$ and can be expressed in terms
 of the $(-2 \epsilon)$-SRs as:
 \begin{equation}
 \tilde g^{\mu \nu}  +\frac{ \tilde \ell^\mu \tilde \ell^\nu}{\mu^2}  \quad \to  \quad  \GH^{AB} \equiv  \GG^{AB} - \QQ^A \QQ^B  \, .
 \label{Eq:Pref}
\end{equation}
The factor $\GH^{AB}$ can be easily accounted for by defining the cut propagator as
\vspace{-0.4cm}
\begin{equation}
\parbox{20mm}{\unitlength=0.20bp%
\begin{feynartspicture}(300,300)(1,1)
\FADiagram{}
\FAProp(4.,10.)(16.,10.)(0.,){/ScalarDash}{0}
\FALabel(5.5,8.93)[t]{\tiny $a,A$}
\FALabel(14.5,8.93)[t]{\tiny $b, B$}
\FAVert(4.,10.){0}
\FAVert(16.,10.){0}
\FAProp(10.,6.)(10.,14.)(0.,){/GhostDash}{0}
\end{feynartspicture}} =  \hat G^{AB}\,\delta^{ab} \, . 
\label{Eq:Rules2a}
\end{equation}

The generalized four-dimensional spinors and polarization vectors defined above
can be used for constructing tree-level amplitudes with full
$\mu$-dependence. Therefore,  in the context of on-shell and
unitarity-based methods,  they  are a simple alternative to approaches 
introducing  explicit higher-dimensional extension of either  the
Dirac~\cite{Giele:2008ve,Ellis:2008ir} or the spinor~\cite{Cheung:2009dc,Davies:2011vt} algebra.

\section{One-loop amplitudes}
\label{sec:4P}

In the next sections we  apply generalized-unitarity methods  within
FDF to some paradigmatic example of one-loop  $2 \to 2$ scattering amplitudes.
Therefore in this section we present the decomposition of the latter ones in terms
of the MIs. 
\smallskip

First, we consider one-loop four-point amplitudes 
with four outgoing  massless particles 
\begin{align}
0 \to 1(p_1) \, 2(p_2) \, 3(p_3)  \, 4(p_4) \,  ,
\label{Eq:ProM}
\end{align}
where $p_i$ is the momentum of the particle $i$. 
In general, a massless four-point one-loop  amplitude  can be
decomposed in terms MIs, as follows 
\begin{subequations}
\begin{align}
A_4 &{}= \frac{1}{(4 \pi)^{2-\epsilon}}  \bigg [ \,    c_{1|2|3|4;\,0}\, I_{1|2|3|4}+\big(c_{12|3|4;\,0}\, I_{12|3|4}   \nn
&{} \quad  +c_{1|2|34;\,0}\, I_{1|2|34}+c_{1|23|4;\,0}\, I_{1|23|4}+c_{2|3|41;\,0}\, I_{2|3|41}\big)   \nn
&{} \quad +\big(c_{12|34;\,0}\, I_{12|34}+c_{23|41;\,0}\, I_{23|41}\big)\bigg ] +  \mathcal{R}\, , \label{Eq:Deco} \\
\mathcal{R} &{}=    \frac{1}{(4 \pi)^{2-\epsilon}} \bigg [ \, c_{1|2|3|4;\,4}\, I_{1|2|3|4}[\mu^{4}]+\big(c_{12|3|4;\,2}\, I_{12|34}[\mu^{2}]  \nn
&{} \quad  +c_{1|2|34;\,2}\, I_{1|2|34}[\mu^{2}]+c_{1|23|4;\,2}\, I_{1|23|4}[\mu^{2}]  \nn
&{} \quad +c_{2|3|41;\,2}\, I_{2|3|41}[\mu^{2}]  \big ) \nn
&{} \quad+ \big (c_{12|34;\,2}\, I_{12|34}[\mu^{2}]+c_{23|41;\,2}\, I_{23|41}[\mu^{2}]  \big )   \bigg ]  \, . \label{Eq:Ratio} 
\end{align}
\label{Eq:Decomposition}
\end{subequations}
\smallskip

We consider also the process involving three gluons, $1,2,3$,  and a Higgs boson,  $H$,
\begin{align}
0 \to 1(p_1) \, 2(p_2) \, 3(p_3)  \, H(p_H) \,  ,
\label{Eq:ProH}
\end{align}
in the large top-mass limit~\cite{Wilczek:1977zn,Dawson:1990zj}. 
The one-loop amplitude for this process is decomposed as follows,
\begin{subequations}
\begin{align}
A_{4,H}&{}= \frac{1}{\left(4\pi\right)^{2-\epsilon}}\bigg[\,\big(c_{1|2|3|H;\,0}\, I_{1|2|3|H}+c_{1|2|H|3;\,0}\, I_{1|2|H|3}\nn
&{}\quad +c_{1|H|2|3;\,0}\, I_{1|H|2|3}\big)+\big(c_{12|3|H;\,0}\, I_{12|3|H}\nn
&{}\quad +c_{12|H|3;\,0}\, I_{12|H|3}+c_{1|23|H;\,0}\, I_{1|23|H}\nn
&{}\quad +c_{1|H|23;\,0}\, I_{1|H|23}+c_{2|H|31;\,0}\, I_{2|H|31}\nn
&{}\quad +c_{H|2|31;\,0}\, I_{H|2|31}+c_{1|2|3H;\,0}\, I_{1|2|3H}\nn
&{}\quad +c_{1|2H|3;\,0}\, I_{1|2H|3}+c_{1H|2|3;\,0}\, I_{1H|2|3}\big)\nn
&{}\quad +\big(c_{12|3H;\,0}\, I_{12|3H}+c_{23|H1;\,0}\, I_{23|H1}\nn
&{}\quad +c_{H2|31;\,0}\, I_{H2|31} \big) + c_{123 | H ;\,0}\, I_{123|H}
\, \bigg] \, +\mathcal{R}_H\, , \\
\mathcal{R}_H&{}= \frac{1}{\left(4\pi\right)^{2-\epsilon}}\bigg[\,\big(c_{1|2|3|H;\,4}\, I_{1|2|3|H}\left[\mu^4\right] \nn
&{}\quad +c_{1|2|H|3;\,4}\, I_{1|2|H|3}\left[\mu^{4}\right] +c_{1|H|2|3;\,4}\, I_{1|H|2|3}\left[\mu^{4}\right]\big)\nn
&{}\quad + \big(c_{12|3|H;\,2}\, I_{12|3|H}\left[\mu^{2}\right] +c_{12|H|3;\,2}\, I_{12|H|3}\left[\mu^{2}\right]\nn
&{}\quad +c_{1|23|H;\,2}\, I_{1|23|H}\left[\mu^{2}\right] +c_{1|H|23;\,2}\, I_{1|H|23}\left[\mu^{2}\right] \nn
&{}\quad +c_{2|H|31;\,2}\, I_{2|H|31}\left[\mu^{2}\right] +c_{H|2|31;\,2}\, I_{H|2|31}\left[\mu^{2}\right] \nn
&{}\quad +c_{1|2|3H;\,2}\, I_{1|2|3H}\left[\mu^{2}\right] +c_{1|2H|3;\,2}\, I_{1|2H|3}\left[\mu^{2}\right] \nn
&{}\quad +c_{1H|2|3;\,2}\, I_{1H|2|3}\left[\mu^{2}\right]\big) \nn
&{}\quad +\big(c_{12|3H;\,2}\, I_{12|3H}\left[\mu^{2}\right] +c_{23|H1;\,2}\, I_{23|H1}\left[\mu^{2}\right] \nn
&{}\quad +c_{H2|31;\,2}\, I_{H2|31}\left[\mu^{2}\right]
+c_{123 | H ;\,2}\, I_{123|H}\left[\mu^{2}\right]
\big)
\,\bigg] \, , 
\end{align}
\label{Eq:DecompositionH}
\end{subequations}
The expressions for the MIs appearing in
Eq.~(\ref{Eq:Decomposition}) and~(\ref{Eq:DecompositionH}) are given in
Appendix~\ref{App:MI}. 

In Eq.~(\ref{Eq:Decomposition}) and~(\ref{Eq:DecompositionH}), the contribution generating
the rational terms have been collected in ${\cal R}$ and ${\cal R}_H$, respectively, hence
distinguished by the so-called cut-constructible terms.  We remark that  within the FDF this distinction 
is pointless and has been performed only to improve the readability of the formulas. Indeed 
within the FDF the two contributions are computed simultaneously from the same cuts.
\smallskip

The coefficients $c$'s entering in the decompositions~(\ref{Eq:Decomposition}) and~(\ref{Eq:DecompositionH})  can 
be obtained by using the generalized unitarity techniques  for quadruple~\cite{Britto:2004nc,Badger:2008cm}, 
triple~\cite{Mastrolia:2006ki,Forde:2007mi,Badger:2008cm}, and double~\cite{Britto:2005ha, Britto:2006sj, Mastrolia:2009dr} cuts.  We observe that single-cut 
techniques~\cite{Kilgore:2007qr,Britto:2009wz, Britto:2010um} are not needed because of the absence of ($d$-dimensional) massive particles in the loop.  In general, the cut  $C_{i_1\cdots i_k}$, 
 defined by the conditions $D_{i_1} =\cdots = D_{i_k}=0$,  allows for the determination of the coefficients $c_{i_1\cdots i_k; \, n}$.

\section{The $\mathbf{gggg}$ amplitude}
\label{sec:gggg}

As a first example we consider the  four-gluon color-ordered helicity amplitude $A_{4}\left(1_{g}^{+},2_{g}^{+},3_{g}^{+},4_{g}^{+}\right)$. 
The latter vanishes at tree-level, while the one-loop contribution is finite, rational and can be obtained from the 
quadruple cut $C_{1|2|3|4}$~\cite{Bern:1991aq,Kunszt:1993sd,Bern:1993qk, Bern:1995db, Brandhuber:2005jw}.
Therefore the relevant tree-level three-point amplitudes  are the ones involving either 
three gluons or  two scalars and one gluon. The tree-level amplitudes
with two gluons and one scalar  should be included as well but they 
are not needed since their cut diagrams vanish because of the $(-2\epsilon)$-SRs, see the discussion below. The tree-level are computed by using  
the color-ordered Feynman rules collected in Appendix~\ref{App:COFR}. 

The general expression of the three-point all-gluon  amplitude is given by 
\vspace{-0.4cm}
\begin{align}
&\parbox{20mm}{\unitlength=0.20bp%
\begin{feynartspicture}(300,300)(1,1)
\FADiagram{}
\FAProp(3.,10.)(10.,10.)(0.,){/Cycles}{0}
\FALabel(5.3,8.93)[t]{\scriptsize $2^{\lambda_2}$}
\FAProp(16.,15.)(10.,10.)(0.,){/Cycles}{0}
\FALabel(12.2273,13.5749)[br]{\scriptsize $\mathbf{1}^{\lambda_1}$}
\FAProp(16.,5.)(10.,10.)(0.,){/Cycles}{0}
\FALabel(12.8873,5.81315)[tr]{\scriptsize $\mathbf{3}^{\lambda_3}$}
\FAVert(10.,10.){0}
\end{feynartspicture}}
= \frac{ig}{\sqrt{2}}\Big[g^{\mu\nu}\left(\mathbf{1}- 2\right)^{\sigma}+g^{\nu\sigma}\left( 2-\mathbf{3} \right)^{\mu}\nonumber \\[-1 ex]
& +g^{\sigma\mu}\left(\mathbf{3}-\mathbf{1}  \right)^{\nu}\Big ]
\varepsilon_{\mu}^{\lambda_1}\left(\mathbf{1}\right)\varepsilon_{\nu}^{\lambda_2}\left(2,  r_{2} \right)
\varepsilon_{\sigma}^{\lambda_3}\left(\mathbf{3}\right) \, .
\label{Eq:3Point}
\end{align}
Generalized massive momenta, carrying dependence on $\mu$, are  denoted by a bold font,  and the polarization of the particle will be the  superscript of 
the corresponding momentum. The momenta are outgoing,
\begin{equation}
\mathbf{1} + 2 + \mathbf{3} =0 \, ,
\label{Eq:MomCon}
\end{equation}
and  in general  $\hat q_{\mathbf 1}$ and  $\hat q_{\mathbf 3}$ can be chosen to be proportional,
\begin{equation}
\hat q_ {\mathbf 3}  = \xi \, \hat q_{\mathbf 1}   \, .
\label{Eq:Prop}
\end{equation}
Moreover the spinors associated to the momenta ${\mathbf j}^\flat$ and $\hat q_{\mathbf j}$
are such that
\begin{equation}
\langle  {\mathbf j}^\flat |  \hat q_{\mathbf j} \rangle = [  \hat q_{\mathbf j}   |  {\mathbf j}^\flat  ]   = \mu \, , \qquad   {\mathbf j} =  {\mathbf 1},  {\mathbf 3} \, .
\end{equation}
The polarization vector associated to a massless momentum $k$ is defined  as~\cite{Dixon:1996wi}
 \begin{align}
 \varepsilon_+^\mu\left (k, r_k \right ) = \frac{\left\langle  r_k \left |\gamma^{\mu}\right| k \right]}{\sqrt{2} \left \langle r_k\, k\right \rangle},  \; \;
\varepsilon_-^\mu\left (k, r_k \right ) = -\frac{\left [ r_k \left |\gamma^{\mu}\right| k \right \rangle}{\sqrt{2} \left [ r_k\, k \right ] }\, ,  
 \end{align}
in terms of an arbitrary reference spinor $r_k$.   We observe that the amplitude~(\ref{Eq:3Point}) is independent of the choice of  $r_2$. 
The proof proceeds along the lines of a similar proof presented in Ref.~\cite{Badger:2005zh}.  A change in the reference momentum  shifts
the amplitude~(\ref{Eq:3Point}) by an amount proportional to 
\begin{align}
& \left [ g^{\mu\nu}\left(\mathbf{1}- 2\right)^{\sigma}+g^{\nu\sigma}\left( 2-\mathbf{3} \right)^{\mu}  +g^{\sigma\mu}\left(\mathbf{3}-\mathbf{1}  \right)^{\nu}\right ] \nn
& \quad \varepsilon_{\mu}^{\lambda_1}\left(\mathbf{1} \right)  \, 2_{\nu} \, 
\varepsilon_{\sigma}^{\lambda_3}\left(\mathbf{3}\right) \, ,
\end{align}
which vanishes owing to momentum conservation, Eq.~(\ref{Eq:MomCon}),  and to the transversality condition~(\ref{Eq:propEps}).

The  explicit expressions of the polarized  amplitudes in the  FDF are:
\vspace{-0.4cm}
\begin{align}
\parbox{20mm}{\unitlength=0.20bp%
\begin{feynartspicture}(300,300)(1,1)
\FADiagram{}
\FAProp(3.,10.)(10.,10.)(0.,){/Cycles}{0}
\FALabel(5.3,8.93)[t]{\scriptsize $2^{+}$}
\FAProp(16.,15.)(10.,10.)(0.,){/Cycles}{0}
\FALabel(12.2273,13.5749)[br]{\scriptsize $\mathbf{1}^{+}$}
\FAProp(16.,5.)(10.,10.)(0.,){/Cycles}{0}
\FALabel(12.8873,5.81315)[tr]{\scriptsize $\mathbf{3}^{+}$}
\FAVert(10.,10.){0}
\end{feynartspicture}}  ={}&  0 \, , \nonumber \\[-4.ex]
\parbox{20mm}{\unitlength=0.20bp%
\begin{feynartspicture}(300,300)(1,1)
\FADiagram{}
\FAProp(3.,10.)(10.,10.)(0.,){/Cycles}{0}
\FALabel(5.3,8.93)[t]{\scriptsize $2^{+}$}
\FAProp(16.,15.)(10.,10.)(0.,){/Cycles}{0}
\FALabel(12.2273,13.5749)[br]{\scriptsize $\mathbf{1}^{+}$}
\FAProp(16.,5.)(10.,10.)(0.,){/Cycles}{0}
\FALabel(12.8873,5.81315)[tr]{\scriptsize $\mathbf{3}^{-}$}
\FAVert(10.,10.){0}
\end{feynartspicture}}  ={}&   i g \left (  \, \frac{[{\mathbf 1}^\flat|2]   [\hat q_{\mathbf 1} |2] }{\mu} + \frac{\langle
   r_2|{\mathbf 1}|2]}{ \langle
   r_2|2\rangle} \right )  \, , \nonumber \\[-4.ex]
\parbox{20mm}{\unitlength=0.20bp%
\begin{feynartspicture}(300,300)(1,1)
\FADiagram{}
\FAProp(3.,10.)(10.,10.)(0.,){/Cycles}{0}
\FALabel(5.3,8.93)[t]{\scriptsize $2^{+}$}
\FAProp(16.,15.)(10.,10.)(0.,){/Cycles}{0}
\FALabel(12.2273,13.5749)[br]{\scriptsize $\mathbf{1}^{0}$}
\FAProp(16.,5.)(10.,10.)(0.,){/Cycles}{0}
\FALabel(12.8873,5.81315)[tr]{\scriptsize $\mathbf{3}^{+}$}
\FAVert(10.,10.){0}
\end{feynartspicture}}  ={}& 0\, ,  \nonumber \\[-4.ex]
\parbox{20mm}{\unitlength=0.20bp%
\begin{feynartspicture}(300,300)(1,1)
\FADiagram{}
\FAProp(3.,10.)(10.,10.)(0.,){/Cycles}{0}
\FALabel(5.3,8.93)[t]{\scriptsize $2^{+}$}
\FAProp(16.,15.)(10.,10.)(0.,){/Cycles}{0}
\FALabel(12.2273,13.5749)[br]{\scriptsize $\mathbf{1}^{0}$}
\FAProp(16.,5.)(10.,10.)(0.,){/Cycles}{0}
\FALabel(12.8873,5.81315)[tr]{\scriptsize $\mathbf{3}^{-}$}
\FAVert(10.,10.){0}
\end{feynartspicture}}  ={}&
   \frac{\sqrt{2} i g\, [\hat q_{\mathbf 1}|2]^2 }{\mu}
   \; ,
\nonumber \\[-4.ex]
\parbox{20mm}{\unitlength=0.20bp%
\begin{feynartspicture}(300,300)(1,1)
\FADiagram{}
\FAProp(3.,10.)(10.,10.)(0.,){/Cycles}{0}
\FALabel(5.3,8.93)[t]{\scriptsize $2^{+}$}
\FAProp(16.,15.)(10.,10.)(0.,){/Cycles}{0}
\FALabel(12.2273,13.5749)[br]{\scriptsize $\mathbf{1}^{-}$}
\FAProp(16.,5.)(10.,10.)(0.,){/Cycles}{0}
\FALabel(12.8873,5.81315)[tr]{\scriptsize $\mathbf{3}^{-}$}
\FAVert(10.,10.){0}
\end{feynartspicture}}  ={}&
ig\frac{\left[\hat{q}_{\boldsymbol{1}}|2\right]\left[\hat{q}_{\boldsymbol{3}}|2\right]\langle\boldsymbol{1}^{\flat}|\boldsymbol{3}^{\flat}\rangle}{\mu^{2}}
   \; ,
\nonumber \\[-4.ex]
\parbox{20mm}{\unitlength=0.20bp%
\begin{feynartspicture}(300,300)(1,1)
\FADiagram{}
\FAProp(3.,10.)(10.,10.)(0.,){/Cycles}{0}
\FALabel(5.3,8.93)[t]{\scriptsize $2^{+}$}
\FAProp(16.,15.)(10.,10.)(0.,){/Cycles}{0}
\FALabel(12.2273,13.5749)[br]{\scriptsize $\mathbf{1}^{0}$}
\FAProp(16.,5.)(10.,10.)(0.,){/Cycles}{0}
\FALabel(12.8873,5.81315)[tr]{\scriptsize $\mathbf{3}^{0}$}
\FAVert(10.,10.){0}
\end{feynartspicture}}  ={}&
-  ig \,  \frac{ \langle r_2|  {\mathbf 1}|2]}{  \langle r_2|2\rangle}  \Big  \{  1  -
\frac{(1+\xi)}{\xi \, \mu^2} \Big [ (1+\xi)\, \mu^2 \nonumber  \\[-4.ex]
& + \xi\, \langle \hat q_{\mathbf 1} |  2| \hat q_{\mathbf 1}]     \Big ] \Big \}
\, .
\end{align}

The three-point amplitude involving a gluon and two scalars is
\vspace{-0.4cm}
\begin{align}
\parbox{20mm}{\unitlength=0.20bp%
\begin{feynartspicture}(300,300)(1,1)
\FADiagram{}
\FAProp(3.,10.)(10.,10.)(0.,){/Cycles}{0}
\FALabel(5.3,8.93)[t]{\scriptsize $2^{+}$}
\FAProp(16.,15.)(10.,10.)(0.,){/ScalarDash}{0}
\FALabel(12.2273,13.5749)[br]{\scriptsize $\mathbf{1}$}
\FAProp(16.,5.)(10.,10.)(0.,){/ScalarDash}{0}
\FALabel(12.8873,5.81315)[tr]{\scriptsize $\mathbf{3}$}
\FAVert(10.,10.){0}
\end{feynartspicture}} &=   \frac{ig}{\sqrt{2}} \left ( \mathbf{3} - \mathbf{1} \right )^\mu \varepsilon_{\mu}^{\lambda_2}\left(2,  r_{2} \right) \GG^{AB} \nonumber \\[-4.ex]
 &=-  ig \,  \frac{ \langle r_2|  {\mathbf 1}|2]}{  \langle r_2|2\rangle} \GG^{AB} \, .
\end{align}

The tree-level amplitudes computed above can be used in the cut
construction of the one-loop amplitude.

In the FDF, the quadruple-cut  $C_{1|2|3|4}$ and the coefficients $c_{1|2|3|4; \;  n}$ can be decomposed into a sum of five contributions,
\begin{align}
C_{1|2|3|4} = \sum_{i=0}^4 \, C^{[i]}_{1|2|3|4} \, ,  \qquad c_{1|2|3|4;\, n} = \sum_{i=0}^4 \, c^{[i]}_{1|2|3|4; \,n} \, ,
\end{align}
where  $C^{[i]}$ ($c^{[i]}$) is  the contribution to the cut
(coefficient) involving $i$ internal scalars.  
In the picture below, internal lines are understood to be on-shell.
The quadruple cuts  read as follows 
\vspace{-0.2cm}
\begin{subequations} 
\begin{align}
C^{[0]}_{1|2|3|4} &{}= 
      \parbox{25mm}{\unitlength=0.24bp%
\begin{feynartspicture}(300,300)(1,1)
\FADiagram{}
%
\FAProp(4.,15.)(6.5,13.5)(0.,){/Cycles}{0}
\FAProp(4.,5.)(6.5,6.5)(0.,){/Cycles}{0}
\FAProp(16.,15.)(13.5,13.5)(0.,){/Cycles}{0}
\FAProp(16.,5.)(13.5,6.5)(0.,){/Cycles}{0}
%
\FAProp(6.5,13.5)(6.5,6.5)(0.,){/Cycles}{0}
\FAProp(13.5,13.5)(6.5,13.5)(0.,){/Cycles}{0}
\FAProp(6.5,6.5)(13.5,6.5)(0.,){/Cycles}{0}
\FAProp(13.5,6.5)(13.5,13.5)(0.,){/Cycles}{0}
%
\FAVert(6.5,13.5){0}
\FAVert(6.5,6.5){0}
\FAVert(13.5,13.5){0}
\FAVert(13.5,6.5){0}
%
%
\FALabel(11.1,4.5)[]{\tiny $+$}
\FALabel(8.9,4.5)[]{\tiny $-$}
\FALabel(4.5, 8.9)[]{\tiny $+$}
\FALabel(4.5, 11.1)[]{\tiny $-$}
\FALabel(8.9, 15.)[]{\tiny $+$}
\FALabel(11.1, 15.)[]{\tiny $-$}
\FALabel(15., 11.1)[]{\tiny $+$}
\FALabel(15., 8.9)[]{\tiny $-$}
%
%
\FALabel(3.,4.)[]{\tiny$1^+$}
\FALabel(3.,16.)[]{\tiny$2^+$}
\FALabel(17.,16.)[]{\tiny$3^+$}
\FALabel(17.,4.)[]{\tiny$4^+$}
\end{feynartspicture}}
    +  \parbox{25mm}{\unitlength=0.24bp%
\begin{feynartspicture}(300,300)(1,1)
\FADiagram{}
%
\FAProp(4.,15.)(6.5,13.5)(0.,){/Cycles}{0}
\FAProp(4.,5.)(6.5,6.5)(0.,){/Cycles}{0}
\FAProp(16.,15.)(13.5,13.5)(0.,){/Cycles}{0}
\FAProp(16.,5.)(13.5,6.5)(0.,){/Cycles}{0}
%
\FAProp(6.5,13.5)(6.5,6.5)(0.,){/Cycles}{0}
\FAProp(13.5,13.5)(6.5,13.5)(0.,){/Cycles}{0}
\FAProp(6.5,6.5)(13.5,6.5)(0.,){/Cycles}{0}
\FAProp(13.5,6.5)(13.5,13.5)(0.,){/Cycles}{0}
%
\FAVert(6.5,13.5){0}
\FAVert(6.5,6.5){0}
\FAVert(13.5,13.5){0}
\FAVert(13.5,6.5){0}
%
%
\FALabel(11.1,4.5)[]{\tiny $-$}
\FALabel(8.9,4.5)[]{\tiny $+$}
\FALabel(4.5, 8.9)[]{\tiny $-$}
\FALabel(4.5, 11.1)[]{\tiny $+$}
\FALabel(8.9, 15.)[]{\tiny $-$}
\FALabel(11.1, 15.)[]{\tiny $+$}
\FALabel(15., 11.1)[]{\tiny $-$}
\FALabel(15., 8.9)[]{\tiny $+$}
%
%
\FALabel(3.,4.)[]{\tiny$1^+$}
\FALabel(3.,16.)[]{\tiny$2^+$}
\FALabel(17.,16.)[]{\tiny$3^+$}
\FALabel(17.,4.)[]{\tiny$4^+$}
\end{feynartspicture}} \nn [-5.5ex]
& +  \parbox{25mm}{\unitlength=0.24bp%
\begin{feynartspicture}(300,300)(1,1)
\FADiagram{}
%
\FAProp(4.,15.)(6.5,13.5)(0.,){/Cycles}{0}
\FAProp(4.,5.)(6.5,6.5)(0.,){/Cycles}{0}
\FAProp(16.,15.)(13.5,13.5)(0.,){/Cycles}{0}
\FAProp(16.,5.)(13.5,6.5)(0.,){/Cycles}{0}
%
\FAProp(6.5,13.5)(6.5,6.5)(0.,){/Cycles}{0}
\FAProp(13.5,13.5)(6.5,13.5)(0.,){/Cycles}{0}
\FAProp(6.5,6.5)(13.5,6.5)(0.,){/Cycles}{0}
\FAProp(13.5,6.5)(13.5,13.5)(0.,){/Cycles}{0}
%
\FAVert(6.5,13.5){0}
\FAVert(6.5,6.5){0}
\FAVert(13.5,13.5){0}
\FAVert(13.5,6.5){0}
%
%
\FALabel(11.1,4.5)[]{\tiny $0$}
\FALabel(8.9,4.5)[]{\tiny $0$}
\FALabel(4.5, 8.9)[]{\tiny $0$}
\FALabel(4.5, 11.1)[]{\tiny $0$}
\FALabel(8.9, 15.)[]{\tiny $0$}
\FALabel(11.1, 15.)[]{\tiny $0$}
\FALabel(15., 11.1)[]{\tiny $0$}
\FALabel(15., 8.9)[]{\tiny $0$}
%
%
\FALabel(3.,4.)[]{\tiny$1^+$}
\FALabel(3.,16.)[]{\tiny$2^+$}
\FALabel(17.,16.)[]{\tiny$3^+$}
\FALabel(17.,4.)[]{\tiny$4^+$}
\end{feynartspicture}}  
\, ,  \label{Will4g} \\[-5.ex] 
C^{[1]}_{1|2|3|4} &{}= \sum_{h_i = \pm, 0}  \,  \mathcal{T}_1  \parbox{25mm}{\unitlength=0.24bp%
\begin{feynartspicture}(300,300)(1,1)
\FADiagram{}
%
\FAProp(4.,15.)(6.5,13.5)(0.,){/Cycles}{0}
\FAProp(4.,5.)(6.5,6.5)(0.,){/Cycles}{0}
\FAProp(16.,15.)(13.5,13.5)(0.,){/Cycles}{0}
\FAProp(16.,5.)(13.5,6.5)(0.,){/Cycles}{0}
%
\FAProp(6.5,13.5)(6.5,6.5)(0.,){/Cycles}{0}
\FAProp(13.5,13.5)(6.5,13.5)(0.,){/Cycles}{0}
\FAProp(6.5,6.5)(13.5,6.5)(0.,){/Cycles}{0}
\FAProp(13.5,6.5)(13.5,13.5)(0.,){/ScalarDash}{0}
%
\FAVert(6.5,13.5){0}
\FAVert(6.5,6.5){0}
\FAVert(13.5,13.5){0}
\FAVert(13.5,6.5){0}
%
%
\FALabel(10.0,4.5)[]{\tiny $-h_1 h_1$}
\FALabel(4.1, 8.9)[]{\tiny $h_2$}
\FALabel(4.1, 11.1)[]{\tiny $-h_2$}
\FALabel(10.0, 15.)[]{\tiny $-h_3  h_3$}
%
%
\FALabel(3.,4.)[]{\tiny$1^+$}
\FALabel(3.,16.)[]{\tiny$2^+$}
\FALabel(17.,16.)[]{\tiny$3^+$}
\FALabel(17.,4.)[]{\tiny$4^+$}
\end{feynartspicture}} +  \cp     \, ,  \label{Eq:Scal1} \\[-5.ex]  
C^{[2]}_{1|2|3|4} &{}=  \sum_{h_i = \pm, 0}  \,  \mathcal{T}^2_1  \parbox{25mm}{\unitlength=0.24bp%
\begin{feynartspicture}(300,300)(1,1)
\FADiagram{}
%
\FAProp(4.,15.)(6.5,13.5)(0.,){/Cycles}{0}
\FAProp(4.,5.)(6.5,6.5)(0.,){/Cycles}{0}
\FAProp(16.,15.)(13.5,13.5)(0.,){/Cycles}{0}
\FAProp(16.,5.)(13.5,6.5)(0.,){/Cycles}{0}
%
\FAProp(6.5,13.5)(6.5,6.5)(0.,){/ScalarDash}{0}
\FAProp(13.5,13.5)(6.5,13.5)(0.,){/Cycles}{0}
\FAProp(6.5,6.5)(13.5,6.5)(0.,){/Cycles}{0}
\FAProp(13.5,6.5)(13.5,13.5)(0.,){/ScalarDash}{0}
%
\FAVert(6.5,13.5){0}
\FAVert(6.5,6.5){0}
\FAVert(13.5,13.5){0}
\FAVert(13.5,6.5){0}
%
%
\FALabel(10.0,4.5)[]{\tiny $-h_1 h_1$}
\FALabel(10.0, 15.)[]{\tiny $-h_2  h_2$}
%
%
\FALabel(3.,4.)[]{\tiny$1^+$}
\FALabel(3.,16.)[]{\tiny$2^+$}
\FALabel(17.,16.)[]{\tiny$3^+$}
\FALabel(17.,4.)[]{\tiny$4^+$}
\end{feynartspicture}}  \nn[-5.5ex]
& \qquad  +   \mathcal{T}_2 \parbox{25mm}{\unitlength=0.24bp%
\begin{feynartspicture}(300,300)(1,1)
\FADiagram{}
%
\FAProp(4.,15.)(6.5,13.5)(0.,){/Cycles}{0}
\FAProp(4.,5.)(6.5,6.5)(0.,){/Cycles}{0}
\FAProp(16.,15.)(13.5,13.5)(0.,){/Cycles}{0}
\FAProp(16.,5.)(13.5,6.5)(0.,){/Cycles}{0}
%
\FAProp(6.5,13.5)(6.5,6.5)(0.,){/Cycles}{0}
\FAProp(13.5,13.5)(6.5,13.5)(0.,){/ScalarDash}{0}
\FAProp(6.5,6.5)(13.5,6.5)(0.,){/Cycles}{0}
\FAProp(13.5,6.5)(13.5,13.5)(0.,){/ScalarDash}{0}
%
\FAVert(6.5,13.5){0}
\FAVert(6.5,6.5){0}
\FAVert(13.5,13.5){0}
\FAVert(13.5,6.5){0}
%
%
\FALabel(10.0,4.5)[]{\tiny $-h_1 h_1$}
\FALabel(4.1, 8.9)[]{\tiny $h_2$}
\FALabel(4.1, 11.1)[]{\tiny $-h_2$}
%
%
\FALabel(3.,4.)[]{\tiny$1^+$}
\FALabel(3.,16.)[]{\tiny$2^+$}
\FALabel(17.,16.)[]{\tiny$3^+$}
\FALabel(17.,4.)[]{\tiny$4^+$}
\end{feynartspicture}}  + \cp   \, ,  \label{Eq:Scal2}   \\[-5.ex]  
C^{[3]}_{1|2|3|4} &{}=  \sum_{h_1 = \pm, 0}  \,    \mathcal{T}_3  \parbox{25mm}{\unitlength=0.24bp%
\begin{feynartspicture}(300,300)(1,1)
\FADiagram{}
%
\FAProp(4.,15.)(6.5,13.5)(0.,){/Cycles}{0}
\FAProp(4.,5.)(6.5,6.5)(0.,){/Cycles}{0}
\FAProp(16.,15.)(13.5,13.5)(0.,){/Cycles}{0}
\FAProp(16.,5.)(13.5,6.5)(0.,){/Cycles}{0}
%
\FAProp(6.5,13.5)(6.5,6.5)(0.,){/ScalarDash}{0}
\FAProp(13.5,13.5)(6.5,13.5)(0.,){/ScalarDash}{0}
\FAProp(6.5,6.5)(13.5,6.5)(0.,){/Cycles}{0}
\FAProp(13.5,6.5)(13.5,13.5)(0.,){/ScalarDash}{0}
%
\FAVert(6.5,13.5){0}
\FAVert(6.5,6.5){0}
\FAVert(13.5,13.5){0}
\FAVert(13.5,6.5){0}
%
%
\FALabel(10.0,4.5)[]{\tiny $-h_1 h_1$}
%
%
\FALabel(3.,4.)[]{\tiny$1^+$}
\FALabel(3.,16.)[]{\tiny$2^+$}
\FALabel(17.,16.)[]{\tiny$3^+$}
\FALabel(17.,4.)[]{\tiny$4^+$}
\end{feynartspicture}}  + \cp   \, ,  \label{Eq:Scal3}  \\[-5.ex]
C^{[4]}_{1|2|3|4} &{}=   \mathcal{T}_4   \parbox{25mm}{\unitlength=0.24bp%
\begin{feynartspicture}(300,300)(1,1)
\FADiagram{}
%
\FAProp(4.,15.)(6.5,13.5)(0.,){/Cycles}{0}
\FAProp(4.,5.)(6.5,6.5)(0.,){/Cycles}{0}
\FAProp(16.,15.)(13.5,13.5)(0.,){/Cycles}{0}
\FAProp(16.,5.)(13.5,6.5)(0.,){/Cycles}{0}
%
\FAProp(6.5,13.5)(6.5,6.5)(0.,){/ScalarDash}{0}
\FAProp(13.5,13.5)(6.5,13.5)(0.,){/ScalarDash}{0}
\FAProp(6.5,6.5)(13.5,6.5)(0.,){/ScalarDash}{0}
\FAProp(13.5,13.5)(13.5,6.5)(0.,){/ScalarDash}{0}
%
\FAVert(6.5,13.5){0}
\FAVert(6.5,6.5){0}
\FAVert(13.5,13.5){0}
\FAVert(13.5,6.5){0}
%
%
%
%
\FALabel(3.,4.)[]{\tiny$1^+$}
\FALabel(3.,16.)[]{\tiny$2^+$}
\FALabel(17.,16.)[]{\tiny$3^+$}
\FALabel(17.,4.)[]{\tiny$4^+$}
\end{feynartspicture}}
\, , 
\label{Eq:AllScalars} 
\end{align}
\label{Eq:AllCut}
\end{subequations}
where  the abbreviation ``c.p."  means ``cyclic permutations of the external particles''.  In Eqs.~(\ref{Eq:AllCut}) , the $(-2\epsilon)$-SR
have been stripped off and collected in the prefactors $\mathcal{T}_i$,
\begin{align}
&\mathcal{T}_1 &{}={}& \QQ^A\GH^{AB} \QQ^B &{}={}& \phantom{-} 0 \,  ,  \nn
&\mathcal{T}_2 &{}={}& \QQ^A \GH^{AB} \GG^{BC}  \GH^{CD} \QQ^D&{}={}& \phantom{-}  0 \,  , \nn
&\mathcal{T}_3 &{}={}& \QQ^A \GH^{AB} \GG^{BC}  \GH^{CD} \GG^{DE} \GH^{EF} \QQ^F&{}={}& \phantom{-} 0  \, , \nn
&\mathcal{T}_4 &{}={}& \mbox{tr} \left (  \GG \,  \GH \,  \GG \,   \GH \,  \GG \,  \GH \,  \GG \,  \GH \right ) &{}={}& - 1 \, .
\end{align}
The prefactors  $\mathcal{T}_1, \ldots , \mathcal{T}_3$ force the cuts~(\ref{Eq:Scal1}) - (\ref{Eq:Scal3}) to vanish identically.
The only cuts contributing, Eqs.~(\ref{Will4g}) and~(\ref{Eq:AllScalars}), lead to the following  coefficients
\begin{align}
c^{[0]}_{1|2|3|4; \; 0} &=0  \, ,
& c^{[0]}_{1|2|3|4; \; 4} &=3 i\frac{\left[12\right]\left[34\right]}{\left\langle 12\right\rangle \left\langle 34\right\rangle } \,  ,   \nn
c^{[4]}_{1|2|3|4; \; 0} &=0 \, ,
& c^{[4]}_{1|2|3|4; \; 4} &= - i\frac{\left[12\right]\left[34\right]}{\left\langle 12\right\rangle \left\langle 34\right\rangle }  \,  .
\end{align}
Therefore the only non-vanishing   coefficient, $c_{1|2|3|4; \; 4}$,    is   
\begin{align}
c_{1|2|3|4; \; 4} &{}= c^{[0]}_{1|2|3|4; \; 4} + c^{[4]}_{1|2|3|4; \; 4} =
2i\frac{\left[12\right]\left[34\right]}{\left\langle 12\right\rangle \left\langle 34\right\rangle } \, .
\end{align}
The color-ordered one-loop amplitude can be obtained from
Eq.~(\ref{Eq:Decomposition}), which in this simple case reduces to 
\begin{eqnarray}
A_{4}\left(1_{g}^{+},2_{g}^{+},3_{g}^{+},4_{g}^{+}\right) &=&
c_{1|2|3|4; \; 4} \ I_{1|2|3|4}[\mu^4]
 \nonumber \\ 
&=& 
 -\frac{i}{48 \, \pi^2} \, \frac{\left[12\right]\left[34\right]}{\left\langle 12\right\rangle \left\langle 34\right\rangle } \,  ,
\label{Eq:Res4G}
\end{eqnarray}
and is  in agreement with the literature~\cite{Kunszt:1993sd}. This example clearly shows the difference between
our computation and the one based on the supersymmetric decomposition~\cite{Brandhuber:2005jw}.  In the  latter one,
the result is uniquely  originated by the complex scalar contribution. Instead  in our procedure 
the result arises from both  the massive gluons and the massive scalars $s_g$.

\section{The $\mathbf{gg q\bar q}$ amplitude}
\label{sec:ggqq}

In this section we apply  generalized-unitarity methods  within the FDF scheme to 
a more involved  $2 \to 2$ process. In particular we show the calculation of the leading color one-loop contribution to the helicity amplitude  
$A_{4}\left(1_{g}^{-},2_{g}^{+},3_{\bar{q}}^{-},4_{q}^{+}\right)$, which at tree-level reads,
\begin{align}
A_{4}^{\text{tree}}&=-i\frac{\left\langle 13\right\rangle ^{3}\left\langle 14\right\rangle }{\left\langle 12\right\rangle \left\langle 23\right\rangle \left\langle 34\right\rangle \left\langle 41\right\rangle }  \, .
\end{align}
The leading-color contribution to a one-loop
amplitude with $n$ particles and two external fermions can be decomposed in terms of primitive amplitudes~\cite{Bern:1994fz}. Following the notation of Ref.~\cite{Davies:2011vt}, we have 
\begin{align}
A_{4}^{1\text{ loop}}  &= A_{4}^{\text{L}}  -\frac{1}{N_c^2}    A_{4}^{\text{R}}  +\frac{N_f}{N_c}  A_{4}^{\text{L}, [1/2]}   + \frac{N_s}{N_c}  A_{4}^{\text{L}, [0]} \, , 
\end{align}
where $N_c$ is the number of colors while $N_f$  ($N_s$) the number of fermions (scalars). For the helicity configuration we consider both $ A_{4}^{\text{L}, [1/2]}$ and
$A_{4}^{\text{L}, [0]}$ vanish, thus we will only  focus on  the contributions of the left-turning amplitude  $A_{4}^{\text{L}} $ and on the right-turning one,  $A_{4}^{\text{R}} $.
The Feynman diagrams leading to the relevant tree-level amplitudes are shown in Appendix~\ref{App:TREE}. They are computed by using the color-ordered Feynman rules
collected in Appendix~\ref{App:COFR}.
\medskip

\paragraph*{Left-turning amplitude --}
The quadruple cut is given by 
\vspace{-0.4cm}
\begin{align}
C^{\mbox{\tiny [L]}}_{1|2|3|4} &{}= 
      \parbox{20mm}{\unitlength=0.20bp%
\begin{feynartspicture}(300,300)(1,1)
\FADiagram{}
%
\FAProp(4.,15.)(6.5,13.5)(0.,){/Cycles}{0}
\FAProp(4.,5.)(6.5,6.5)(0.,){/Cycles}{0}
\FAProp(16.,15.)(13.5,13.5)(0.,){/Straight}{1}
\FAProp(16.,5.)(13.5,6.5)(0.,){/Straight}{-1}
%
\FAProp(6.5,13.5)(6.5,6.5)(0.,){/Cycles}{0}
\FAProp(13.5,13.5)(6.5,13.5)(0.,){/Cycles}{0}
\FAProp(6.5,6.5)(13.5,6.5)(0.,){/Cycles}{0}
\FAProp(13.5,13.5)(13.5,6.5)(0.,){/Straight}{1}
%
\FAVert(6.5,13.5){0}
\FAVert(6.5,6.5){0}
\FAVert(13.5,13.5){0}
\FAVert(13.5,6.5){0}
%
%
\FALabel(11.1,4.5)[]{\tiny $+$}
\FALabel(8.9,4.5)[]{\tiny $-$}
\FALabel(4.5, 8.9)[]{\tiny $+$}
\FALabel(4.5, 11.1)[]{\tiny $-$}
\FALabel(8.9, 15.)[]{\tiny $+$}
\FALabel(11.1, 15.)[]{\tiny $-$}
\FALabel(15., 11.1)[]{\tiny $\pm$}
\FALabel(15., 8.9)[]{\tiny $\mp$}
%
%
\FALabel(3.,4.)[]{\tiny$1$}
\FALabel(3.,16.)[]{\tiny$2$}
\FALabel(17.,16.)[]{\tiny$3$}
\FALabel(17.,4.)[]{\tiny$4$}
\end{feynartspicture}}
   + \parbox{20mm}{\unitlength=0.20bp%
\begin{feynartspicture}(300,300)(1,1)
\FADiagram{}
%
\FAProp(4.,15.)(6.5,13.5)(0.,){/Cycles}{0}
\FAProp(4.,5.)(6.5,6.5)(0.,){/Cycles}{0}
\FAProp(16.,15.)(13.5,13.5)(0.,){/Straight}{1}
\FAProp(16.,5.)(13.5,6.5)(0.,){/Straight}{-1}
%
\FAProp(6.5,13.5)(6.5,6.5)(0.,){/Cycles}{0}
\FAProp(13.5,13.5)(6.5,13.5)(0.,){/Cycles}{0}
\FAProp(6.5,6.5)(13.5,6.5)(0.,){/Cycles}{0}
\FAProp(13.5,13.5)(13.5,6.5)(0.,){/Straight}{1}
%
\FAVert(6.5,13.5){0}
\FAVert(6.5,6.5){0}
\FAVert(13.5,13.5){0}
\FAVert(13.5,6.5){0}
%
%
\FALabel(11.1,4.5)[]{\tiny $-$}
\FALabel(8.9,4.5)[]{\tiny $+$}
\FALabel(4.5, 8.9)[]{\tiny $-$}
\FALabel(4.5, 11.1)[]{\tiny $+$}
\FALabel(8.9, 15.)[]{\tiny $-$}
\FALabel(11.1, 15.)[]{\tiny $+$}
\FALabel(15., 11.1)[]{\tiny $\pm$}
\FALabel(15., 8.9)[]{\tiny $\mp$}
%
%
\FALabel(3.,4.)[]{\tiny$1$}
\FALabel(3.,16.)[]{\tiny$2$}
\FALabel(17.,16.)[]{\tiny$3$}
\FALabel(17.,4.)[]{\tiny$4$}
\end{feynartspicture}}    \nn[-5.5ex]
&{}+\parbox{20mm}{\unitlength=0.20bp%
\begin{feynartspicture}(300,300)(1,1)
\FADiagram{}
%
\FAProp(4.,15.)(6.5,13.5)(0.,){/Cycles}{0}
\FAProp(4.,5.)(6.5,6.5)(0.,){/Cycles}{0}
\FAProp(16.,15.)(13.5,13.5)(0.,){/Straight}{1}
\FAProp(16.,5.)(13.5,6.5)(0.,){/Straight}{-1}
%
\FAProp(6.5,13.5)(6.5,6.5)(0.,){/Cycles}{0}
\FAProp(13.5,13.5)(6.5,13.5)(0.,){/Cycles}{0}
\FAProp(6.5,6.5)(13.5,6.5)(0.,){/Cycles}{0}
\FAProp(13.5,13.5)(13.5,6.5)(0.,){/Straight}{1}
%
\FAVert(6.5,13.5){0}
\FAVert(6.5,6.5){0}
\FAVert(13.5,13.5){0}
\FAVert(13.5,6.5){0}
%
%
\FALabel(11.1,4.5)[]{\tiny $0$}
\FALabel(8.9,4.5)[]{\tiny $0$}
\FALabel(4.5, 8.9)[]{\tiny $0$}
\FALabel(4.5, 11.1)[]{\tiny $0$}
\FALabel(8.9, 15.)[]{\tiny $0$}
\FALabel(11.1, 15.)[]{\tiny $0$}
\FALabel(15., 11.1)[]{\tiny $\pm$}
\FALabel(15., 8.9)[]{\tiny $\mp$}
%
%
\FALabel(3.,4.)[]{\tiny$1$}
\FALabel(3.,16.)[]{\tiny$2$}
\FALabel(17.,16.)[]{\tiny$3$}
\FALabel(17.,4.)[]{\tiny$4$}
\end{feynartspicture}}
   + \parbox{20mm}{\unitlength=0.20bp%
\begin{feynartspicture}(300,300)(1,1)
\FADiagram{}
%
\FAProp(4.,15.)(6.5,13.5)(0.,){/Cycles}{0}
\FAProp(4.,5.)(6.5,6.5)(0.,){/Cycles}{0}
\FAProp(16.,15.)(13.5,13.5)(0.,){/Straight}{1}
\FAProp(16.,5.)(13.5,6.5)(0.,){/Straight}{-1}
%
\FAProp(6.5,13.5)(6.5,6.5)(0.,){/ScalarDash}{0}
\FAProp(13.5,13.5)(6.5,13.5)(0.,){/ScalarDash}{0}
\FAProp(6.5,6.5)(13.5,6.5)(0.,){/ScalarDash}{0}
\FAProp(13.5,13.5)(13.5,6.5)(0.,){/Straight}{1}
%
\FAVert(6.5,13.5){0}
\FAVert(6.5,6.5){0}
\FAVert(13.5,13.5){0}
\FAVert(13.5,6.5){0}
%
%
%
%
\FALabel(3.,4.)[]{\tiny$1$}
\FALabel(3.,16.)[]{\tiny$2$}
\FALabel(17.,16.)[]{\tiny$3$}
\FALabel(17.,4.)[]{\tiny$4$}
\end{feynartspicture}}    \, , \nn[-2ex]
c^{\mbox{\tiny [L]}}_{1|2|3|4; \; 0} &{}=  \frac{1}{2}A_{4}^{\text{tree}}\left(1-\frac{s_{14}^{3}}{s_{13}^{3}}\right)s_{12}s_{14}   \, , \nn
c^{\mbox{\tiny [L]}}_{1|2|3|4; \; 4}  &{}= 0 \, .
\end{align}
The first two cut diagrams  contribute to both the  cut-constructible  and to the rational part, while the last two cut diagrams cancel against each other.

The triple cuts are given by  
\vspace{-0.4cm}
\begin{subequations}
\begin{align}
C^{\mbox{\tiny [L]}}_{12|3|4} &{}=
      \parbox{20mm}{\unitlength=0.20bp%
\begin{feynartspicture}(300,300)(1,1)
\FADiagram{}
%
\FAProp(4.,11.5)(6.5,10.0)(0.,){/Cycles}{0}
\FAProp(6.5,10.0)(4.,8.5)(0.,){/Cycles}{0}
\FAProp(16.,15.)(13.5,13.5)(0.,){/Straight}{1}
\FAProp(16.,5.)(13.5,6.5)(0.,){/Straight}{-1}
%
\FAProp(13.5,13.5)(6.5,10.0)(0.,){/Cycles}{0}
\FAProp(6.5,10.0)(13.5,6.5)(0.,){/Cycles}{0}
\FAProp(13.5,13.5)(13.5,6.5)(0.,){/Straight}{1}
%
%
\FAVert(13.5,13.5){0}
\FAVert(13.5,6.5){0}
\FAVert(6.5,10.){0}
%
%
\FALabel(10.5667, 6.4667)[]{\tiny $-$}
\FALabel(8.23333, 7.6333)[]{\tiny $+$}
\FALabel(8.23333, 12.3667)[]{\tiny $-$}
\FALabel(10.5667, 13.5333)[]{\tiny $+$}
\FALabel(15., 11.1)[]{\tiny $\pm$}
\FALabel(15., 8.9)[]{\tiny $\mp$}
%
%
\FALabel(3.1,8.)[]{\tiny$1$}
\FALabel(3.1,12.)[]{\tiny$2$}
\FALabel(17.,16.)[]{\tiny$3$}
\FALabel(17.,4.)[]{\tiny$4$}
\end{feynartspicture}}
   + \parbox{20mm}{\unitlength=0.20bp%
\begin{feynartspicture}(300,300)(1,1)
\FADiagram{}
%
\FAProp(4.,11.5)(6.5,10.0)(0.,){/Cycles}{0}
\FAProp(6.5,10.0)(4.,8.5)(0.,){/Cycles}{0}
\FAProp(16.,15.)(13.5,13.5)(0.,){/Straight}{1}
\FAProp(16.,5.)(13.5,6.5)(0.,){/Straight}{-1}
%
\FAProp(13.5,13.5)(6.5,10.0)(0.,){/Cycles}{0}
\FAProp(6.5,10.0)(13.5,6.5)(0.,){/Cycles}{0}
\FAProp(13.5,13.5)(13.5,6.5)(0.,){/Straight}{1}
%
%
\FAVert(13.5,13.5){0}
\FAVert(13.5,6.5){0}
\FAVert(6.5,10.){0}
%
%
\FALabel(10.5667, 6.4667)[]{\tiny $+$}
\FALabel(8.23333, 7.6333)[]{\tiny $-$}
\FALabel(8.23333, 12.3667)[]{\tiny $+$}
\FALabel(10.5667, 13.5333)[]{\tiny $-$}
\FALabel(15., 11.1)[]{\tiny $\pm$}
\FALabel(15., 8.9)[]{\tiny $\mp$}
%
%
\FALabel(3.1,8.)[]{\tiny$1$}
\FALabel(3.1,12.)[]{\tiny$2$}
\FALabel(17.,16.)[]{\tiny$3$}
\FALabel(17.,4.)[]{\tiny$4$}
\end{feynartspicture}}
   + \parbox{20mm}{\unitlength=0.20bp%
\begin{feynartspicture}(300,300)(1,1)
\FADiagram{}
%
\FAProp(4.,11.5)(6.5,10.0)(0.,){/Cycles}{0}
\FAProp(6.5,10.0)(4.,8.5)(0.,){/Cycles}{0}
\FAProp(16.,15.)(13.5,13.5)(0.,){/Straight}{1}
\FAProp(16.,5.)(13.5,6.5)(0.,){/Straight}{-1}
%
\FAProp(13.5,13.5)(6.5,10.0)(0.,){/Cycles}{0}
\FAProp(6.5,10.0)(13.5,6.5)(0.,){/Cycles}{0}
\FAProp(13.5,13.5)(13.5,6.5)(0.,){/Straight}{1}
%
%
\FAVert(13.5,13.5){0}
\FAVert(13.5,6.5){0}
\FAVert(6.5,10.){0}
%
%
\FALabel(10.5667, 6.4667)[]{\tiny $0$}
\FALabel(8.23333, 7.6333)[]{\tiny $0$}
\FALabel(8.23333, 12.3667)[]{\tiny $+$}
\FALabel(10.5667, 13.5333)[]{\tiny $-$}
\FALabel(15., 11.1)[]{\tiny $\pm$}
\FALabel(15., 8.9)[]{\tiny $\mp$}
%
%
\FALabel(3.1,8.)[]{\tiny$1$}
\FALabel(3.1,12.)[]{\tiny$2$}
\FALabel(17.,16.)[]{\tiny$3$}
\FALabel(17.,4.)[]{\tiny$4$}
\end{feynartspicture}}  \nn[-5.5ex]
&{}+\parbox{20mm}{\unitlength=0.20bp%
\begin{feynartspicture}(300,300)(1,1)
\FADiagram{}
%
\FAProp(4.,11.5)(6.5,10.0)(0.,){/Cycles}{0}
\FAProp(6.5,10.0)(4.,8.5)(0.,){/Cycles}{0}
\FAProp(16.,15.)(13.5,13.5)(0.,){/Straight}{1}
\FAProp(16.,5.)(13.5,6.5)(0.,){/Straight}{-1}
%
\FAProp(13.5,13.5)(6.5,10.0)(0.,){/Cycles}{0}
\FAProp(6.5,10.0)(13.5,6.5)(0.,){/Cycles}{0}
\FAProp(13.5,13.5)(13.5,6.5)(0.,){/Straight}{1}
%
%
\FAVert(13.5,13.5){0}
\FAVert(13.5,6.5){0}
\FAVert(6.5,10.){0}
%
%
\FALabel(10.5667, 6.4667)[]{\tiny $+$}
\FALabel(8.23333, 7.6333)[]{\tiny $-$}
\FALabel(8.23333, 12.3667)[]{\tiny $0$}
\FALabel(10.5667, 13.5333)[]{\tiny $0$}
\FALabel(15., 11.1)[]{\tiny $\pm$}
\FALabel(15., 8.9)[]{\tiny $\mp$}
%
%
\FALabel(3.1,8.)[]{\tiny$1$}
\FALabel(3.1,12.)[]{\tiny$2$}
\FALabel(17.,16.)[]{\tiny$3$}
\FALabel(17.,4.)[]{\tiny$4$}
\end{feynartspicture}}
   + \parbox{20mm}{\unitlength=0.20bp%
\begin{feynartspicture}(300,300)(1,1)
\FADiagram{}
%
\FAProp(4.,11.5)(6.5,10.0)(0.,){/Cycles}{0}
\FAProp(6.5,10.0)(4.,8.5)(0.,){/Cycles}{0}
\FAProp(16.,15.)(13.5,13.5)(0.,){/Straight}{1}
\FAProp(16.,5.)(13.5,6.5)(0.,){/Straight}{-1}
%
\FAProp(13.5,13.5)(6.5,10.0)(0.,){/Cycles}{0}
\FAProp(6.5,10.0)(13.5,6.5)(0.,){/Cycles}{0}
\FAProp(13.5,13.5)(13.5,6.5)(0.,){/Straight}{1}
%
%
\FAVert(13.5,13.5){0}
\FAVert(13.5,6.5){0}
\FAVert(6.5,10.){0}
%
%
\FALabel(10.5667, 6.4667)[]{\tiny $0$}
\FALabel(8.23333, 7.6333)[]{\tiny $0$}
\FALabel(8.23333, 12.3667)[]{\tiny $0$}
\FALabel(10.5667, 13.5333)[]{\tiny $0$}
\FALabel(15., 11.1)[]{\tiny $\pm$}
\FALabel(15., 8.9)[]{\tiny $\mp$}
%
%
\FALabel(3.1,8.)[]{\tiny$1$}
\FALabel(3.1,12.)[]{\tiny$2$}
\FALabel(17.,16.)[]{\tiny$3$}
\FALabel(17.,4.)[]{\tiny$4$}
\end{feynartspicture}}
   + \parbox{20mm}{\unitlength=0.20bp%
\begin{feynartspicture}(300,300)(1,1)
\FADiagram{}
%
\FAProp(4.,11.5)(6.5,10.0)(0.,){/Cycles}{0}
\FAProp(6.5,10.0)(4.,8.5)(0.,){/Cycles}{0}
\FAProp(16.,15.)(13.5,13.5)(0.,){/Straight}{1}
\FAProp(16.,5.)(13.5,6.5)(0.,){/Straight}{-1}
%
\FAProp(13.5,13.5)(6.5,10.0)(0.,){/ScalarDash}{0}
\FAProp(6.5,10.0)(13.5,6.5)(0.,){/ScalarDash}{0}
\FAProp(13.5,13.5)(13.5,6.5)(0.,){/Straight}{1}
%
%
\FAVert(13.5,13.5){0}
\FAVert(13.5,6.5){0}
\FAVert(6.5,10.){0}
%
%
\FALabel(15., 11.1)[]{\tiny $\pm$}
\FALabel(15., 8.9)[]{\tiny $\mp$}
%
%
\FALabel(3.1,8.)[]{\tiny$1$}
\FALabel(3.1,12.)[]{\tiny$2$}
\FALabel(17.,16.)[]{\tiny$3$}
\FALabel(17.,4.)[]{\tiny$4$}
\end{feynartspicture}}   \, , \nn[-2.ex]
c^{\mbox{\tiny [L]}}_{12|3|4; \; 0} &{}=   \frac{1}{2}A_{4}^{\text{tree}}\left(1-\frac{s_{14}^{3}}{s_{13}^{3}}\right)s_{12} \, , \nn
c^{\mbox{\tiny [L]}}_{12|3|4; \; 2} &{}=   \frac{1}{2}A_{4}^{\text{tree}}\left(2-\frac{s_{12}^{2}}{s_{13}^{2}}\right) \, ; \\
C^{\mbox{\tiny [L]}}_{1|2|34} &{}=
      \parbox{20mm}{\unitlength=0.20bp%
\begin{feynartspicture}(300,300)(1,1)
\FADiagram{}
%
\FAProp(4.,15.)(6.5,13.5)(0.,){/Cycles}{0}
\FAProp(4.,5.)(6.5,6.5)(0.,){/Cycles}{0}
\FAProp(16.,11.5)(13.5,10.0)(0.,){/Straight}{1}
\FAProp(16.,8.5)(13.5,10.0)(0.,){/Straight}{-1}
%
\FAProp(6.5,13.5)(6.5,6.5)(0.,){/Cycles}{0}
\FAProp(13.5,10.0)(6.5,13.5)(0.,){/Cycles}{0}
\FAProp(6.5,6.5)(13.5,10.0)(0.,){/Cycles}{0}
%
\FAVert(6.5,13.5){0}
\FAVert(6.5,6.5){0}
\FAVert(13.5,10.0){0}
%
%
\FALabel(4.5, 8.9)[]{\tiny $+$}
\FALabel(4.5, 11.1)[]{\tiny $-$}
\FALabel(9.43333, 13.5333)[]{\tiny $+$}
\FALabel(11.7667, 12.3667)[]{\tiny $-$}
\FALabel(11.7667, 7.6333)[]{\tiny $+$}
\FALabel(9.43333, 6.4667)[]{\tiny $-$}
%
%
\FALabel(3.,4.)[]{\tiny$1$}
\FALabel(3.,16.)[]{\tiny$2$}
\FALabel(17.,12.)[]{\tiny$3$}
\FALabel(17.,8.)[]{\tiny$4$}
\end{feynartspicture}}
   + \parbox{20mm}{\unitlength=0.20bp%
\begin{feynartspicture}(300,300)(1,1)
\FADiagram{}
%
\FAProp(4.,15.)(6.5,13.5)(0.,){/Cycles}{0}
\FAProp(4.,5.)(6.5,6.5)(0.,){/Cycles}{0}
\FAProp(16.,11.5)(13.5,10.0)(0.,){/Straight}{1}
\FAProp(16.,8.5)(13.5,10.0)(0.,){/Straight}{-1}
%
\FAProp(6.5,13.5)(6.5,6.5)(0.,){/Cycles}{0}
\FAProp(13.5,10.0)(6.5,13.5)(0.,){/Cycles}{0}
\FAProp(6.5,6.5)(13.5,10.0)(0.,){/Cycles}{0}
%
\FAVert(6.5,13.5){0}
\FAVert(6.5,6.5){0}
\FAVert(13.5,10.0){0}
%
%
\FALabel(4.5, 8.9)[]{\tiny $-$}
\FALabel(4.5, 11.1)[]{\tiny $+$}
\FALabel(9.43333, 13.5333)[]{\tiny $-$}
\FALabel(11.7667, 12.3667)[]{\tiny $+$}
\FALabel(11.7667, 7.6333)[]{\tiny $-$}
\FALabel(9.43333, 6.4667)[]{\tiny $+$}
%
%
\FALabel(3.,4.)[]{\tiny$1$}
\FALabel(3.,16.)[]{\tiny$2$}
\FALabel(17.,12.)[]{\tiny$3$}
\FALabel(17.,8.)[]{\tiny$4$}
\end{feynartspicture}}  \nn[-5.5ex]
      &{}+\parbox{20mm}{\unitlength=0.20bp%
\begin{feynartspicture}(300,300)(1,1)
\FADiagram{}
%
\FAProp(4.,15.)(6.5,13.5)(0.,){/Cycles}{0}
\FAProp(4.,5.)(6.5,6.5)(0.,){/Cycles}{0}
\FAProp(16.,11.5)(13.5,10.0)(0.,){/Straight}{1}
\FAProp(16.,8.5)(13.5,10.0)(0.,){/Straight}{-1}
%
\FAProp(6.5,13.5)(6.5,6.5)(0.,){/Cycles}{0}
\FAProp(13.5,10.0)(6.5,13.5)(0.,){/Cycles}{0}
\FAProp(6.5,6.5)(13.5,10.0)(0.,){/Cycles}{0}
%
\FAVert(6.5,13.5){0}
\FAVert(6.5,6.5){0}
\FAVert(13.5,10.0){0}
%
%
\FALabel(4.5, 8.9)[]{\tiny $0$}
\FALabel(4.5, 11.1)[]{\tiny $0$}
\FALabel(9.43333, 13.5333)[]{\tiny $0$}
\FALabel(11.7667, 12.3667)[]{\tiny $0$}
\FALabel(11.7667, 7.6333)[]{\tiny $0$}
\FALabel(9.43333, 6.4667)[]{\tiny $0$}
%
%
\FALabel(3.,4.)[]{\tiny$1$}
\FALabel(3.,16.)[]{\tiny$2$}
\FALabel(17.,12.)[]{\tiny$3$}
\FALabel(17.,8.)[]{\tiny$4$}
\end{feynartspicture}}
   + \parbox{20mm}{\unitlength=0.20bp%
\begin{feynartspicture}(300,300)(1,1)
\FADiagram{}
%
\FAProp(4.,15.)(6.5,13.5)(0.,){/Cycles}{0}
\FAProp(4.,5.)(6.5,6.5)(0.,){/Cycles}{0}
\FAProp(16.,11.5)(13.5,10.0)(0.,){/Straight}{1}
\FAProp(16.,8.5)(13.5,10.0)(0.,){/Straight}{-1}
%
\FAProp(6.5,13.5)(6.5,6.5)(0.,){/ScalarDash}{0}
\FAProp(13.5,10.0)(6.5,13.5)(0.,){/ScalarDash}{0}
\FAProp(6.5,6.5)(13.5,10.0)(0.,){/ScalarDash}{0}
%
\FAVert(6.5,13.5){0}
\FAVert(6.5,6.5){0}
\FAVert(13.5,10.0){0}
%
%
%
%
\FALabel(3.,4.)[]{\tiny$1$}
\FALabel(3.,16.)[]{\tiny$2$}
\FALabel(17.,12.)[]{\tiny$3$}
\FALabel(17.,8.)[]{\tiny$4$}
\end{feynartspicture}} \, , \nn[-2.ex]
c^{\mbox{\tiny [L]}}_{1|2|34; \; 0} &{}=  -\frac{1}{2}A_{4}^{\text{tree}}\left(1+\frac{s_{14}{}^{3}}{s_{13}{}^{3}}\right)s_{12}  \, , \nn
c^{\mbox{\tiny [L]}}_{1|2|34; \; 2} &{}=  -\frac{1}{2}A_{4}^{\text{tree}}\frac{s_{12}^{2}}{s_{13}^{2}}  \, ;  \\
\label{Eq:LT123}
C^{\mbox{\tiny [L]}}_{1|23|4} &{}=
      \parbox{20mm}{\unitlength=0.20bp%
\begin{feynartspicture}(300,300)(1,1)
\FADiagram{}
%
\FAProp(7.5,15.)(10.0,13.5)(0.,){/Cycles}{0}
\FAProp(4.,5.)(6.5,6.5)(0.,){/Cycles}{0}
\FAProp(12.5,15.)(10.0,13.5)(0.,){/Straight}{1}
\FAProp(16.,5.)(13.5,6.5)(0.,){/Straight}{-1}
%
\FAProp(10.0,13.5)(6.5,6.5)(0.,){/Cycles}{0}
\FAProp(6.5,6.5)(13.5,6.5)(0.,){/Cycles}{0}
\FAProp(10.0,13.5)(13.5,6.5)(0.,){/Straight}{1}
%
\FAVert(10.0,13.5){0}
\FAVert(6.5,6.5){0}
\FAVert(10.0,13.5){0}
\FAVert(13.5,6.5){0}
%
%
\FALabel(11.1,4.5)[]{\tiny $+$}
\FALabel(8.9,4.5)[]{\tiny $-$}
\FALabel(6.06667, 9.63333)[]{\tiny $+$}
\FALabel(7.23333, 11.9667)[]{\tiny $-$}
\FALabel(12.7667, 11.9667)[]{\tiny $+$}
\FALabel(13.9333, 9.63333)[]{\tiny $-$}
%
%
\FALabel(3.,4.)[]{\tiny$1$}
\FALabel(6.5,16.)[]{\tiny$2$}
\FALabel(13.5,16.0)[]{\tiny$3$}
\FALabel(17.,4.)[]{\tiny$4$}
\end{feynartspicture}}
   + \parbox{20mm}{\unitlength=0.20bp%
\begin{feynartspicture}(300,300)(1,1)
\FADiagram{}
%
\FAProp(7.5,15.)(10.0,13.5)(0.,){/Cycles}{0}
\FAProp(4.,5.)(6.5,6.5)(0.,){/Cycles}{0}
\FAProp(12.5,15.)(10.0,13.5)(0.,){/Straight}{1}
\FAProp(16.,5.)(13.5,6.5)(0.,){/Straight}{-1}
%
\FAProp(10.0,13.5)(6.5,6.5)(0.,){/Cycles}{0}
\FAProp(6.5,6.5)(13.5,6.5)(0.,){/Cycles}{0}
\FAProp(10.0,13.5)(13.5,6.5)(0.,){/Straight}{1}
%
\FAVert(10.0,13.5){0}
\FAVert(6.5,6.5){0}
\FAVert(10.0,13.5){0}
\FAVert(13.5,6.5){0}
%
%
\FALabel(11.1,4.5)[]{\tiny $0$}
\FALabel(8.9,4.5)[]{\tiny $0$}
\FALabel(6.06667, 9.63333)[]{\tiny $0$}
\FALabel(7.23333, 11.9667)[]{\tiny $0$}
\FALabel(12.7667, 11.9667)[]{\tiny $-$}
\FALabel(13.9333, 9.63333)[]{\tiny $+$}
%
%
\FALabel(3.,4.)[]{\tiny$1$}
\FALabel(6.5,16.)[]{\tiny$2$}
\FALabel(13.5,16.0)[]{\tiny$3$}
\FALabel(17.,4.)[]{\tiny$4$}
\end{feynartspicture}}
   + \parbox{20mm}{\unitlength=0.20bp%
\begin{feynartspicture}(300,300)(1,1)
\FADiagram{}
%
\FAProp(7.5,15.)(10.0,13.5)(0.,){/Cycles}{0}
\FAProp(4.,5.)(6.5,6.5)(0.,){/Cycles}{0}
\FAProp(12.5,15.)(10.0,13.5)(0.,){/Straight}{1}
\FAProp(16.,5.)(13.5,6.5)(0.,){/Straight}{-1}
%
\FAProp(10.0,13.5)(6.5,6.5)(0.,){/ScalarDash}{0}
\FAProp(6.5,6.5)(13.5,6.5)(0.,){/ScalarDash}{0}
\FAProp(10.0,13.5)(13.5,6.5)(0.,){/Straight}{1}
%
\FAVert(10.0,13.5){0}
\FAVert(6.5,6.5){0}
\FAVert(10.0,13.5){0}
\FAVert(13.5,6.5){0}
%
%
%
\FALabel(12.7667, 11.9667)[]{\tiny $-$}
\FALabel(13.9333, 9.63333)[]{\tiny $+$}
%
%
\FALabel(3.,4.)[]{\tiny$1$}
\FALabel(6.5,16.)[]{\tiny$2$}
\FALabel(13.5,16.0)[]{\tiny$3$}
\FALabel(17.,4.)[]{\tiny$4$}
\end{feynartspicture}}  \, , \nn[-2.ex]
c^{\mbox{\tiny [L]}}_{1|23|4; \; 0} &{}=-\frac{1}{2}A_{4}^{\text{tree}}\left(1+\frac{s_{14}^{3}}{s_{13}^{3}}\right)s_{14} \, , \nn
c^{\mbox{\tiny [L]}}_{1|23|4; \; 2} &{}=  -\frac{1}{2}A_{4}^{\text{tree}}\frac{s_{14}s_{12}}{s_{13}^{2}}  \, ;  \\
C^{\mbox{\tiny [L]}}_{2|3|41} &{}=
      \parbox{20mm}{\unitlength=0.20bp%
\begin{feynartspicture}(300,300)(1,1)
\FADiagram{}
%
\FAProp(4.,15.)(6.5,13.5)(0.,){/Cycles}{0}
\FAProp(7.5,5.)(10.,6.5)(0.,){/Cycles}{0}
\FAProp(16.,15.)(13.5,13.5)(0.,){/Straight}{1}
\FAProp(12.5,5.)(10.0,6.5)(0.,){/Straight}{-1}
%
\FAProp(6.5,13.5)(10.0,6.5)(0.,){/Cycles}{0}
\FAProp(13.5,13.5)(6.5,13.5)(0.,){/Cycles}{0}
\FAProp(13.5,13.5)(10.0,6.5)(0.,){/Straight}{1}
%
\FAVert(6.5,13.5){0}
\FAVert(13.5,13.5){0}
\FAVert(10.0,6.5){0}
%
%
\FALabel(8.9, 15.)[]{\tiny $+$}
\FALabel(11.1, 15.)[]{\tiny $-$}
\FALabel(13.9333, 10.3667)[]{\tiny $+$}
\FALabel(12.7667, 8.03333)[]{\tiny $-$}
\FALabel(6.0667, 10.3667)[]{\tiny $+$}
\FALabel(7.2333, 8.03333)[]{\tiny $+$}
%
%
\FALabel(6.5,4.)[]{\tiny$1$}
\FALabel(3.,16.)[]{\tiny$2$}
\FALabel(17.,16.)[]{\tiny$3$}
\FALabel(13.5,4.)[]{\tiny$4$}
\end{feynartspicture}}
   + \parbox{20mm}{\unitlength=0.20bp%
\begin{feynartspicture}(300,300)(1,1)
\FADiagram{}
%
\FAProp(4.,15.)(6.5,13.5)(0.,){/Cycles}{0}
\FAProp(7.5,5.)(10.,6.5)(0.,){/Cycles}{0}
\FAProp(16.,15.)(13.5,13.5)(0.,){/Straight}{1}
\FAProp(12.5,5.)(10.0,6.5)(0.,){/Straight}{-1}
%
\FAProp(6.5,13.5)(10.0,6.5)(0.,){/Cycles}{0}
\FAProp(13.5,13.5)(6.5,13.5)(0.,){/Cycles}{0}
\FAProp(13.5,13.5)(10.0,6.5)(0.,){/Straight}{1}
%
\FAVert(6.5,13.5){0}
\FAVert(13.5,13.5){0}
\FAVert(10.0,6.5){0}
%
%
\FALabel(8.9, 15.)[]{\tiny $0$}
\FALabel(11.1, 15.)[]{\tiny $0$}
\FALabel(13.9333, 10.3667)[]{\tiny $-$}
\FALabel(12.7667, 8.03333)[]{\tiny $+$}
\FALabel(6.0667, 10.3667)[]{\tiny $0$}
\FALabel(7.2333, 8.03333)[]{\tiny $0$}
%
%
\FALabel(6.5,4.)[]{\tiny$1$}
\FALabel(3.,16.)[]{\tiny$2$}
\FALabel(17.,16.)[]{\tiny$3$}
\FALabel(13.5,4.)[]{\tiny$4$}
\end{feynartspicture}}
   +  \parbox{20mm}{\unitlength=0.20bp%
\begin{feynartspicture}(300,300)(1,1)
\FADiagram{}
%
\FAProp(4.,15.)(6.5,13.5)(0.,){/Cycles}{0}
\FAProp(7.5,5.)(10.,6.5)(0.,){/Cycles}{0}
\FAProp(16.,15.)(13.5,13.5)(0.,){/Straight}{1}
\FAProp(12.5,5.)(10.0,6.5)(0.,){/Straight}{-1}
%
\FAProp(6.5,13.5)(10.0,6.5)(0.,){/ScalarDash}{0}
\FAProp(13.5,13.5)(6.5,13.5)(0.,){/ScalarDash}{0}
\FAProp(13.5,13.5)(10.0,6.5)(0.,){/Straight}{1}
%
\FAVert(6.5,13.5){0}
\FAVert(13.5,13.5){0}
\FAVert(10.0,6.5){0}
%
%
\FALabel(13.9333, 10.3667)[]{\tiny $-$}
\FALabel(12.7667, 8.03333)[]{\tiny $+$}
%
%
%
\FALabel(6.5,4.)[]{\tiny$1$}
\FALabel(3.,16.)[]{\tiny$2$}
\FALabel(17.,16.)[]{\tiny$3$}
\FALabel(13.5,4.)[]{\tiny$4$}
\end{feynartspicture}}  \, , \nn[-2.ex]
c^{\mbox{\tiny [L]}}_{2|3|41; \; 0} &{}= -\frac{1}{2}A_{4}^{\text{tree}}\left(1+\frac{s_{14}^{3}}{s_{13}^{3}}\right)s_{14}  \, , \nn
c^{\mbox{\tiny [L]}}_{2|3|41; \; 2} &{}=  -\frac{1}{2}A_{4}^{\text{tree}}\frac{s_{14}s_{12}}{s_{13}^{2}}  \, .
\end{align}
\end{subequations}
In all the triple cuts the last two cut diagrams cancel against each other. In the cut $C^{\mbox{\tiny [L]}}_{12|3|4}$, Eq.~(\ref{Eq:LT123}),  the third cut diagram exactly compensates the contribution of the fourth one. 

The double cuts read as follows 
\vspace{-0.4cm}
\begin{subequations}
\begin{align}
C^{\mbox{\tiny [L]}}_{12|34} &{}=
      \parbox{20mm}{\unitlength=0.20bp%
\begin{feynartspicture}(300,300)(1,1)
\FADiagram{}
\FAProp(4.,11.5)(6.5,10.0)(0.,){/Cycles}{0}
\FAProp(6.5,10.0)(4.,8.5)(0.,){/Cycles}{0}
\FAProp(16.,11.5)(13.5,10.0)(0.,){/Straight}{1}
\FAProp(16.,8.5)(13.5,10.0)(0.,){/Straight}{-1}
\FAVert(6.5,10.){0}
\FAVert(13.5,10.0){0}
\FALabel(3.1,8.)[]{\tiny$1$}
\FALabel(3.1,12.)[]{\tiny$2$}
\FALabel(17.,12.)[]{\tiny$3$}
\FALabel(17.,8.)[]{\tiny$4$}
\FAProp(6.5,10.)(13.5,10.0)(0.8,){/Cycles}{0}
\FAProp(6.5,10.)(13.5,10.0)(-0.8,){/Cycles}{0}
\FALabel(11.1,5.0)[]{\tiny $+$}
\FALabel(8.9,5.0)[]{\tiny $-$}
\FALabel(8.9, 14.5)[]{\tiny $+$}
\FALabel(11.1, 14.5)[]{\tiny $-$}
\end{feynartspicture}}
   + \parbox{20mm}{\unitlength=0.20bp%
\begin{feynartspicture}(300,300)(1,1)
\FADiagram{}
\FAProp(4.,11.5)(6.5,10.0)(0.,){/Cycles}{0}
\FAProp(6.5,10.0)(4.,8.5)(0.,){/Cycles}{0}
\FAProp(16.,11.5)(13.5,10.0)(0.,){/Straight}{1}
\FAProp(16.,8.5)(13.5,10.0)(0.,){/Straight}{-1}
\FAVert(6.5,10.){0}
\FAVert(13.5,10.0){0}
\FALabel(3.1,8.)[]{\tiny$1$}
\FALabel(3.1,12.)[]{\tiny$2$}
\FALabel(17.,12.)[]{\tiny$3$}
\FALabel(17.,8.)[]{\tiny$4$}
\FAProp(6.5,10.)(13.5,10.0)(0.8,){/Cycles}{0}
\FAProp(6.5,10.)(13.5,10.0)(-0.8,){/Cycles}{0}
\FALabel(11.1,5.0)[]{\tiny $0$}
\FALabel(8.9,5.0)[]{\tiny $0$}
\FALabel(8.9, 14.5)[]{\tiny $+$}
\FALabel(11.1, 14.5)[]{\tiny $-$}
\end{feynartspicture}}
   + \parbox{20mm}{\unitlength=0.20bp%
\begin{feynartspicture}(300,300)(1,1)
\FADiagram{}
\FAProp(4.,11.5)(6.5,10.0)(0.,){/Cycles}{0}
\FAProp(6.5,10.0)(4.,8.5)(0.,){/Cycles}{0}
\FAProp(16.,11.5)(13.5,10.0)(0.,){/Straight}{1}
\FAProp(16.,8.5)(13.5,10.0)(0.,){/Straight}{-1}
\FAVert(6.5,10.){0}
\FAVert(13.5,10.0){0}
\FALabel(3.1,8.)[]{\tiny$1$}
\FALabel(3.1,12.)[]{\tiny$2$}
\FALabel(17.,12.)[]{\tiny$3$}
\FALabel(17.,8.)[]{\tiny$4$}
\FAProp(6.5,10.)(13.5,10.0)(0.8,){/Cycles}{0}
\FAProp(6.5,10.)(13.5,10.0)(-0.8,){/Cycles}{0}
\FALabel(11.1,5.0)[]{\tiny $+$}
\FALabel(8.9,5.0)[]{\tiny $-$}
\FALabel(8.9, 14.5)[]{\tiny $0$}
\FALabel(11.1, 14.5)[]{\tiny $0$}
\end{feynartspicture}}   \nn[-5.5ex]
&{}+  \parbox{20mm}{\unitlength=0.20bp%
\begin{feynartspicture}(300,300)(1,1)
\FADiagram{}
\FAProp(4.,11.5)(6.5,10.0)(0.,){/Cycles}{0}
\FAProp(6.5,10.0)(4.,8.5)(0.,){/Cycles}{0}
\FAProp(16.,11.5)(13.5,10.0)(0.,){/Straight}{1}
\FAProp(16.,8.5)(13.5,10.0)(0.,){/Straight}{-1}
\FAVert(6.5,10.){0}
\FAVert(13.5,10.0){0}
\FALabel(3.1,8.)[]{\tiny$1$}
\FALabel(3.1,12.)[]{\tiny$2$}
\FALabel(17.,12.)[]{\tiny$3$}
\FALabel(17.,8.)[]{\tiny$4$}
\FAProp(6.5,10.)(13.5,10.0)(0.8,){/Cycles}{0}
\FAProp(6.5,10.)(13.5,10.0)(-0.8,){/Cycles}{0}
\FALabel(11.1,5.0)[]{\tiny $0$}
\FALabel(8.9,5.0)[]{\tiny $0$}
\FALabel(8.9, 14.5)[]{\tiny $0$}
\FALabel(11.1, 14.5)[]{\tiny $0$}
\end{feynartspicture}}
   + \parbox{20mm}{\unitlength=0.20bp%
\begin{feynartspicture}(300,300)(1,1)
\FADiagram{}
\FAProp(4.,11.5)(6.5,10.0)(0.,){/Cycles}{0}
\FAProp(6.5,10.0)(4.,8.5)(0.,){/Cycles}{0}
\FAProp(16.,11.5)(13.5,10.0)(0.,){/Straight}{1}
\FAProp(16.,8.5)(13.5,10.0)(0.,){/Straight}{-1}
\FAVert(6.5,10.){0}
\FAVert(13.5,10.0){0}
\FALabel(3.1,8.)[]{\tiny$1$}
\FALabel(3.1,12.)[]{\tiny$2$}
\FALabel(17.,12.)[]{\tiny$3$}
\FALabel(17.,8.)[]{\tiny$4$}
\FAProp(6.5,10.)(13.5,10.0)(0.8,){/ScalarDash}{0}
\FAProp(6.5,10.)(13.5,10.0)(-0.8,){/ScalarDash}{0}
%
\end{feynartspicture}} \, , \nn[-2ex]
c^{\mbox{\tiny [L]}}_{12|34; \; 0} &{}= A_{4}^{\text{tree}}\frac{s_{14}}{s_{13}}\left(\frac{s_{14}}{s_{13}}-\frac{1}{2}\right)  \, , \nn
c^{\mbox{\tiny [L]}}_{12|34; \; 2} &{}= 0  \, ;  \\
C^{\mbox{\tiny [L]}}_{23|41} &{}=
      \parbox{20mm}{\unitlength=0.20bp%
\begin{feynartspicture}(300,300)(1,1)
\FADiagram{}
\FAProp(7.5,15.)(10.0,13.5)(0.,){/Cycles}{0}
\FAProp(12.5,15.)(10.0,13.5)(0.,){/Straight}{1}
\FAProp(7.5,5.)(10.,6.5)(0.,){/Cycles}{0}
\FAProp(12.5,5.)(10.0,6.5)(0.,){/Straight}{-1}
\FAVert(10.0,13.5){0}
\FAVert(10.0,6.5){0}
\FALabel(6.5,16.)[]{\tiny$2$}
\FALabel(13.5,16.0)[]{\tiny$3$}
\FALabel(6.5,4.)[]{\tiny$1$}
\FALabel(13.5,4.)[]{\tiny$4$}
\FAProp(10.0,6.5)(10.0,13.5)(0.8,){/Straight}{-1}
\FAProp(10.0,6.5)(10.0,13.5)(-0.8,){/Cycles}{0}
\FALabel(5.0, 8.9)[]{\tiny $+$}
\FALabel(5.0, 11.1)[]{\tiny $-$}
\FALabel(14.5, 11.1)[]{\tiny $+$}
\FALabel(14.5, 8.9)[]{\tiny $-$}
\end{feynartspicture}}
   + \parbox{20mm}{\unitlength=0.20bp%
\begin{feynartspicture}(300,300)(1,1)
\FADiagram{}
\FAProp(7.5,15.)(10.0,13.5)(0.,){/Cycles}{0}
\FAProp(12.5,15.)(10.0,13.5)(0.,){/Straight}{1}
\FAProp(7.5,5.)(10.,6.5)(0.,){/Cycles}{0}
\FAProp(12.5,5.)(10.0,6.5)(0.,){/Straight}{-1}
\FAVert(10.0,13.5){0}
\FAVert(10.0,6.5){0}
\FALabel(6.5,16.)[]{\tiny$2$}
\FALabel(13.5,16.0)[]{\tiny$3$}
\FALabel(6.5,4.)[]{\tiny$1$}
\FALabel(13.5,4.)[]{\tiny$4$}
\FAProp(10.0,6.5)(10.0,13.5)(0.8,){/Straight}{-1}
\FAProp(10.0,6.5)(10.0,13.5)(-0.8,){/Cycles}{0}
\FALabel(5.0, 8.9)[]{\tiny $0$}
\FALabel(5.0, 11.1)[]{\tiny $0$}
\FALabel(14.5, 11.1)[]{\tiny $+$}
\FALabel(14.5, 8.9)[]{\tiny $-$}
\end{feynartspicture}}   \nn[-5.5ex]
&{}+ \parbox{20mm}{\unitlength=0.20bp%
\begin{feynartspicture}(300,300)(1,1)
\FADiagram{}
\FAProp(7.5,15.)(10.0,13.5)(0.,){/Cycles}{0}
\FAProp(12.5,15.)(10.0,13.5)(0.,){/Straight}{1}
\FAProp(7.5,5.)(10.,6.5)(0.,){/Cycles}{0}
\FAProp(12.5,5.)(10.0,6.5)(0.,){/Straight}{-1}
\FAVert(10.0,13.5){0}
\FAVert(10.0,6.5){0}
\FALabel(6.5,16.)[]{\tiny$2$}
\FALabel(13.5,16.0)[]{\tiny$3$}
\FALabel(6.5,4.)[]{\tiny$1$}
\FALabel(13.5,4.)[]{\tiny$4$}
\FAProp(10.0,6.5)(10.0,13.5)(0.8,){/Straight}{-1}
\FAProp(10.0,6.5)(10.0,13.5)(-0.8,){/Cycles}{0}
\FALabel(5.0, 8.9)[]{\tiny $0$}
\FALabel(5.0, 11.1)[]{\tiny $0$}
\FALabel(14.5, 11.1)[]{\tiny $-$}
\FALabel(14.5, 8.9)[]{\tiny $+$}
\end{feynartspicture}}
		  + \parbox{20mm}{\unitlength=0.20bp%
\begin{feynartspicture}(300,300)(1,1)
\FADiagram{}
\FAProp(7.5,15.)(10.0,13.5)(0.,){/Cycles}{0}
\FAProp(12.5,15.)(10.0,13.5)(0.,){/Straight}{1}
\FAProp(7.5,5.)(10.,6.5)(0.,){/Cycles}{0}
\FAProp(12.5,5.)(10.0,6.5)(0.,){/Straight}{-1}
\FAVert(10.0,13.5){0}
\FAVert(10.0,6.5){0}
\FALabel(6.5,16.)[]{\tiny$2$}
\FALabel(13.5,16.0)[]{\tiny$3$}
\FALabel(6.5,4.)[]{\tiny$1$}
\FALabel(13.5,4.)[]{\tiny$4$}
\FAProp(10.0,6.5)(10.0,13.5)(0.8,){/Straight}{-1}
\FAProp(10.0,6.5)(10.0,13.5)(-0.8,){/ScalarDash}{0}
%
\FALabel(14.5, 11.1)[]{\tiny $-$}
\FALabel(14.5, 8.9)[]{\tiny $+$}
\end{feynartspicture}}\, , \nn[-2ex]
   c^{\mbox{\tiny [L]}}_{23|41; \; 0} &{}=A_{4}^{\text{tree}}\left(\frac{3}{2}-\frac{s_{14}^{2}}{s_{13}^{2}}+\frac{1}{2}\frac{s_{14}}{s_{13}}\right) \, , \nn
c^{\mbox{\tiny [L]}}_{23|41; \; 2} &{}=  0 \, .
\end{align}
\end{subequations}
In both cases the  last two diagrams  cancel against each other. In the case of the cut $C^{\mbox{\tiny [L]}}_{13}$ the second and the third diagram cancel as well.
\bigskip

\paragraph*{Right-turning amplitude --}
The quadruple cut is given by 
\vspace{-0.4cm}
\begin{align}
C^{\mbox{\tiny [R]}}_{1|2|3|4} &{}=
   \parbox{20mm}{\unitlength=0.20bp%
\begin{feynartspicture}(300,300)(1,1)
\FADiagram{}
%
\FAProp(4.,15.)(6.5,13.5)(0.,){/Cycles}{0}
\FAProp(4.,5.)(6.5,6.5)(0.,){/Cycles}{0}
\FAProp(16.,15.)(13.5,13.5)(0.,){/Straight}{1}
\FAProp(16.,5.)(13.5,6.5)(0.,){/Straight}{-1}
%
\FAProp(6.5,13.5)(6.5,6.5)(0.,){/Straight}{1}
\FAProp(13.5,13.5)(6.5,13.5)(0.,){/Straight}{1}
\FAProp(6.5,6.5)(13.5,6.5)(0.,){/Straight}{1}
\FAProp(13.5,6.5)(13.5,13.5)(0.,){/Cycles}{0}
%
\FAVert(6.5,13.5){0}
\FAVert(6.5,6.5){0}
\FAVert(13.5,13.5){0}
\FAVert(13.5,6.5){0}
%
%
\FALabel(11.1,4.5)[]{\tiny $-$}
\FALabel(8.9,4.5)[]{\tiny $+$}
\FALabel(4.5, 8.9)[]{\tiny $\mp$}
\FALabel(4.5, 11.1)[]{\tiny $\pm$}
\FALabel(8.9, 15.)[]{\tiny $-$}
\FALabel(11.1, 15.)[]{\tiny $+$}
\FALabel(15., 11.1)[]{\tiny $-$}
\FALabel(15., 8.9)[]{\tiny $+$}
%
%
\FALabel(3.,4.)[]{\tiny$1$}
\FALabel(3.,16.)[]{\tiny$2$}
\FALabel(17.,16.)[]{\tiny$3$}
\FALabel(17.,4.)[]{\tiny$4$}
\end{feynartspicture}}
+ \parbox{20mm}{\unitlength=0.20bp%
\begin{feynartspicture}(300,300)(1,1)
\FADiagram{}
%
\FAProp(4.,15.)(6.5,13.5)(0.,){/Cycles}{0}
\FAProp(4.,5.)(6.5,6.5)(0.,){/Cycles}{0}
\FAProp(16.,15.)(13.5,13.5)(0.,){/Straight}{1}
\FAProp(16.,5.)(13.5,6.5)(0.,){/Straight}{-1}
%
\FAProp(6.5,13.5)(6.5,6.5)(0.,){/Straight}{1}
\FAProp(13.5,13.5)(6.5,13.5)(0.,){/Straight}{1}
\FAProp(6.5,6.5)(13.5,6.5)(0.,){/Straight}{1}
\FAProp(13.5,6.5)(13.5,13.5)(0.,){/Cycles}{0}
%
\FAVert(6.5,13.5){0}
\FAVert(6.5,6.5){0}
\FAVert(13.5,13.5){0}
\FAVert(13.5,6.5){0}
%
%
\FALabel(11.1,4.5)[]{\tiny $+$}
\FALabel(8.9,4.5)[]{\tiny $-$}
\FALabel(4.5, 8.9)[]{\tiny $\mp$}
\FALabel(4.5, 11.1)[]{\tiny $\pm$}
\FALabel(8.9, 15.)[]{\tiny $+$}
\FALabel(11.1, 15.)[]{\tiny $-$}
\FALabel(15., 11.1)[]{\tiny $0$}
\FALabel(15., 8.9)[]{\tiny $0$}
%
%
\FALabel(3.,4.)[]{\tiny$1$}
\FALabel(3.,16.)[]{\tiny$2$}
\FALabel(17.,16.)[]{\tiny$3$}
\FALabel(17.,4.)[]{\tiny$4$}
\end{feynartspicture}} 
+ \parbox{20mm}{\unitlength=0.20bp%
\begin{feynartspicture}(300,300)(1,1)
\FADiagram{}
%
\FAProp(4.,15.)(6.5,13.5)(0.,){/Cycles}{0}
\FAProp(4.,5.)(6.5,6.5)(0.,){/Cycles}{0}
\FAProp(16.,15.)(13.5,13.5)(0.,){/Straight}{1}
\FAProp(16.,5.)(13.5,6.5)(0.,){/Straight}{-1}
%
\FAProp(6.5,13.5)(6.5,6.5)(0.,){/Straight}{1}
\FAProp(13.5,13.5)(6.5,13.5)(0.,){/Straight}{1}
\FAProp(6.5,6.5)(13.5,6.5)(0.,){/Straight}{1}
\FAProp(13.5,6.5)(13.5,13.5)(0.,){/ScalarDash}{0}
%
\FAVert(6.5,13.5){0}
\FAVert(6.5,6.5){0}
\FAVert(13.5,13.5){0}
\FAVert(13.5,6.5){0}
%
%
\FALabel(11.1,4.5)[]{\tiny $+$}
\FALabel(8.9,4.5)[]{\tiny $-$}
\FALabel(4.5, 8.9)[]{\tiny $\mp$}
\FALabel(4.5, 11.1)[]{\tiny $\pm$}
\FALabel(8.9, 15.)[]{\tiny $+$}
\FALabel(11.1, 15.)[]{\tiny $-$}
%
%
\FALabel(3.,4.)[]{\tiny$1$}
\FALabel(3.,16.)[]{\tiny$2$}
\FALabel(17.,16.)[]{\tiny$3$}
\FALabel(17.,4.)[]{\tiny$4$}
\end{feynartspicture}}\, , \nn[-2.ex]
c^{\mbox{\tiny [R]}}_{1|2|3|4; \; 0} &{}=-\frac{1}{2}A_{4}^{\text{tree}}\frac{s_{12}^{3}}{s_{13}^{3}}s_{12}s_{14}\, , \nn
c^{\mbox{\tiny [R]}}_{1|2|3|4; \; 4} &{} =0  \, .
\end{align}
The first helicity  configuration contributes only to the cut-constructible part  while  the second one cancels  against the box with internal scalars. 

The triple cuts are given by
\vspace{-0.4cm}
\begin{subequations}
\begin{align}
C^{\mbox{\tiny [R]}}_{12|3|4} &{}=
      \parbox{20mm}{\unitlength=0.20bp%
\begin{feynartspicture}(300,300)(1,1)
\FADiagram{}
%
\FAProp(4.,11.5)(6.5,10.0)(0.,){/Cycles}{0}
\FAProp(6.5,10.0)(4.,8.5)(0.,){/Cycles}{0}
\FAProp(16.,15.)(13.5,13.5)(0.,){/Straight}{1}
\FAProp(16.,5.)(13.5,6.5)(0.,){/Straight}{-1}
%
\FAProp(13.5,13.5)(6.5,10.0)(0.,){/Straight}{1}
\FAProp(6.5,10.0)(13.5,6.5)(0.,){/Straight}{1}
\FAProp(13.5,6.5)(13.5,13.5)(0.,){/Cycles}{0}
%
%
\FAVert(13.5,13.5){0}
\FAVert(13.5,6.5){0}
\FAVert(6.5,10.){0}
%
%
\FALabel(10.5667, 6.4667)[]{\tiny $-$}
\FALabel(8.23333, 7.6333)[]{\tiny $+$}
\FALabel(8.23333, 12.3667)[]{\tiny $-$}
\FALabel(10.5667, 13.5333)[]{\tiny $+$}
\FALabel(15., 11.1)[]{\tiny $\pm$}
\FALabel(15., 8.9)[]{\tiny $\mp$}
%
%
\FALabel(3.1,8.)[]{\tiny$1$}
\FALabel(3.1,12.)[]{\tiny$2$}
\FALabel(17.,16.)[]{\tiny$3$}
\FALabel(17.,4.)[]{\tiny$4$}
\end{feynartspicture}}
   + \parbox{20mm}{\unitlength=0.20bp%
\begin{feynartspicture}(300,300)(1,1)
\FADiagram{}
%
\FAProp(4.,11.5)(6.5,10.0)(0.,){/Cycles}{0}
\FAProp(6.5,10.0)(4.,8.5)(0.,){/Cycles}{0}
\FAProp(16.,15.)(13.5,13.5)(0.,){/Straight}{1}
\FAProp(16.,5.)(13.5,6.5)(0.,){/Straight}{-1}
%
\FAProp(13.5,13.5)(6.5,10.0)(0.,){/Straight}{1}
\FAProp(6.5,10.0)(13.5,6.5)(0.,){/Straight}{1}
\FAProp(13.5,6.5)(13.5,13.5)(0.,){/Cycles}{0}
%
%
\FAVert(13.5,13.5){0}
\FAVert(13.5,6.5){0}
\FAVert(6.5,10.){0}
%
%
\FALabel(10.5667, 6.4667)[]{\tiny $+$}
\FALabel(8.23333, 7.6333)[]{\tiny $-$}
\FALabel(8.23333, 12.3667)[]{\tiny $+$}
\FALabel(10.5667, 13.5333)[]{\tiny $-$}
\FALabel(15., 11.1)[]{\tiny $+$}
\FALabel(15., 8.9)[]{\tiny $-$}
%
%
\FALabel(3.1,8.)[]{\tiny$1$}
\FALabel(3.1,12.)[]{\tiny$2$}
\FALabel(17.,16.)[]{\tiny$3$}
\FALabel(17.,4.)[]{\tiny$4$}
\end{feynartspicture}}  \nn[-5.5ex]
  &{}+ \parbox{20mm}{\unitlength=0.20bp%
\begin{feynartspicture}(300,300)(1,1)
\FADiagram{}
%
\FAProp(4.,11.5)(6.5,10.0)(0.,){/Cycles}{0}
\FAProp(6.5,10.0)(4.,8.5)(0.,){/Cycles}{0}
\FAProp(16.,15.)(13.5,13.5)(0.,){/Straight}{1}
\FAProp(16.,5.)(13.5,6.5)(0.,){/Straight}{-1}
%
\FAProp(13.5,13.5)(6.5,10.0)(0.,){/Straight}{1}
\FAProp(6.5,10.0)(13.5,6.5)(0.,){/Straight}{1}
\FAProp(13.5,6.5)(13.5,13.5)(0.,){/Cycles}{0}
%
%
\FAVert(13.5,13.5){0}
\FAVert(13.5,6.5){0}
\FAVert(6.5,10.){0}
%
%
\FALabel(10.5667, 6.4667)[]{\tiny $\mp$}
\FALabel(8.23333, 7.6333)[]{\tiny $\pm$}
\FALabel(8.23333, 12.3667)[]{\tiny $\mp$}
\FALabel(10.5667, 13.5333)[]{\tiny $\pm$}
\FALabel(15., 11.1)[]{\tiny $0$}
\FALabel(15., 8.9)[]{\tiny $0$}
%
%
\FALabel(3.1,8.)[]{\tiny$1$}
\FALabel(3.1,12.)[]{\tiny$2$}
\FALabel(17.,16.)[]{\tiny$3$}
\FALabel(17.,4.)[]{\tiny$4$}
\end{feynartspicture}}
   + \parbox{20mm}{\unitlength=0.20bp%
\begin{feynartspicture}(300,300)(1,1)
\FADiagram{}
%
\FAProp(4.,11.5)(6.5,10.0)(0.,){/Cycles}{0}
\FAProp(6.5,10.0)(4.,8.5)(0.,){/Cycles}{0}
\FAProp(16.,15.)(13.5,13.5)(0.,){/Straight}{1}
\FAProp(16.,5.)(13.5,6.5)(0.,){/Straight}{-1}
%
\FAProp(13.5,13.5)(6.5,10.0)(0.,){/Straight}{1}
\FAProp(6.5,10.0)(13.5,6.5)(0.,){/Straight}{1}
\FAProp(13.5,6.5)(13.5,13.5)(0.,){/ScalarDash}{0}
%
%
\FAVert(13.5,13.5){0}
\FAVert(13.5,6.5){0}
\FAVert(6.5,10.){0}
%
%
\FALabel(10.5667, 6.4667)[]{\tiny $\mp$}
\FALabel(8.23333, 7.6333)[]{\tiny $\pm$}
\FALabel(8.23333, 12.3667)[]{\tiny $\mp$}
\FALabel(10.5667, 13.5333)[]{\tiny $\pm$}
%
%
\FALabel(3.1,8.)[]{\tiny$1$}
\FALabel(3.1,12.)[]{\tiny$2$}
\FALabel(17.,16.)[]{\tiny$3$}
\FALabel(17.,4.)[]{\tiny$4$}
\end{feynartspicture}} \, , \nn[-2.ex] 
c^{\mbox{\tiny [R]}}_{12|3|4; \; 0} &{}= -\frac{1}{2}A_{4}^{\text{tree}}\left(2+\frac{s_{12}^{3}}{s_{13}^{3}}\right)s_{12} \, , \nn
c^{\mbox{\tiny [R]}}_{12|3|4; \; 2} &{} = -\frac{1}{2}A_{4}^{\text{tree}}\left(1+\frac{s_{14}^{2}}{s_{13}^{2}}\right)  \, ;  \\
C^{\mbox{\tiny [R]}}_{1|2|34} &{} =
      \parbox{20mm}{\unitlength=0.20bp%
\begin{feynartspicture}(300,300)(1,1)
\FADiagram{}
%
\FAProp(4.,15.)(6.5,13.5)(0.,){/Cycles}{0}
\FAProp(4.,5.)(6.5,6.5)(0.,){/Cycles}{0}
\FAProp(16.,11.5)(13.5,10.0)(0.,){/Straight}{1}
\FAProp(16.,8.5)(13.5,10.0)(0.,){/Straight}{-1}
%
\FAProp(6.5,13.5)(6.5,6.5)(0.,){/Straight}{1}
\FAProp(13.5,10.0)(6.5,13.5)(0.,){/Straight}{1}
\FAProp(6.5,6.5)(13.5,10.0)(0.,){/Straight}{1}
%
\FAVert(6.5,13.5){0}
\FAVert(6.5,6.5){0}
\FAVert(13.5,10.0){0}
%
%
\FALabel(4.5, 8.9)[]{\tiny $-$}
\FALabel(4.5, 11.1)[]{\tiny $+$}
\FALabel(9.43333, 13.5333)[]{\tiny $-$}
\FALabel(11.7667, 12.3667)[]{\tiny $+$}
\FALabel(11.7667, 7.6333)[]{\tiny $-$}
\FALabel(9.43333, 6.4667)[]{\tiny $+$}
%
%
\FALabel(3.,4.)[]{\tiny$1$}
\FALabel(3.,16.)[]{\tiny$2$}
\FALabel(17.,12.)[]{\tiny$3$}
\FALabel(17.,8.)[]{\tiny$4$}
\end{feynartspicture}}
   + \parbox{20mm}{\unitlength=0.20bp%
\begin{feynartspicture}(300,300)(1,1)
\FADiagram{}
%
\FAProp(4.,15.)(6.5,13.5)(0.,){/Cycles}{0}
\FAProp(4.,5.)(6.5,6.5)(0.,){/Cycles}{0}
\FAProp(16.,11.5)(13.5,10.0)(0.,){/Straight}{1}
\FAProp(16.,8.5)(13.5,10.0)(0.,){/Straight}{-1}
%
\FAProp(6.5,13.5)(6.5,6.5)(0.,){/Straight}{1}
\FAProp(13.5,10.0)(6.5,13.5)(0.,){/Straight}{1}
\FAProp(6.5,6.5)(13.5,10.0)(0.,){/Straight}{1}
%
\FAVert(6.5,13.5){0}
\FAVert(6.5,6.5){0}
\FAVert(13.5,10.0){0}
%
%
\FALabel(4.5, 8.9)[]{\tiny $+$}
\FALabel(4.5, 11.1)[]{\tiny $-$}
\FALabel(9.43333, 13.5333)[]{\tiny $+$}
\FALabel(11.7667, 12.3667)[]{\tiny $-$}
\FALabel(11.7667, 7.6333)[]{\tiny $+$}
\FALabel(9.43333, 6.4667)[]{\tiny $-$}
%
%
\FALabel(3.,4.)[]{\tiny$1$}
\FALabel(3.,16.)[]{\tiny$2$}
\FALabel(17.,12.)[]{\tiny$3$}
\FALabel(17.,8.)[]{\tiny$4$}
\end{feynartspicture}}  \, , \nn[-2.ex]
c^{\mbox{\tiny [R]}}_{1|2|34; \; 0} &{}=  -\frac{1}{2}A_{4}^{\text{tree}}\frac{s_{12}^{3}}{s_{13}^{3}}s_{12} \, , \nn
c^{\mbox{\tiny [R]}}_{1|2|34; \; 2} &{} =- \frac{1}{2}A_{4}^{\text{tree}}\frac{s_{12}}{s_{13}}\left(1-\frac{s_{14}}{s_{13}}\right) \, ;  \\
C^{\mbox{\tiny [R]}}_{1|23|4} &{}=
      \parbox{20mm}{\unitlength=0.20bp%
\begin{feynartspicture}(300,300)(1,1)
\FADiagram{}
%
\FAProp(7.5,15.)(10.0,13.5)(0.,){/Cycles}{0}
\FAProp(4.,5.)(6.5,6.5)(0.,){/Cycles}{0}
\FAProp(12.5,15.)(10.0,13.5)(0.,){/Straight}{1}
\FAProp(16.,5.)(13.5,6.5)(0.,){/Straight}{-1}
%
\FAProp(10.0,13.5)(6.5,6.5)(0.,){/Straight}{1}
\FAProp(6.5,6.5)(13.5,6.5)(0.,){/Straight}{1}
\FAProp(13.5,6.5)(10.0,13.5)(0.,){/Cycles}{0}
%
\FAVert(10.0,13.5){0}
\FAVert(6.5,6.5){0}
\FAVert(10.0,13.5){0}
\FAVert(13.5,6.5){0}
%
%
\FALabel(11.1,4.5)[]{\tiny $-$}
\FALabel(8.9,4.5)[]{\tiny $+$}
\FALabel(6.06667, 9.63333)[]{\tiny $-$}
\FALabel(7.23333, 11.9667)[]{\tiny $+$}
\FALabel(12.7667, 11.9667)[]{\tiny $-$}
\FALabel(13.9333, 9.63333)[]{\tiny $+$}
%
%
\FALabel(3.,4.)[]{\tiny$1$}
\FALabel(6.5,16.)[]{\tiny$2$}
\FALabel(13.5,16.0)[]{\tiny$3$}
\FALabel(17.,4.)[]{\tiny$4$}
\end{feynartspicture}}
   + \parbox{20mm}{\unitlength=0.20bp%
\begin{feynartspicture}(300,300)(1,1)
\FADiagram{}
%
\FAProp(7.5,15.)(10.0,13.5)(0.,){/Cycles}{0}
\FAProp(4.,5.)(6.5,6.5)(0.,){/Cycles}{0}
\FAProp(12.5,15.)(10.0,13.5)(0.,){/Straight}{1}
\FAProp(16.,5.)(13.5,6.5)(0.,){/Straight}{-1}
%
\FAProp(10.0,13.5)(6.5,6.5)(0.,){/Straight}{1}
\FAProp(6.5,6.5)(13.5,6.5)(0.,){/Straight}{1}
\FAProp(13.5,6.5)(10.0,13.5)(0.,){/Cycles}{0}
%
\FAVert(10.0,13.5){0}
\FAVert(6.5,6.5){0}
\FAVert(10.0,13.5){0}
\FAVert(13.5,6.5){0}
%
%
\FALabel(11.1,4.5)[]{\tiny $+$}
\FALabel(8.9,4.5)[]{\tiny $-$}
\FALabel(6.06667, 9.63333)[]{\tiny $+$}
\FALabel(7.23333, 11.9667)[]{\tiny $-$}
\FALabel(12.7667, 11.9667)[]{\tiny $0$}
\FALabel(13.9333, 9.63333)[]{\tiny $0$}
%
%
\FALabel(3.,4.)[]{\tiny$1$}
\FALabel(6.5,16.)[]{\tiny$2$}
\FALabel(13.5,16.0)[]{\tiny$3$}
\FALabel(17.,4.)[]{\tiny$4$}
\end{feynartspicture}}
   + \parbox{20mm}{\unitlength=0.20bp%
\begin{feynartspicture}(300,300)(1,1)
\FADiagram{}
%
\FAProp(7.5,15.)(10.0,13.5)(0.,){/Cycles}{0}
\FAProp(4.,5.)(6.5,6.5)(0.,){/Cycles}{0}
\FAProp(12.5,15.)(10.0,13.5)(0.,){/Straight}{1}
\FAProp(16.,5.)(13.5,6.5)(0.,){/Straight}{-1}
%
\FAProp(10.0,13.5)(6.5,6.5)(0.,){/Straight}{1}
\FAProp(6.5,6.5)(13.5,6.5)(0.,){/Straight}{1}
\FAProp(13.5,6.5)(10.0,13.5)(0.,){/ScalarDash}{0}
%
\FAVert(10.0,13.5){0}
\FAVert(6.5,6.5){0}
\FAVert(10.0,13.5){0}
\FAVert(13.5,6.5){0}
%
%
\FALabel(11.1,4.5)[]{\tiny $+$}
\FALabel(8.9,4.5)[]{\tiny $-$}
\FALabel(6.06667, 9.63333)[]{\tiny $+$}
\FALabel(7.23333, 11.9667)[]{\tiny $-$}
%
%
%
\FALabel(3.,4.)[]{\tiny$1$}
\FALabel(6.5,16.)[]{\tiny$2$}
\FALabel(13.5,16.0)[]{\tiny$3$}
\FALabel(17.,4.)[]{\tiny$4$}
\end{feynartspicture}}  \, , \nn[-2.ex]
c^{\mbox{\tiny [R]}}_{1|23|4; \; 0} &{}=  -\frac{1}{2}A_{4}^{\text{tree}}\frac{s_{12}^{3}}{s_{13}^{3}}s_{14}  \, , \nn
c^{\mbox{\tiny [R]}}_{1|23|4; \; 2} &{} = - \frac{1}{2}A_{4}^{\text{tree}}\frac{s_{12}s_{14}}{s_{13}^{2}}  \, ;  \\
C^{\mbox{\tiny [R]}}_{2|3|41} &{}=
      \parbox{20mm}{\unitlength=0.20bp%
\begin{feynartspicture}(300,300)(1,1)
\FADiagram{}
%
\FAProp(4.,15.)(6.5,13.5)(0.,){/Cycles}{0}
\FAProp(7.5,5.)(10.,6.5)(0.,){/Cycles}{0}
\FAProp(16.,15.)(13.5,13.5)(0.,){/Straight}{1}
\FAProp(12.5,5.)(10.0,6.5)(0.,){/Straight}{-1}
%
\FAProp(6.5,13.5)(10.0,6.5)(0.,){/Straight}{1}
\FAProp(13.5,13.5)(6.5,13.5)(0.,){/Straight}{1}
\FAProp(10.0,6.5)(13.5,13.5)(0.,){/Cycles}{0}
%
\FAVert(6.5,13.5){0}
\FAVert(13.5,13.5){0}
\FAVert(10.0,6.5){0}
%
%
\FALabel(8.9, 15.)[]{\tiny $-$}
\FALabel(11.1, 15.)[]{\tiny $+$}
\FALabel(13.9333, 10.3667)[]{\tiny $-$}
\FALabel(12.7667, 8.03333)[]{\tiny $+$}
\FALabel(6.0667, 10.3667)[]{\tiny $+$}
\FALabel(7.2333, 8.03333)[]{\tiny $-$}
%
%
\FALabel(6.5,4.)[]{\tiny$1$}
\FALabel(3.,16.)[]{\tiny$2$}
\FALabel(17.,16.)[]{\tiny$3$}
\FALabel(13.5,4.)[]{\tiny$4$}
\end{feynartspicture}}
   + \parbox{20mm}{\unitlength=0.20bp%
\begin{feynartspicture}(300,300)(1,1)
\FADiagram{}
%
\FAProp(4.,15.)(6.5,13.5)(0.,){/Cycles}{0}
\FAProp(7.5,5.)(10.,6.5)(0.,){/Cycles}{0}
\FAProp(16.,15.)(13.5,13.5)(0.,){/Straight}{1}
\FAProp(12.5,5.)(10.0,6.5)(0.,){/Straight}{-1}
%
\FAProp(6.5,13.5)(10.0,6.5)(0.,){/Straight}{1}
\FAProp(13.5,13.5)(6.5,13.5)(0.,){/Straight}{1}
\FAProp(10.0,6.5)(13.5,13.5)(0.,){/Cycles}{0}
%
\FAVert(6.5,13.5){0}
\FAVert(13.5,13.5){0}
\FAVert(10.0,6.5){0}
%
%
\FALabel(8.9, 15.)[]{\tiny $+$}
\FALabel(11.1, 15.)[]{\tiny $-$}
\FALabel(13.9333, 10.3667)[]{\tiny $0$}
\FALabel(12.7667, 8.03333)[]{\tiny $0$}
\FALabel(6.0667, 10.3667)[]{\tiny $-$}
\FALabel(7.2333, 8.03333)[]{\tiny $+$}
%
%
\FALabel(6.5,4.)[]{\tiny$1$}
\FALabel(3.,16.)[]{\tiny$2$}
\FALabel(17.,16.)[]{\tiny$3$}
\FALabel(13.5,4.)[]{\tiny$4$}
\end{feynartspicture}}
   + \parbox{20mm}{\unitlength=0.20bp%
\begin{feynartspicture}(300,300)(1,1)
\FADiagram{}
%
\FAProp(4.,15.)(6.5,13.5)(0.,){/Cycles}{0}
\FAProp(7.5,5.)(10.,6.5)(0.,){/Cycles}{0}
\FAProp(16.,15.)(13.5,13.5)(0.,){/Straight}{1}
\FAProp(12.5,5.)(10.0,6.5)(0.,){/Straight}{-1}
%
\FAProp(6.5,13.5)(10.0,6.5)(0.,){/Straight}{1}
\FAProp(13.5,13.5)(6.5,13.5)(0.,){/Straight}{1}
\FAProp(10.0,6.5)(13.5,13.5)(0.,){/ScalarDash}{0}
%
\FAVert(6.5,13.5){0}
\FAVert(13.5,13.5){0}
\FAVert(10.0,6.5){0}
%
%
\FALabel(8.9, 15.)[]{\tiny $+$}
\FALabel(11.1, 15.)[]{\tiny $-$}
%
\FALabel(6.0667, 10.3667)[]{\tiny $-$}
\FALabel(7.2333, 8.03333)[]{\tiny $+$}
%
%
\FALabel(6.5,4.)[]{\tiny$1$}
\FALabel(3.,16.)[]{\tiny$2$}
\FALabel(17.,16.)[]{\tiny$3$}
\FALabel(13.5,4.)[]{\tiny$4$}
\end{feynartspicture}} \, , \nn[-2.ex]
c^{\mbox{\tiny [R]}}_{2|3|41; \; 0} &{}=    -\frac{1}{2}A_{4}^{\text{tree}}\frac{s_{12}^{3}}{s_{13}^{3}}s_{14} \, , \nn
c^{\mbox{\tiny [R]}}_{2|3|41; \; 2} &{} = -  \frac{1}{2}A_{4}^{\text{tree}}\frac{s_{12}s_{14}}{s_{13}^{2}} \, .
\end{align}
\end{subequations}
In the case of the cuts $C^{\mbox{\tiny [R]}}_{12|3|4}$ and $C^{\mbox{\tiny [R]}}_{1|2|34}$ the first diagram gives  contributions to the both  cut-constructible and the rational part,
while the second one contributes to the rational part only. In the cuts $C^{\mbox{\tiny [R]}}_{12|3|4}$, $C^{\mbox{\tiny [R]}}_{1|23|4}$ and $C^{\mbox{\tiny [R]}}_{2|3|41}$ the last two diagrams cancel against
each other, i.e. the scalar contribution exactly compensates the contribution of the longitudinal polarization of the gluon.
The double cuts are
\vspace{-0.4cm}
\begin{subequations}
\begin{align}
C^{\mbox{\tiny [R]}}_{12|34} &{} =
      \parbox{20mm}{\unitlength=0.20bp%
\begin{feynartspicture}(300,300)(1,1)
\FADiagram{}
\FAProp(4.,11.5)(6.5,10.0)(0.,){/Cycles}{0}
\FAProp(6.5,10.0)(4.,8.5)(0.,){/Cycles}{0}
\FAProp(16.,11.5)(13.5,10.0)(0.,){/Straight}{1}
\FAProp(16.,8.5)(13.5,10.0)(0.,){/Straight}{-1}
\FAVert(6.5,10.){0}
\FAVert(13.5,10.0){0}
\FALabel(3.1,8.)[]{\tiny$1$}
\FALabel(3.1,12.)[]{\tiny$2$}
\FALabel(17.,12.)[]{\tiny$3$}
\FALabel(17.,8.)[]{\tiny$4$}
\FAProp(6.5,10.)(13.5,10.0)(0.8,){/Straight}{1}
\FAProp(6.5,10.)(13.5,10.0)(-0.8,){/Straight}{-1}
\FALabel(11.1,5.0)[]{\tiny $-$}
\FALabel(8.9,5.0)[]{\tiny $+$}
\FALabel(8.9, 14.5)[]{\tiny $-$}
\FALabel(11.1, 14.5)[]{\tiny $+$}
\end{feynartspicture}} \, , \nn[-2.ex]
c^{\mbox{\tiny [R]}}_{12|34; \; 0} &{}=  A_{4}^{\text{tree}}\left[\frac{s_{12}}{s_{13}}\left(\frac{s_{14}}{s_{13}}+\frac{3}{2}\right)+\frac{3}{2}\right]\, , \nn
c^{\mbox{\tiny [R]}}_{12|34; \; 2} &{} =  0 \, ;  \\
C^{\mbox{\tiny [R]}}_{23|41} &{}=
      \parbox{20mm}{\unitlength=0.20bp%
\begin{feynartspicture}(300,300)(1,1)
\FADiagram{}
\FAProp(7.5,15.)(10.0,13.5)(0.,){/Cycles}{0}
\FAProp(12.5,15.)(10.0,13.5)(0.,){/Straight}{1}
\FAProp(7.5,5.)(10.,6.5)(0.,){/Cycles}{0}
\FAProp(12.5,5.)(10.0,6.5)(0.,){/Straight}{-1}
\FAVert(10.0,13.5){0}
\FAVert(10.0,6.5){0}
\FALabel(6.5,16.)[]{\tiny$2$}
\FALabel(13.5,16.0)[]{\tiny$3$}
\FALabel(6.5,4.)[]{\tiny$1$}
\FALabel(13.5,4.)[]{\tiny$4$}
\FAProp(10.0,6.5)(10.0,13.5)(0.8,){/Cycles}{0}
\FAProp(10.0,6.5)(10.0,13.5)(-0.8,){/Straight}{-1}
\FALabel(5.0, 8.9)[]{\tiny $-$}
\FALabel(5.0, 11.1)[]{\tiny $+$}
\FALabel(14.5, 11.1)[]{\tiny $-$}
\FALabel(14.5, 8.9)[]{\tiny $+$}
\end{feynartspicture}}
   + \parbox{20mm}{\unitlength=0.20bp%
\begin{feynartspicture}(300,300)(1,1)
\FADiagram{}
\FAProp(7.5,15.)(10.0,13.5)(0.,){/Cycles}{0}
\FAProp(12.5,15.)(10.0,13.5)(0.,){/Straight}{1}
\FAProp(7.5,5.)(10.,6.5)(0.,){/Cycles}{0}
\FAProp(12.5,5.)(10.0,6.5)(0.,){/Straight}{-1}
\FAVert(10.0,13.5){0}
\FAVert(10.0,6.5){0}
\FALabel(6.5,16.)[]{\tiny$2$}
\FALabel(13.5,16.0)[]{\tiny$3$}
\FALabel(6.5,4.)[]{\tiny$1$}
\FALabel(13.5,4.)[]{\tiny$4$}
\FAProp(10.0,6.5)(10.0,13.5)(0.8,){/Cycles}{0}
\FAProp(10.0,6.5)(10.0,13.5)(-0.8,){/Straight}{-1}
\FALabel(5.0, 8.9)[]{\tiny $+$}
\FALabel(5.0, 11.1)[]{\tiny $-$}
\FALabel(14.5, 11.1)[]{\tiny $0$}
\FALabel(14.5, 8.9)[]{\tiny $0$}
\end{feynartspicture}}
+ \parbox{20mm}{\unitlength=0.20bp%
\begin{feynartspicture}(300,300)(1,1)
\FADiagram{}
\FAProp(7.5,15.)(10.0,13.5)(0.,){/Cycles}{0}
\FAProp(12.5,15.)(10.0,13.5)(0.,){/Straight}{1}
\FAProp(7.5,5.)(10.,6.5)(0.,){/Cycles}{0}
\FAProp(12.5,5.)(10.0,6.5)(0.,){/Straight}{-1}
\FAVert(10.0,13.5){0}
\FAVert(10.0,6.5){0}
\FALabel(6.5,16.)[]{\tiny$2$}
\FALabel(13.5,16.0)[]{\tiny$3$}
\FALabel(6.5,4.)[]{\tiny$1$}
\FALabel(13.5,4.)[]{\tiny$4$}
\FAProp(10.0,6.5)(10.0,13.5)(0.8,){/ScalarDash}{0}
\FAProp(10.0,6.5)(10.0,13.5)(-0.8,){/Straight}{-1}
\FALabel(5.0, 8.9)[]{\tiny $+$}
\FALabel(5.0, 11.1)[]{\tiny $-$}
%
\end{feynartspicture}} \, , \nn[-2.ex]
c^{\mbox{\tiny [R]}}_{23|41; \; 0} &{}=  -A_{4}^{\text{tree}}\frac{s_{12}}{s_{13}}\left(\frac{s_{14}}{s_{13}}+\frac{3}{2}\right)\, , \nn
c^{\mbox{\tiny [R]}}_{23|41; \; 2} &{} =  0 \, .
\end{align}
\end{subequations}
For the cut $C^{\mbox{\tiny [R]}}_{24}$, the first diagram contributes to the cut-constructible part only  while  the second one is cancelled by the diagram with an internal scalar.
\bigskip

\paragraph*{Leading-color amplitude --} The leading color amplitude can be obtained 
from  the decomposition~(\ref{Eq:Decomposition}) by using  the coefficients
 \begin{align}
 c_{i_1\cdots i_k; \, n }  =  c^{\mbox{\tiny [L]}}_{i_1\cdots i_k; \, n } -\frac{1}{N_c^2}  c^{\mbox{\tiny [R]}}_{i_1\cdots i_k; \, n } \, , 
 \end{align} 
and the explicit expression of the MIs, Eq.~(\ref{Eq:MIproM}). The result agrees with the one presented in Ref.~\cite{Kunszt:1993sd}.

\section{The $\mathbf{gggH}$ amplitude}
\label{sec:gggH} 
In this section, we show the calculation of the leading color one-loop
contribution to the helicity amplitude
$A_{4}\left(1_{g}^{-},2_{g}^{+},3_{g}^{+},H\right)$ in the heavy
top-mass limit. This example allows us to show how the FDF scheme can
be applied in the context of an effective theory, where the Higgs boson
couples directly to the gluon.
The Feynman rules for the Higgs-gluon and Higgs-scalar couplings in the 
FDF  are given in Appendix~\ref{App:COFR}. They are used to
compute the tree-level amplitudes sewn along the cuts.
The tree-level amplitudes are not shown,  but they can be easily 
obtained by using a construction similar to the one used in Appendix~\ref{App:TREE}.
In the following we  present directly the determination of the coefficients 
by means of generalized unitarity methods.

The leading-order contribution reads as follows
\begin{align}
A_{4,H}^{\mbox{\tiny tree}}&=i\frac{\left[23\right]^{4}}{\left[12\right]\left[23\right]\left[31\right]} \, .
\end{align}
The quadruple cuts are given by:
\vspace{-0.4cm}
\begin{subequations}
\begin{align}
C_{1|2|3|H} &{}= 
      \parbox{20mm}{\unitlength=0.20bp%
\begin{feynartspicture}(300,300)(1,1)
\FADiagram{}
%
\FAProp(4.,15.)(6.5,13.5)(0.,){/Cycles}{0}
\FAProp(4.,5.)(6.5,6.5)(0.,){/Cycles}{0}
\FAProp(16.,15.)(13.5,13.5)(0.,){/Cycles}{0}
\FAProp(16.,5.)(13.5,6.5)(0.,){/ScalarDash}{0}
%
\FAProp(6.5,13.5)(6.5,6.5)(0.,){/Cycles}{0}
\FAProp(13.5,13.5)(6.5,13.5)(0.,){/Cycles}{0}
\FAProp(6.5,6.5)(13.5,6.5)(0.,){/Cycles}{0}
\FAProp(13.5,13.5)(13.5,6.5)(0.,){/Cycles}{0}
%
\FAVert(6.5,13.5){0}
\FAVert(6.5,6.5){0}
\FAVert(13.5,13.5){0}
\FAVert(13.5,6.5){0}
%
%
%
%
\FALabel(3.,4.)[]{\tiny$1^-$}
\FALabel(3.,16.)[]{\tiny$2^+$}
\FALabel(17.,16.)[]{\tiny$3^+$}
\FALabel(17.,4.)[]{\tiny$H$}
\end{feynartspicture}
}
+     \parbox{20mm}{\unitlength=0.20bp%
\begin{feynartspicture}(300,300)(1,1)
\FADiagram{}
%
\FAProp(4.,15.)(6.5,13.5)(0.,){/Cycles}{0}
\FAProp(4.,5.)(6.5,6.5)(0.,){/Cycles}{0}
\FAProp(16.,15.)(13.5,13.5)(0.,){/Cycles}{0}
\FAProp(16.,5.)(13.5,6.5)(0.,){/ScalarDash}{0}
%
\FAProp(6.5,13.5)(6.5,6.5)(0.,){/ScalarDash}{0}
\FAProp(13.5,13.5)(6.5,13.5)(0.,){/ScalarDash}{0}
\FAProp(6.5,6.5)(13.5,6.5)(0.,){/ScalarDash}{0}
\FAProp(13.5,13.5)(13.5,6.5)(0.,){/ScalarDash}{0}
%
\FAVert(6.5,13.5){0}
\FAVert(6.5,6.5){0}
\FAVert(13.5,13.5){0}
\FAVert(13.5,6.5){0}
%
%
%
%
\FALabel(3.,4.)[]{\tiny$1^-$}
\FALabel(3.,16.)[]{\tiny$2^+$}
\FALabel(17.,16.)[]{\tiny$3^+$}
\FALabel(17.,4.)[]{\tiny$H$}
\end{feynartspicture}
}   \, , \nn[-2.ex]
c_{1|2|3|H;\,0}&{}=-\frac{1}{2}A_{4,H}^{\mbox{\tiny tree}}s_{12}s_{23} \, , \nn
c_{1|2|3|H;\,4}&{}=0 \, ;  \\
C_{1|2|H|3} &{}= 
      \parbox{20mm}{\unitlength=0.20bp%
\begin{feynartspicture}(300,300)(1,1)
\FADiagram{}
%
\FAProp(4.,15.)(6.5,13.5)(0.,){/Cycles}{0}
\FAProp(4.,5.)(6.5,6.5)(0.,){/Cycles}{0}
\FAProp(16.,15.)(13.5,13.5)(0.,){/ScalarDash}{0}
\FAProp(16.,5.)(13.5,6.5)(0.,){/Cycles}{0}
%
\FAProp(6.5,13.5)(6.5,6.5)(0.,){/Cycles}{0}
\FAProp(13.5,13.5)(6.5,13.5)(0.,){/Cycles}{0}
\FAProp(6.5,6.5)(13.5,6.5)(0.,){/Cycles}{0}
\FAProp(13.5,13.5)(13.5,6.5)(0.,){/Cycles}{0}
%
\FAVert(6.5,13.5){0}
\FAVert(6.5,6.5){0}
\FAVert(13.5,13.5){0}
\FAVert(13.5,6.5){0}
%
%
%
%
\FALabel(3.,4.)[]{\tiny$1^-$}
\FALabel(3.,16.)[]{\tiny$2^+$}
\FALabel(17.,16.)[]{\tiny$H$}
\FALabel(17.,4.)[]{\tiny$3^+$}
\end{feynartspicture}
}
+     \parbox{20mm}{\unitlength=0.20bp%
\begin{feynartspicture}(300,300)(1,1)
\FADiagram{}
%
\FAProp(4.,15.)(6.5,13.5)(0.,){/Cycles}{0}
\FAProp(4.,5.)(6.5,6.5)(0.,){/Cycles}{0}
\FAProp(16.,15.)(13.5,13.5)(0.,){/ScalarDash}{0}
\FAProp(16.,5.)(13.5,6.5)(0.,){/Cycles}{0}
%
\FAProp(6.5,13.5)(6.5,6.5)(0.,){/ScalarDash}{0}
\FAProp(13.5,13.5)(6.5,13.5)(0.,){/ScalarDash}{0}
\FAProp(6.5,6.5)(13.5,6.5)(0.,){/ScalarDash}{0}
\FAProp(13.5,13.5)(13.5,6.5)(0.,){/ScalarDash}{0}
%
\FAVert(6.5,13.5){0}
\FAVert(6.5,6.5){0}
\FAVert(13.5,13.5){0}
\FAVert(13.5,6.5){0}
%
%
%
%
\FALabel(3.,4.)[]{\tiny$1^-$}
\FALabel(3.,16.)[]{\tiny$2^+$}
\FALabel(17.,16.)[]{\tiny$H$}
\FALabel(17.,4.)[]{\tiny$3^+$}
\end{feynartspicture}
}   \, , \nn[-2.ex]
c_{1|2|H|3;\,0}&{}=-\frac{1}{2}A_{4,H}^{\mbox{\tiny tree}}s_{13}s_{12}\, , \nn
c_{1|2|H|3;\,4}&{}=0  \, ;  \\
C_{1|H|2|3} &{}= 
      \parbox{20mm}{\unitlength=0.20bp%
\begin{feynartspicture}(300,300)(1,1)
\FADiagram{}
%
\FAProp(4.,15.)(6.5,13.5)(0.,){/ScalarDash}{0}
\FAProp(4.,5.)(6.5,6.5)(0.,){/Cycles}{0}
\FAProp(16.,15.)(13.5,13.5)(0.,){/Cycles}{0}
\FAProp(16.,5.)(13.5,6.5)(0.,){/Cycles}{0}
%
\FAProp(6.5,13.5)(6.5,6.5)(0.,){/Cycles}{0}
\FAProp(13.5,13.5)(6.5,13.5)(0.,){/Cycles}{0}
\FAProp(6.5,6.5)(13.5,6.5)(0.,){/Cycles}{0}
\FAProp(13.5,13.5)(13.5,6.5)(0.,){/Cycles}{0}
%
\FAVert(6.5,13.5){0}
\FAVert(6.5,6.5){0}
\FAVert(13.5,13.5){0}
\FAVert(13.5,6.5){0}
%
%
%
%
\FALabel(3.,4.)[]{\tiny$1^-$}
\FALabel(3.,16.)[]{\tiny$H$}
\FALabel(17.,16.)[]{\tiny$2^+$}
\FALabel(17.,4.)[]{\tiny$3^+$}
\end{feynartspicture}
}
+     \parbox{20mm}{\unitlength=0.20bp%
\begin{feynartspicture}(300,300)(1,1)
\FADiagram{}
%
\FAProp(4.,15.)(6.5,13.5)(0.,){/ScalarDash}{0}
\FAProp(4.,5.)(6.5,6.5)(0.,){/Cycles}{0}
\FAProp(16.,15.)(13.5,13.5)(0.,){/Cycles}{0}
\FAProp(16.,5.)(13.5,6.5)(0.,){/Cycles}{0}
%
\FAProp(6.5,13.5)(6.5,6.5)(0.,){/ScalarDash}{0}
\FAProp(13.5,13.5)(6.5,13.5)(0.,){/ScalarDash}{0}
\FAProp(6.5,6.5)(13.5,6.5)(0.,){/ScalarDash}{0}
\FAProp(13.5,13.5)(13.5,6.5)(0.,){/ScalarDash}{0}
%
\FAVert(6.5,13.5){0}
\FAVert(6.5,6.5){0}
\FAVert(13.5,13.5){0}
\FAVert(13.5,6.5){0}
%
%
%
%
\FALabel(3.,4.)[]{\tiny$1^-$}
\FALabel(3.,16.)[]{\tiny$H$}
\FALabel(17.,16.)[]{\tiny$2^+$}
\FALabel(17.,4.)[]{\tiny$3^+$}
\end{feynartspicture}
}   \, , \nn[-2.ex]
c_{1|H|2|3;\,0}&{}=-\frac{1}{2}A_{4,H}^{\mbox{\tiny tree}}s_{23}s_{13}\, , \nn
c_{1|H|2|3;\,4}&{}=0 \, .
\end{align}
\label{Eq:coeffF}
\end{subequations}
The triple cuts with  two massive channels are
\vspace{-0.4cm}
\begin{subequations}
\begin{align}
C_{12|3|H} &{}= 
      \parbox{20mm}{\unitlength=0.20bp%
\begin{feynartspicture}(300,300)(1,1)
\FADiagram{}
%
\FAProp(4.,11.5)(6.5,10.0)(0.,){/Cycles}{0}
\FAProp(6.5,10.0)(4.,8.5)(0.,){/Cycles}{0}
\FAProp(16.,15.)(13.5,13.5)(0.,){/Cycles}{0}
\FAProp(16.,5.)(13.5,6.5)(0.,){/ScalarDash}{0}
%
\FAProp(13.5,13.5)(6.5,10.0)(0.,){/Cycles}{0}
\FAProp(6.5,10.0)(13.5,6.5)(0.,){/Cycles}{0}
\FAProp(13.5,13.5)(13.5,6.5)(0.,){/Cycles}{0}
%
%
\FAVert(13.5,13.5){0}
\FAVert(13.5,6.5){0}
\FAVert(6.5,10.){0}
%
%
%
%
\FALabel(3.1,8.)[]{\tiny$1^-$}
\FALabel(3.1,12.)[]{\tiny$2^+$}
\FALabel(17.,16.)[]{\tiny$3^+$}
\FALabel(17.,4.)[]{\tiny$H$}
\end{feynartspicture}}
+    \parbox{20mm}{\unitlength=0.20bp%
\begin{feynartspicture}(300,300)(1,1)
\FADiagram{}
%
\FAProp(4.,11.5)(6.5,10.0)(0.,){/Cycles}{0}
\FAProp(6.5,10.0)(4.,8.5)(0.,){/Cycles}{0}
\FAProp(16.,15.)(13.5,13.5)(0.,){/Cycles}{0}
\FAProp(16.,5.)(13.5,6.5)(0.,){/ScalarDash}{0}
%
\FAProp(13.5,13.5)(6.5,10.0)(0.,){/ScalarDash}{0}
\FAProp(6.5,10.0)(13.5,6.5)(0.,){/ScalarDash}{0}
\FAProp(13.5,13.5)(13.5,6.5)(0.,){/ScalarDash}{0}
%
%
\FAVert(13.5,13.5){0}
\FAVert(13.5,6.5){0}
\FAVert(6.5,10.){0}
%
%
%
%
\FALabel(3.1,8.)[]{\tiny$1^-$}
\FALabel(3.1,12.)[]{\tiny$2^+$}
\FALabel(17.,16.)[]{\tiny$3^+$}
\FALabel(17.,4.)[]{\tiny$H$}
\end{feynartspicture}}   \, , \nn[-2.ex]
c_{12|3|H;\,0} &{}=\frac{1}{2}A_{4,H}^{\mbox{\tiny tree}}\left(s_{13}+s_{23}\right)\, , \nn
c_{12|3|H;\,2} &{}=0  \, ;  \\
C_{12|H|3} &{}= 
      \parbox{20mm}{\unitlength=0.20bp%
\begin{feynartspicture}(300,300)(1,1)
\FADiagram{}
%
\FAProp(4.,11.5)(6.5,10.0)(0.,){/Cycles}{0}
\FAProp(6.5,10.0)(4.,8.5)(0.,){/Cycles}{0}
\FAProp(16.,15.)(13.5,13.5)(0.,){/ScalarDash}{0}
\FAProp(16.,5.)(13.5,6.5)(0.,){/Cycles}{0}
%
\FAProp(13.5,13.5)(6.5,10.0)(0.,){/Cycles}{0}
\FAProp(6.5,10.0)(13.5,6.5)(0.,){/Cycles}{0}
\FAProp(13.5,13.5)(13.5,6.5)(0.,){/Cycles}{0}
%
%
\FAVert(13.5,13.5){0}
\FAVert(13.5,6.5){0}
\FAVert(6.5,10.){0}
%
%
%
%
\FALabel(3.1,8.)[]{\tiny$1^-$}
\FALabel(3.1,12.)[]{\tiny$2^+$}
\FALabel(17.,16.)[]{\tiny$H$}
\FALabel(17.,4.)[]{\tiny$3^+$}
\end{feynartspicture}}
+    \parbox{20mm}{\unitlength=0.20bp%
\begin{feynartspicture}(300,300)(1,1)
\FADiagram{}
%
\FAProp(4.,11.5)(6.5,10.0)(0.,){/Cycles}{0}
\FAProp(6.5,10.0)(4.,8.5)(0.,){/Cycles}{0}
\FAProp(16.,15.)(13.5,13.5)(0.,){/ScalarDash}{0}
\FAProp(16.,5.)(13.5,6.5)(0.,){/Cycles}{0}
%
\FAProp(13.5,13.5)(6.5,10.0)(0.,){/ScalarDash}{0}
\FAProp(6.5,10.0)(13.5,6.5)(0.,){/ScalarDash}{0}
\FAProp(13.5,13.5)(13.5,6.5)(0.,){/ScalarDash}{0}
%
%
\FAVert(13.5,13.5){0}
\FAVert(13.5,6.5){0}
\FAVert(6.5,10.){0}
%
%
%
%
\FALabel(3.1,8.)[]{\tiny$1^-$}
\FALabel(3.1,12.)[]{\tiny$2^+$}
\FALabel(17.,16.)[]{\tiny$H$}
\FALabel(17.,4.)[]{\tiny$3^+$}
\end{feynartspicture}}   \, , \nn[-2.ex]
c_{12|H|3;\,0}&{}=\frac{1}{2}A_{4,H}^{\mbox{\tiny tree}}\left(s_{13}+s_{23}\right)\, , \nn
c_{12|H|3;\,2}&{}=0 \, ;   \\
C_{1|23|H} &{}= 
      \parbox{20mm}{\unitlength=0.20bp%
\begin{feynartspicture}(300,300)(1,1)
\FADiagram{}
%
\FAProp(7.5,15.)(10.0,13.5)(0.,){/Cycles}{0}
\FAProp(4.,5.)(6.5,6.5)(0.,){/Cycles}{0}
\FAProp(12.5,15.)(10.0,13.5)(0.,){/Cycles}{0}
\FAProp(16.,5.)(13.5,6.5)(0.,){/ScalarDash}{0}
%
\FAProp(10.0,13.5)(6.5,6.5)(0.,){/Cycles}{0}
\FAProp(6.5,6.5)(13.5,6.5)(0.,){/Cycles}{0}
\FAProp(10.0,13.5)(13.5,6.5)(0.,){/Cycles}{0}
%
\FAVert(10.0,13.5){0}
\FAVert(6.5,6.5){0}
\FAVert(10.0,13.5){0}
\FAVert(13.5,6.5){0}
%
%
%
%
%
\FALabel(3.,4.)[]{\tiny$1^-$}
\FALabel(6.5,16.)[]{\tiny$2^+$}
\FALabel(13.5,16.0)[]{\tiny$3^+$}
\FALabel(17.,4.)[]{\tiny$H$}
\end{feynartspicture}}
+    \parbox{20mm}{\unitlength=0.20bp%
\begin{feynartspicture}(300,300)(1,1)
\FADiagram{}
%
\FAProp(7.5,15.)(10.0,13.5)(0.,){/Cycles}{0}
\FAProp(4.,5.)(6.5,6.5)(0.,){/Cycles}{0}
\FAProp(12.5,15.)(10.0,13.5)(0.,){/Cycles}{0}
\FAProp(16.,5.)(13.5,6.5)(0.,){/ScalarDash}{0}
%
\FAProp(10.0,13.5)(6.5,6.5)(0.,){/ScalarDash}{0}
\FAProp(6.5,6.5)(13.5,6.5)(0.,){/ScalarDash}{0}
\FAProp(10.0,13.5)(13.5,6.5)(0.,){/ScalarDash}{0}
%
\FAVert(10.0,13.5){0}
\FAVert(6.5,6.5){0}
\FAVert(10.0,13.5){0}
\FAVert(13.5,6.5){0}
%
%
%
%
%
\FALabel(3.,4.)[]{\tiny$1^-$}
\FALabel(6.5,16.)[]{\tiny$2^+$}
\FALabel(13.5,16.0)[]{\tiny$3^+$}
\FALabel(17.,4.)[]{\tiny$H$}
\end{feynartspicture}}   \, , \nn[-2.ex]
c_{1|23|H;\,0}&{}=\frac{1}{2}A_{4,H}^{\mbox{\tiny tree}}\left(s_{12}+s_{13}\right)\, , \nn
c_{1|23|H;\,2}&{}=0 \, ;   \\
C_{1|H|23} &{}= 
      \parbox{20mm}{\unitlength=0.20bp%
\begin{feynartspicture}(300,300)(1,1)
\FADiagram{}
%
\FAProp(7.5,15.)(10.0,13.5)(0.,){/Cycles}{0}
\FAProp(4.,5.)(6.5,6.5)(0.,){/ScalarDash}{0}
\FAProp(12.5,15.)(10.0,13.5)(0.,){/Cycles}{0}
\FAProp(16.,5.)(13.5,6.5)(0.,){/Cycles}{0}
%
\FAProp(10.0,13.5)(6.5,6.5)(0.,){/Cycles}{0}
\FAProp(6.5,6.5)(13.5,6.5)(0.,){/Cycles}{0}
\FAProp(10.0,13.5)(13.5,6.5)(0.,){/Cycles}{0}
%
\FAVert(10.0,13.5){0}
\FAVert(6.5,6.5){0}
\FAVert(10.0,13.5){0}
\FAVert(13.5,6.5){0}
%
%
%
%
%
\FALabel(3.,4.)[]{\tiny$H$}
\FALabel(6.5,16.)[]{\tiny$2^+$}
\FALabel(13.5,16.0)[]{\tiny$3^+$}
\FALabel(17.,4.)[]{\tiny$1^-$}
\end{feynartspicture}}
+    \parbox{20mm}{\unitlength=0.20bp%
\begin{feynartspicture}(300,300)(1,1)
\FADiagram{}
%
\FAProp(7.5,15.)(10.0,13.5)(0.,){/Cycles}{0}
\FAProp(4.,5.)(6.5,6.5)(0.,){/ScalarDash}{0}
\FAProp(12.5,15.)(10.0,13.5)(0.,){/Cycles}{0}
\FAProp(16.,5.)(13.5,6.5)(0.,){/Cycles}{0}
%
\FAProp(10.0,13.5)(6.5,6.5)(0.,){/ScalarDash}{0}
\FAProp(6.5,6.5)(13.5,6.5)(0.,){/ScalarDash}{0}
\FAProp(10.0,13.5)(13.5,6.5)(0.,){/ScalarDash}{0}
%
\FAVert(10.0,13.5){0}
\FAVert(6.5,6.5){0}
\FAVert(10.0,13.5){0}
\FAVert(13.5,6.5){0}
%
%
%
%
%
\FALabel(3.,4.)[]{\tiny$H$}
\FALabel(6.5,16.)[]{\tiny$2^+$}
\FALabel(13.5,16.0)[]{\tiny$3^+$}
\FALabel(17.,4.)[]{\tiny$1^-$}
\end{feynartspicture}}   \, , \nn[-2.ex]
c_{1|H|23;\,0}&{}=\frac{1}{2}A_{4,H}^{\mbox{\tiny tree}}\left(s_{12}+s_{13}\right)\, , \nn
c_{1|H|23;\,2}&{}=0  \, ;  \\
C_{2|H|31} &{}= 
      \parbox{20mm}{\unitlength=0.20bp%
\begin{feynartspicture}(300,300)(1,1)
\FADiagram{}
%
\FAProp(4.,15.)(6.5,13.5)(0.,){/Cycles}{0}
\FAProp(7.5,5.)(10.,6.5)(0.,){/Cycles}{0}
\FAProp(16.,15.)(13.5,13.5)(0.,){/ScalarDash}{0}
\FAProp(12.5,5.)(10.0,6.5)(0.,){/Cycles}{0}
%
\FAProp(6.5,13.5)(10.0,6.5)(0.,){/Cycles}{0}
\FAProp(13.5,13.5)(6.5,13.5)(0.,){/Cycles}{0}
\FAProp(13.5,13.5)(10.0,6.5)(0.,){/Cycles}{0}
%
\FAVert(6.5,13.5){0}
\FAVert(13.5,13.5){0}
\FAVert(10.0,6.5){0}
%
%
%
%
%
\FALabel(6.5,4.)[]{\tiny$1^-$}
\FALabel(3.,16.)[]{\tiny$2^+$}
\FALabel(17.,16.)[]{\tiny$H$}
\FALabel(13.5,4.)[]{\tiny$3^+$}
\end{feynartspicture}}
+    \parbox{20mm}{\unitlength=0.20bp%
\begin{feynartspicture}(300,300)(1,1)
\FADiagram{}
%
\FAProp(4.,15.)(6.5,13.5)(0.,){/Cycles}{0}
\FAProp(7.5,5.)(10.,6.5)(0.,){/Cycles}{0}
\FAProp(16.,15.)(13.5,13.5)(0.,){/ScalarDash}{0}
\FAProp(12.5,5.)(10.0,6.5)(0.,){/Cycles}{0}
%
\FAProp(6.5,13.5)(10.0,6.5)(0.,){/ScalarDash}{0}
\FAProp(13.5,13.5)(6.5,13.5)(0.,){/ScalarDash}{0}
\FAProp(13.5,13.5)(10.0,6.5)(0.,){/ScalarDash}{0}
%
\FAVert(6.5,13.5){0}
\FAVert(13.5,13.5){0}
\FAVert(10.0,6.5){0}
%
%
%
%
%
\FALabel(6.5,4.)[]{\tiny$1^-$}
\FALabel(3.,16.)[]{\tiny$2^+$}
\FALabel(17.,16.)[]{\tiny$H$}
\FALabel(13.5,4.)[]{\tiny$3^+$}
\end{feynartspicture}}   \, , \nn[-2.ex]
c_{2|H|31\,0}&{}=\frac{1}{2}A_{4,H}^{\mbox{\tiny tree}}\left(s_{12}+s_{23}\right)\, , \nn
c_{2|H|31\,2}&{}=0 \, ;   \\
C_{H|2|31} &{}= 
      \parbox{20mm}{\unitlength=0.20bp%
\begin{feynartspicture}(300,300)(1,1)
\FADiagram{}
%
\FAProp(4.,15.)(6.5,13.5)(0.,){/ScalarDash}{0}
\FAProp(7.5,5.)(10.,6.5)(0.,){/Cycles}{0}
\FAProp(16.,15.)(13.5,13.5)(0.,){/Cycles}{0}
\FAProp(12.5,5.)(10.0,6.5)(0.,){/Cycles}{0}
%
\FAProp(6.5,13.5)(10.0,6.5)(0.,){/Cycles}{0}
\FAProp(13.5,13.5)(6.5,13.5)(0.,){/Cycles}{0}
\FAProp(13.5,13.5)(10.0,6.5)(0.,){/Cycles}{0}
%
\FAVert(6.5,13.5){0}
\FAVert(13.5,13.5){0}
\FAVert(10.0,6.5){0}
%
%
%
%
%
\FALabel(6.5,4.)[]{\tiny$1^-$}
\FALabel(3.,16.)[]{\tiny$H$}
\FALabel(17.,16.)[]{\tiny$2^+$}
\FALabel(13.5,4.)[]{\tiny$3^+$}
\end{feynartspicture}}
+    \parbox{20mm}{\unitlength=0.20bp%
\begin{feynartspicture}(300,300)(1,1)
\FADiagram{}
%
\FAProp(4.,15.)(6.5,13.5)(0.,){/ScalarDash}{0}
\FAProp(7.5,5.)(10.,6.5)(0.,){/Cycles}{0}
\FAProp(16.,15.)(13.5,13.5)(0.,){/Cycles}{0}
\FAProp(12.5,5.)(10.0,6.5)(0.,){/Cycles}{0}
%
\FAProp(6.5,13.5)(10.0,6.5)(0.,){/ScalarDash}{0}
\FAProp(13.5,13.5)(6.5,13.5)(0.,){/ScalarDash}{0}
\FAProp(13.5,13.5)(10.0,6.5)(0.,){/ScalarDash}{0}
%
\FAVert(6.5,13.5){0}
\FAVert(13.5,13.5){0}
\FAVert(10.0,6.5){0}
%
%
%
%
%
\FALabel(6.5,4.)[]{\tiny$1^-$}
\FALabel(3.,16.)[]{\tiny$H$}
\FALabel(17.,16.)[]{\tiny$2^+$}
\FALabel(13.5,4.)[]{\tiny$3^+$}
\end{feynartspicture}}   \, , \nn[-2.ex]
c_{H|2|31;\,0}&{}=\frac{1}{2}A_{4,H}^{\mbox{\tiny tree}}\left(s_{12}+s_{23}\right)\, , \nn
c_{H|2|31;\,2}&{}=0 \, ;
\end{align}
\end{subequations}
while the ones  with one massive channel only read as follows:
\vspace{-0.4cm}
\begin{subequations}
\begin{align}
C_{1|2|3H} &{}= 
      \parbox{20mm}{\unitlength=0.20bp%
\begin{feynartspicture}(300,300)(1,1)
\FADiagram{}
%
\FAProp(4.,15.)(6.5,13.5)(0.,){/Cycles}{0}
\FAProp(4.,5.)(6.5,6.5)(0.,){/Cycles}{0}
\FAProp(16.,11.5)(13.5,10.0)(0.,){/Cycles}{0}
\FAProp(16.,8.5)(13.5,10.0)(0.,){/ScalarDash}{0}
%
\FAProp(6.5,13.5)(6.5,6.5)(0.,){/Cycles}{0}
\FAProp(13.5,10.0)(6.5,13.5)(0.,){/Cycles}{0}
\FAProp(6.5,6.5)(13.5,10.0)(0.,){/Cycles}{0}
%
\FAVert(6.5,13.5){0}
\FAVert(6.5,6.5){0}
\FAVert(13.5,10.0){0}
%
%
%
%
\FALabel(3.,4.)[]{\tiny$1^-$}
\FALabel(3.,16.)[]{\tiny$2^+$}
\FALabel(17.,12.)[]{\tiny$3^+$}
\FALabel(17.,8.)[]{\tiny$H$}
\end{feynartspicture}}
+    \parbox{20mm}{\unitlength=0.20bp%
\begin{feynartspicture}(300,300)(1,1)
\FADiagram{}
%
\FAProp(4.,15.)(6.5,13.5)(0.,){/Cycles}{0}
\FAProp(4.,5.)(6.5,6.5)(0.,){/Cycles}{0}
\FAProp(16.,11.5)(13.5,10.0)(0.,){/Cycles}{0}
\FAProp(16.,8.5)(13.5,10.0)(0.,){/ScalarDash}{0}
%
\FAProp(6.5,13.5)(6.5,6.5)(0.,){/ScalarDash}{0}
\FAProp(13.5,10.0)(6.5,13.5)(0.,){/ScalarDash}{0}
\FAProp(6.5,6.5)(13.5,10.0)(0.,){/ScalarDash}{0}
%
\FAVert(6.5,13.5){0}
\FAVert(6.5,6.5){0}
\FAVert(13.5,10.0){0}
%
%
%
%
\FALabel(3.,4.)[]{\tiny$1^-$}
\FALabel(3.,16.)[]{\tiny$2^+$}
\FALabel(17.,12.)[]{\tiny$3^+$}
\FALabel(17.,8.)[]{\tiny$H$}
\end{feynartspicture}}   \, , \nn[-2.ex]
c_{1|2|3H;\,0}&{}=0\, , \nn
c_{1|2|3H;\,0}&{}=0  \, ;   \\
C_{1|2H|3} &{}= 
      \parbox{20mm}{\unitlength=0.20bp%
\begin{feynartspicture}(300,300)(1,1)
\FADiagram{}
%
\FAProp(4.,15.)(6.5,13.5)(0.,){/Cycles}{0}
\FAProp(4.,5.)(6.5,6.5)(0.,){/Cycles}{0}
\FAProp(16.,11.5)(13.5,10.0)(0.,){/Cycles}{0}
\FAProp(16.,8.5)(13.5,10.0)(0.,){/ScalarDash}{0}
%
\FAProp(6.5,13.5)(6.5,6.5)(0.,){/Cycles}{0}
\FAProp(13.5,10.0)(6.5,13.5)(0.,){/Cycles}{0}
\FAProp(6.5,6.5)(13.5,10.0)(0.,){/Cycles}{0}
%
\FAVert(6.5,13.5){0}
\FAVert(6.5,6.5){0}
\FAVert(13.5,10.0){0}
%
%
%
%
\FALabel(3.,4.)[]{\tiny$3^+$}
\FALabel(3.,16.)[]{\tiny$1^-$}
\FALabel(17.,12.)[]{\tiny$2^+$}
\FALabel(17.,8.)[]{\tiny$H$}
\end{feynartspicture}}
+    \parbox{20mm}{\unitlength=0.20bp%
\begin{feynartspicture}(300,300)(1,1)
\FADiagram{}
%
\FAProp(4.,15.)(6.5,13.5)(0.,){/Cycles}{0}
\FAProp(4.,5.)(6.5,6.5)(0.,){/Cycles}{0}
\FAProp(16.,11.5)(13.5,10.0)(0.,){/Cycles}{0}
\FAProp(16.,8.5)(13.5,10.0)(0.,){/ScalarDash}{0}
%
\FAProp(6.5,13.5)(6.5,6.5)(0.,){/ScalarDash}{0}
\FAProp(13.5,10.0)(6.5,13.5)(0.,){/ScalarDash}{0}
\FAProp(6.5,6.5)(13.5,10.0)(0.,){/ScalarDash}{0}
%
\FAVert(6.5,13.5){0}
\FAVert(6.5,6.5){0}
\FAVert(13.5,10.0){0}
%
%
%
%
\FALabel(3.,4.)[]{\tiny$3^+$}
\FALabel(3.,16.)[]{\tiny$1^-$}
\FALabel(17.,12.)[]{\tiny$2^+$}
\FALabel(17.,8.)[]{\tiny$H$}
\end{feynartspicture}}   \, , \nn[-2.ex]
c_{1|2H|3;\,0}&{}=0\, , \nn
c_{1|2H|3;\,0}&{}=0  \, ;  \\
C_{1H|2|3} &{}= 
      \parbox{20mm}{\unitlength=0.20bp%
\begin{feynartspicture}(300,300)(1,1)
\FADiagram{}
%
\FAProp(4.,15.)(6.5,13.5)(0.,){/Cycles}{0}
\FAProp(4.,5.)(6.5,6.5)(0.,){/Cycles}{0}
\FAProp(16.,11.5)(13.5,10.0)(0.,){/Cycles}{0}
\FAProp(16.,8.5)(13.5,10.0)(0.,){/ScalarDash}{0}
%
\FAProp(6.5,13.5)(6.5,6.5)(0.,){/Cycles}{0}
\FAProp(13.5,10.0)(6.5,13.5)(0.,){/Cycles}{0}
\FAProp(6.5,6.5)(13.5,10.0)(0.,){/Cycles}{0}
%
\FAVert(6.5,13.5){0}
\FAVert(6.5,6.5){0}
\FAVert(13.5,10.0){0}
%
%
%
%
\FALabel(3.,4.)[]{\tiny$2^+$}
\FALabel(3.,16.)[]{\tiny$3^+$}
\FALabel(17.,12.)[]{\tiny$1^-$}
\FALabel(17.,8.)[]{\tiny$H$}
\end{feynartspicture}}
+    \parbox{20mm}{\unitlength=0.20bp%
\begin{feynartspicture}(300,300)(1,1)
\FADiagram{}
%
\FAProp(4.,15.)(6.5,13.5)(0.,){/Cycles}{0}
\FAProp(4.,5.)(6.5,6.5)(0.,){/Cycles}{0}
\FAProp(16.,11.5)(13.5,10.0)(0.,){/Cycles}{0}
\FAProp(16.,8.5)(13.5,10.0)(0.,){/ScalarDash}{0}
%
\FAProp(6.5,13.5)(6.5,6.5)(0.,){/ScalarDash}{0}
\FAProp(13.5,10.0)(6.5,13.5)(0.,){/ScalarDash}{0}
\FAProp(6.5,6.5)(13.5,10.0)(0.,){/ScalarDash}{0}
%
\FAVert(6.5,13.5){0}
\FAVert(6.5,6.5){0}
\FAVert(13.5,10.0){0}
%
%
%
%
\FALabel(3.,4.)[]{\tiny$2^+$}
\FALabel(3.,16.)[]{\tiny$3^+$}
\FALabel(17.,12.)[]{\tiny$1^-$}
\FALabel(17.,8.)[]{\tiny$H$}
\end{feynartspicture}}   \, , \nn[-2.ex]
c_{1H|2|3;\,0}&{}=0\, , \nn
c_{1H|2|3;\,2}&{}=-2A_{4,H}^{\mbox{\tiny tree}}\frac{s_{12}s_{13}}{s_{23}^{2}}  \, .
\end{align}
\end{subequations}
Finally the double cuts are given by:
\vspace{-0.4cm}
\begin{subequations}
\begin{align}
C_{12|3H} &{}= 
      \parbox{20mm}{\unitlength=0.20bp%
\begin{feynartspicture}(300,300)(1,1)
\FADiagram{}
\FAProp(4.,11.5)(6.5,10.0)(0.,){/Cycles}{0}
\FAProp(6.5,10.0)(4.,8.5)(0.,){/Cycles}{0}
\FAProp(16.,11.5)(13.5,10.0)(0.,){/Cycles}{0}
\FAProp(16.,8.5)(13.5,10.0)(0.,){/ScalarDash}{0}
\FAVert(6.5,10.){0}
\FAVert(13.5,10.0){0}
\FALabel(3.1,8.)[]{\tiny$1^-$}
\FALabel(3.1,12.)[]{\tiny$2^+$}
\FALabel(17.,12.)[]{\tiny$3^+$}
\FALabel(17.,8.)[]{\tiny$H$}
\FAProp(6.5,10.)(13.5,10.0)(0.8,){/Cycles}{0}
\FAProp(6.5,10.)(13.5,10.0)(-0.8,){/Cycles}{0}
%
\end{feynartspicture}}
+    \parbox{20mm}{\unitlength=0.20bp%
\begin{feynartspicture}(300,300)(1,1)
\FADiagram{}
\FAProp(4.,11.5)(6.5,10.0)(0.,){/Cycles}{0}
\FAProp(6.5,10.0)(4.,8.5)(0.,){/Cycles}{0}
\FAProp(16.,11.5)(13.5,10.0)(0.,){/Cycles}{0}
\FAProp(16.,8.5)(13.5,10.0)(0.,){/ScalarDash}{0}
\FAVert(6.5,10.){0}
\FAVert(13.5,10.0){0}
\FALabel(3.1,8.)[]{\tiny$1^-$}
\FALabel(3.1,12.)[]{\tiny$2^+$}
\FALabel(17.,12.)[]{\tiny$3^+$}
\FALabel(17.,8.)[]{\tiny$H$}
\FAProp(6.5,10.)(13.5,10.0)(0.8,){/ScalarDash}{0}
\FAProp(6.5,10.)(13.5,10.0)(-0.8,){/ScalarDash}{0}
%
\end{feynartspicture}}   \, , \nn[-2.ex]
c_{12|3H;\,0}&{}=0\, , \nn
c_{12|3H;\,2}&{}=0  \, ;   \\
C_{23|H1} &{}= 
      \parbox{20mm}{\unitlength=0.20bp%
\begin{feynartspicture}(300,300)(1,1)
\FADiagram{}
\FAProp(7.5,15.)(10.0,13.5)(0.,){/Cycles}{0}
\FAProp(12.5,15.)(10.0,13.5)(0.,){/Cycles}{0}
\FAProp(7.5,5.)(10.,6.5)(0.,){/Cycles}{0}
\FAProp(12.5,5.)(10.0,6.5)(0.,){/ScalarDash}{0}
\FAVert(10.0,13.5){0}
\FAVert(10.0,6.5){0}
\FALabel(6.5,16.)[]{\tiny$2^+$}
\FALabel(13.5,16.0)[]{\tiny$3^+$}
\FALabel(6.5,4.)[]{\tiny$1^-$}
\FALabel(13.5,4.)[]{\tiny$H$}
\FAProp(10.0,6.5)(10.0,13.5)(0.8,){/Cycles}{0}
\FAProp(10.0,6.5)(10.0,13.5)(-0.8,){/Cycles}{0}
%
%
\end{feynartspicture}}
+    \parbox{20mm}{\unitlength=0.20bp%
\begin{feynartspicture}(300,300)(1,1)
\FADiagram{}
\FAProp(7.5,15.)(10.0,13.5)(0.,){/Cycles}{0}
\FAProp(12.5,15.)(10.0,13.5)(0.,){/Cycles}{0}
\FAProp(7.5,5.)(10.,6.5)(0.,){/Cycles}{0}
\FAProp(12.5,5.)(10.0,6.5)(0.,){/ScalarDash}{0}
\FAVert(10.0,13.5){0}
\FAVert(10.0,6.5){0}
\FALabel(6.5,16.)[]{\tiny$2^+$}
\FALabel(13.5,16.0)[]{\tiny$3^+$}
\FALabel(6.5,4.)[]{\tiny$1^-$}
\FALabel(13.5,4.)[]{\tiny$H$}
\FAProp(10.0,6.5)(10.0,13.5)(0.8,){/ScalarDash}{0}
\FAProp(10.0,6.5)(10.0,13.5)(-0.8,){/ScalarDash}{0}
%
%
\end{feynartspicture}}   \, , \nn[-2.ex]
c_{23|H1;\,0}&{}=0\, , \nn
c_{23|H1;\,2}&{}=4A_{4,H}^{\mbox{\tiny tree}}\frac{s_{12}s_{13}}{s_{23}^{3}}  \, ;   \\
C_{H2|31} &{}= 
      \parbox{20mm}{\unitlength=0.20bp%
\begin{feynartspicture}(300,300)(1,1)
\FADiagram{}
\FAProp(7.5,15.)(10.0,13.5)(0.,){/Cycles}{0}
\FAProp(12.5,15.)(10.0,13.5)(0.,){/Cycles}{0}
\FAProp(7.5,5.)(10.,6.5)(0.,){/Cycles}{0}
\FAProp(12.5,5.)(10.0,6.5)(0.,){/ScalarDash}{0}
\FAVert(10.0,13.5){0}
\FAVert(10.0,6.5){0}
\FALabel(6.5,16.)[]{\tiny$3^+$}
\FALabel(13.5,16.0)[]{\tiny$1^-$}
\FALabel(6.5,4.)[]{\tiny$2^+$}
\FALabel(13.5,4.)[]{\tiny$H$}
\FAProp(10.0,6.5)(10.0,13.5)(0.8,){/Cycles}{0}
\FAProp(10.0,6.5)(10.0,13.5)(-0.8,){/Cycles}{0}
%
%
\end{feynartspicture}}
+    \parbox{20mm}{\unitlength=0.20bp%
\begin{feynartspicture}(300,300)(1,1)
\FADiagram{}
\FAProp(7.5,15.)(10.0,13.5)(0.,){/Cycles}{0}
\FAProp(12.5,15.)(10.0,13.5)(0.,){/Cycles}{0}
\FAProp(7.5,5.)(10.,6.5)(0.,){/Cycles}{0}
\FAProp(12.5,5.)(10.0,6.5)(0.,){/ScalarDash}{0}
\FAVert(10.0,13.5){0}
\FAVert(10.0,6.5){0}
\FALabel(6.5,16.)[]{\tiny$3^+$}
\FALabel(13.5,16.0)[]{\tiny$1^-$}
\FALabel(6.5,4.)[]{\tiny$2^+$}
\FALabel(13.5,4.)[]{\tiny$H$}
\FAProp(10.0,6.5)(10.0,13.5)(0.8,){/ScalarDash}{0}
\FAProp(10.0,6.5)(10.0,13.5)(-0.8,){/ScalarDash}{0}
%
%
\end{feynartspicture}}   \, , \nn[-2.ex]
c_{H2|31;\,0}&{}=0\, , \nn
c_{H2|31;\,2}&{}=0 \, .
\end{align}
\label{Eq:coeffL}
\end{subequations}
The cut $C_{123 | H}$ does not give any contribution.
\smallskip

Finally, the one-loop amplitude can be obtained by using the coefficients collected in Eqs.~(\ref{Eq:coeffF}) - (\ref{Eq:coeffL}) and the decomposition~(\ref{Eq:DecompositionH}).
The result agrees with the literature~\cite{Schmidt:1997wr}.

\section{Generalized Open Loop}
\label{sec:ol}

The FDF of $d$-dimensional one-loop amplitudes  is compatible 
with methods generating recursively  the integrands of one-loop amplitudes~ \cite{vanHameren:2009vq,Heinrich:2010ax}
and  leads to the {\it complete  reconstruction}  of the  numerator of Feynman integrands 
as a polynomial in the loop variables, $\ell^\nu$ and $\mu$.  Our scheme allows for a generalization of the   
current implementations of these techniques~\cite{Cascioli:2011va,Hirschi:2011pa,Actis:2012qn}. 
Indeed,  currently the latter can reconstruct only the four-dimensional part the numerator of the integrands,
which is polynomial in $\ell^\nu$ only.  In the following we focus on the generalization of the  {\it open-loop} technique~\cite{Cascioli:2011va} 
within the FDF scheme.

Tree-level and one-loop amplitudes,  ${\cal M}$  and $\delta{\cal M}$,  can be obtained as a sum of Feynman diagrams
\begin{align}
{\cal M} = \sum_{\mbox{\tiny diag}}\, {\cal M} ^{(\mbox{\tiny diag})} \qquad \delta  {\cal M} = \sum_{\mbox{\tiny diag}}\, \delta {\cal M} ^{(\mbox{\tiny diag})}  \, .
\end{align}
The color factor $ \mathcal{C}$  and the  ($-2\epsilon$)-SRs  term $\mathcal{T}$  factorize, thus they  can be stripped off  each diagram
\begin{align}
  {\cal M} ^{(\mbox{\tiny diag})} &=   {\cal C} ^{(\mbox{\tiny diag})}  \,  {\cal A} ^{(\mbox{\tiny diag})} \nn
 \delta {\cal M} ^{(\mbox{\tiny diag})}  &=   {\cal C} ^{(\mbox{\tiny diag})}  \,{\cal T} ^{(\mbox{\tiny diag})}  \,  {\cal A} ^{(\mbox{\tiny diag})}   \, .
\end{align}
The color structures are computed once, as described
in~\cite{Cascioli:2011va}. 
The computation of the   ($-2\epsilon$)-SRs prefactors $\mathcal{T}$
turns out to be even easier,  since they enter  only in the  one-loop
diagrams  and  can be computed once and for all. In the  't Hooft-Feynman gauge 
they can be either $0$ or $1$.

The recursive construction of the color-stripped tree-level diagrams, ${\cal A} ^{(\mbox{\tiny diag})}$, 
is not affected by the new Feynman particles and Feynman rules, which enter at loop-level only.

The one-loop color-stripped  diagram $\delta {\cal A} ^{(\mbox{\tiny diag})}$, characterized by 
a given topology ${\cal I}_n$, is constructed by $n$ tree-level topologies $i_1, \ldots , i_n$,
connected to the loop. The numerator of of the one-loop diagram can be expressed as 
\begin{align}
\mathcal{N}\left( \mathcal{I}_n , \ell, \mu \right ) = \sum_{j=0}^R \sum_{a=0}^{R-j}  \, \mathcal{N}^{[a]}_{\nu_1 \cdots \nu_j} \left( \mathcal{I}_n  \right ) \ell^{\nu_1}\cdots   \ell^{\nu_j} \, \mu^a \, , 
\end{align}
where $R$ is its  rank. The diagram is obtained by performing the integration over the $d$-dimensional loop momentum:
\begin{align}
\delta {\cal A} ^{(\mbox{\tiny diag})} = \sum_{j=0}^R \sum_{a=0}^{R-j}  \, \mathcal{N}^{[a]}_{\nu_1 \cdots \nu_j} \left( \mathcal{I}_n  \right ) \,  I_n^{\nu_1 \cdots \nu_j}\left [\mu^a \right ]\,  ,
 \end{align}
where
\begin{align}
 I_n^{\nu_1 \cdots \nu_j}\left [  \mu^a  \right ] \equiv \int \,  d^d \bar \ell \; \frac{\, \ell^{\nu_1}\ \cdots  \ell^{\nu_j} \, \mu^a}{D_0\cdots D_{n-1}} \, .
\end{align}
The starting point of the open-loop technique is to cut a propagator and to remove the denominators.  The open numerator can be expressed 
in terms of the tree-level topology $i_n$ and a one loop topology ${\cal I}_{n-1}$:
\begin{align}
\mathcal{N}^{\beta}_{\phantom{\beta} \alpha}\left( \mathcal{I}_n , \ell, \mu \right ) &= X^\beta_{\gamma \delta} \left ( {\cal I}_n,  i_n  , {\cal I}_{n-1}\right ) \, \nn
& \quad \mathcal{N}^{\gamma}_{\phantom{\gamma} \alpha}\left( \mathcal{I}_{n-1} , \ell, \mu \right )  \omega^{\delta}\left (i_n\right )\, ,
\end{align}
where $\omega^\delta$ is the expression related to the tree-level topology $i_n$. The vertices $X^\beta_{\gamma \delta}$ 
are obtained by the FDF Feynman rules, Eqs.~(\ref{Eq:FR4}), and can be written as follows,
\begin{align}
X^\beta_{\gamma \delta} =  Y^\beta_{\gamma \delta} + \ell^\nu \, Z^\beta_{\nu; \, \gamma \delta} +  \mu \, W^\beta_{\gamma \delta}  \, .
\end{align}
Therefore the tensor coefficients of the covariant decomposition
\begin{align*}
\mathcal{N}^{\beta}_{\phantom{\beta} \alpha}\left( \mathcal{I}_n , \ell, \mu \right ) &=  \sum_{j=0}^R \sum_{a=0}^{R-j}  \,
\mathcal{N}^{[a]\, \beta}_{\nu_1 \cdots \nu_j ; \,  \alpha}\left( \mathcal{I}_n  \right )  \, \ell^{\nu_1} \cdots  \ell^{\nu_j}  \, \mu^a 
\end{align*}
are obtained by the recursive relation
\begin{align}
&\mathcal{N}^{[a]\, \beta}_{\nu_1 \cdots \nu_j ; \,  \alpha}\left( \mathcal{I}_n  \right )  = \big [ 
Y^{\beta}_{\gamma \delta} \, \mathcal{N}^{[a]\, \beta}_{\nu_1 \cdots \nu_j ; \,  \alpha}\left( \mathcal{I}_{n-1}  \right )  \nn
& \quad +  Z^{\beta}_{\nu_1; \, \gamma \delta}\,  \mathcal{N}^{[a]\, \beta}_{\nu_2 \cdots \nu_j ; \,  \alpha}\left( \mathcal{I}_{n-1}  \right )  \nn
& \quad + W^{\beta}_{\gamma \delta}\,  \mathcal{N}^{[a-1]\, \beta}_{\nu_1 \cdots \nu_j ; \,  \alpha}\left( \mathcal{I}_{n-1}  \right )  
\big ] \, \omega^\delta(i_n) \, .
\end{align}
The recursive generation of integrands within the FDF
can be suitably combined  with public codes  like {\sc Samurai}~\cite{Mastrolia:2010nb} and
{\sc  Ninja}~\cite{Mastrolia:2012bu,Peraro:2014cba}, which can reduce
integrands keeping the full dependence on the loop variables $\ell^\nu$ and $\mu$.
Moreover it can improve the generation of the $d$-dimensional integrands
performed by the packages  {\sc GoSam}~\cite{Cullen:2011ac} and {\sc FormCalc}~\cite{Hahn:1998yk}.
The latter  are public codes dedicated to the automatic evaluation of one-loop multi-particle
scattering amplitudes, and they  already support the FDH regularization scheme.

\section{Conclusions}

We introduced a four-dimensional formulation (FDF) of the $d$-dimensional
regularization of one-loop scattering amplitudes.
Within our FDF, particles that propagate inside the loop 
are represented by massive particles regularizing the divergences.
Their interactions are described by  generalized four-dimensional 
Feynman rules. They  include selection rules   accounting 
for the regularization of the amplitudes.
In particular, massless spin-1 particles in $d$-dimensions were 
represented in four-dimensions by a combination of massive spin-one
particle  and a scalar particle.
Fermions in $d$-dimensions were represented by four-dimensional
fermions obeying the Dirac equation for tachyonic particles.
The integrands of one-loop amplitudes in the FDF  and in the FDH scheme 
differ by spurious terms which vanish upon integration over the loop momentum.
Therefore the two schemes are  equivalent.
\smallskip

In the  FDF, the  polarization and helicity  states
of the particles inside the loop admit an explicit four-dimensional
representation, allowing for  a complete,
four-dimensional, unitarity-based construction
of $d$-dimensional amplitudes.  The  
application of generalized-unitarity methods within the FDF
has been  described in detail by computing
the NLO QCD corrections to helicity amplitudes of
the processes  $gg \to q{\bar q}$, and  $gg \to gH$. \\
Mutual cancellations among the contributions of the longitudinal gluons and the ones of the scalar particles 
suggest a  connection among them that deserves further investigations. 
\smallskip

The FDF Feynman rules  are compatible with methods generating recursively the integrands of one-loop amplitudes.
In this context we have proposed a generalization to the open loop method, which 
allows for a complete reconstruction of the integrand, currently limited to four dimensions only.
\smallskip

The FDF approach is suitable for analytic as
well as numerical implementation. Its main asset  is the
use  of purely  four-dimensional ingredients for the complete
reconstruction of dimensionally-regulated  one-loop amplitudes.
We plan to investigate its applicability  beyond one loop. In particular
we  aim at using explicit four-dimensional representations to avoid
 the complications emerging  from the formal  manipulations 
 of the $(d-4)$-dimensional degrees of freedom.

\begin{acknowledgments}
We wish to thank Francesco Buciuni for cross-checking parts of the results. 
A.R.F. and W.J.T.  thank the Max-Planck-Institute for Physics in Munich for the kind hospitality at several stages of this project. 
For the same reasons, E.M. wishes to  thank the Department of Mathematics and Physics of the University of Salento. \\
A.R.F.  is partially supported by the UNAL-DIB grant N. 20629 of  the  
``Convocatoria del programa nacional de proyectos para el fortalecimiento de la 
investigaci\`on, la creaci\`on y la innovaci\`on en posgrados de la Universidad  Nacional de Colombia 2013-2015". \\
The work of   P.M.  is  supported by the Alexander von Humboldt Foundation, in the framework of the Sofja Kovaleskaja Award Project
``Advanced Mathematical Methods for Particle Physics'', endowed by the German Federal Ministry of Education and Research.  \\
W.J.T. is supported by Fondazione Cassa di Risparmio di Padova e Rovigo (CARIPARO).
\end{acknowledgments}

\appendix
\section{One-loop equivalence}
\label{App:Equiv}

In this Appendix we show that, at one loop,    the FDH
scheme defined by Eqs.~(\ref{Eq:OrthoGs}) -- (\ref{Eq:Gamma02}) is equivalent to 
the one defined by Eqs.~(\ref{Eq:OrthoGs}) -- (\ref{Eq:Gamma01}) and~(\ref{Eq:Gamma03}).

In the two approaches the only differences
may arise from the manipulations  of the
$-2\epsilon$ components 
of the Dirac matrices contracted among each others.
Therefore potential differences in their 
predictions  can only be rational contributions  of
divergent diagrams involving at least an open fermion line.
The loop-dependent part of the integrand of a one-loop diagram is a sum of integrands  of the type
\begin{align}
\mathcal{I}_{\, r,a,k}  &\equiv \frac{\ell^{\mu_1} \cdots \ell^{\mu_r} \; (\mu^2)^a}{D_{i_1}\cdots D_{i_k}}  \, , \nn
 D_j &\equiv (\ell+p_j)^2 -m_j - \mu^2 \, .
\label{}
\end{align}
An integrand $\mathcal{I}_{\, r,a,k}$ leads to a  divergent integral if it  satisfies the conditions
\begin{align}
4 + r + 2\, a  - 2 \, k \ge 0 \, .
\label{Eq:CondDIV}
\end{align}
At one loop in QCD  the diagrams involving at least an open fermion line and integrands fulfilling the conditions~(\ref{Eq:CondDIV}) are
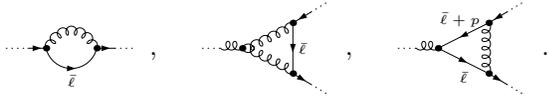
\vspace{-0.2cm}
 \begin{align}
 \parbox{20mm}{\unitlength=0.20bp%
\begin{feynartspicture}(300,300)(1,1)
\FADiagram{}
\FAProp(4.,10.0)(6.5,10.0)(0.,){/Straight}{1}
\FAProp(1.,10.0)(4.,10.0)(0.,){/GhostDash}{0}
\FAProp(16.,10.0)(13.5,10.0)(0.,){/Straight}{-1}
\FAProp(16.,10.0)(19.0,10.0)(0.,){/GhostDash}{0}
\FAVert(6.5,10.){0}
\FAVert(13.5,10.0){0}
\FAProp(6.5,10.)(13.5,10.0)(0.8,){/Straight}{1}
\FAProp(6.5,10.)(13.5,10.0)(-0.8,){/Cycles}{0}
\FALabel(10.0,5.5)[]{\tiny $\bar \ell$}
\end{feynartspicture}} \; ,   \quad
 \parbox{20mm}{\unitlength=0.20bp%
\begin{feynartspicture}(300,300)(1,1)
\FADiagram{}
%
\FAProp(3.5,10.0)(6.5,10.0)(0.,){/Cycles}{0}
\FAProp(1.,10.0)(3.5,10.0)(0.,){/GhostDash}{0}
\FAProp(16.,15.)(13.5,13.5)(0.,){/Straight}{1}
\FAProp(16.,5.)(13.5,6.5)(0.,){/Straight}{-1}
%
\FAProp(13.5,13.5)(6.5,10.0)(0.,){/Cycles}{0}
\FAProp(6.5,10.0)(13.5,6.5)(0.,){/Cycles}{0}
\FAProp(13.5,6.5)(13.5,13.5)(0.,){/Straight}{-1}
\FALabel(14.8,10.0)[]{\tiny $\bar \ell$}
%
%
\FAVert(13.5,13.5){0}
\FAVert(13.5,6.5){0}
\FAVert(6.5,10.){0}
%
\FAProp(16.,15.)(18.5, 16.5)(0.,){/GhostDash}{0}
\FAProp(16.,5.)(18.5, 3.5)(0.,){/GhostDash}{0}
\end{feynartspicture}}\; ,  \quad \parbox{20mm}{\unitlength=0.20bp%
\begin{feynartspicture}(300,300)(1,1)
\FADiagram{}
%
\FAProp(3.5,10.0)(6.5,10.0)(0.,){/Cycles}{0}
\FAProp(1.,10.0)(3.5,10.0)(0.,){/GhostDash}{0}
\FAProp(16.,15.)(13.5,13.5)(0.,){/Straight}{1}
\FAProp(16.,5.)(13.5,6.5)(0.,){/Straight}{-1}
%
\FAProp(13.5,13.5)(6.5,10.0)(0.,){/Straight}{1}
\FAProp(6.5,10.0)(13.5,6.5)(0.,){/Straight}{1}
\FAProp(13.5,6.5)(13.5,13.5)(0.,){/Cycles}{0}
%
%
\FAVert(13.5,13.5){0}
\FAVert(13.5,6.5){0}
\FAVert(6.5,10.){0}
%
\FAProp(16.,15.)(18.5, 16.5)(0.,){/GhostDash}{0}
\FAProp(16.,5.)(18.5, 3.5)(0.,){/GhostDash}{0}
\FALabel(9.5,14.0)[]{\tiny $\bar \ell +p$}
\FALabel(10.0,6.0)[]{\tiny $\bar \ell$}
\end{feynartspicture}} \; . \nonumber
 \end{align}
For these diagrams,  the numerators  obtained by using  the two schemes  differ by terms of the type
\begin{align}
&\cdots \tilde \gamma^\alpha (\slashed \ell + \tilde{\slashed \ell} + m) \tilde \gamma_{\alpha} \cdots \; ,  \nn
& \cdots \tilde \gamma^\alpha (\slashed \ell + \tilde{\slashed \ell} + m) \gamma^\mu  (\slashed \ell +\slashed p +  \tilde{\slashed \ell} + m)  \tilde \gamma_{\alpha} \cdots \; ,
\label{Eq:ExtraT}
\end{align}
where ``$\cdots$" represent four dimensional spinorial objects.   In the FDH scheme it 
is easy to show that the terms~(\ref{Eq:ExtraT}) vanish  in  the $d_s \to 4$ limit, while in 
the other scheme   they vanish as a consequence of  Eq.~(\ref{Eq:Gamma03}).
Therefore the two sets of prescriptions  lead to the same integrand.  

The FDF fulfills the prescriptions~(\ref{Eq:OrthoGs}) -- (\ref{Eq:Gamma01}) and~(\ref{Eq:Gamma03}), 
thus, at one loop, it  leads  to the same   amplitudes of the FDH scheme.

\section{Proof of the completeness relations } 
\label{App:Completeness}

In this Appendix we show that the generalized spinors~(\ref{Eq:SpinorF}) 
fulfill  the completeness relation~(\ref{Eq:CompF4}).  For later convenience
we define the chirality projectors
\begin{align}
\omega_{\pm} = \frac{\mathbb{I}\pm \gamma^5}{2}\;  ,
\end{align}
and we show that:
\begin{subequations}
\begin{align}
& \frac{| q_\ell   ] [\ell^{\flat}   |- |l^{\flat}  ] [q_\ell  | }{[\ell^{\flat}q_\ell ]}  = \nn
 ={}&\frac{ |q_\ell  ] \langle q_\ell \, \ell^{\flat}  \rangle  [\ell^{\flat}  | + |\ell^{\flat}  ] \langle \ell^{\flat}\, q_\ell  \rangle  [q_\ell  | }{2\ell^{\flat}\cdot q_\ell}   \nn
  ={}& \frac{(  |q_\ell  ] \langle q_\ell   |)  ( |\ell^{\flat}  \rangle  [\ell^{\flat}  |  ) + ( |\ell^{\flat}  ] \langle \ell^{\flat}  |  ) ( |q_\ell  \rangle  [q_\ell  |  ) }{2\ell^{\flat}\cdot q_\ell}  \nn
  ={}&\frac{\omega_- \slashed q_\ell\omega_+\slashed \ell^{\flat} + \omega_-  \slashed \ell^\flat \omega_+\slashed q_\ell }{2\ell^{\flat}\cdot q_\ell} \nn
  ={}& \frac{\omega_-^2 \{ \slashed q_\ell \, \slashed \ell^{\flat}  \}}{2\ell^{\flat}\cdot q_\ell} =\omega_- \; ,
  \end{align}
  and similarly 
  \begin{align}
\frac{ |\ell^{\flat}  \rangle  \langle q_\ell  |- |q_\ell  \rangle  \langle \ell^{\flat}  |}{ \langle q_\ell \, \ell^{\flat}  \rangle }  = \omega_+  \, .
  \end{align} 
 \label{Eq:Prope}
 \end{subequations}
Using Eqs.~(\ref{Eq:Prope}) we get
\begin{align}
& \sum_{\lambda=\pm}  u_{\lambda} (\ell  )\bar{u}_{\lambda} (\ell  ) =\nn
\stackrel{\phantom{\mbox{\small Eq.} \, (B2)}}{=}{}&    \Big ( | \ell^{\flat}  \rangle +\frac{ (m-i\mu  )}{ [\ell^{\flat}\, q_\ell  ]} |q_\ell  ]  \Big)  
         \Big ( [\ell^{\flat}  | + \frac{ (m+i\mu  )}{ \langle q_\ell\, \ell^{\flat}  \rangle } \langle q_\ell  |  \Big )+\nn
&  \Big ( |\ell^{\flat}  ]+\frac{ (m+i\mu  )}{ \langle \ell^{\flat}\, q_\ell  \rangle } |q_\ell  \rangle   \Big )  \Big ( \langle \ell^{\flat}  |+\frac{ (m-i\mu  )}{ [q_\ell \, \ell^{\flat}  ]}
 [q_\ell  |  \Big)\nn
\stackrel{\phantom{\mbox{\small Eq.} \, (B2)}}{=}{}& \slashed \ell^{\flat}+\frac{m^{2}+\mu^{2}}{2\ell^{\flat}\cdot q_\ell}\slashed q_\ell+  (m-i\mu  ) \frac{|q_\ell  ] [\ell^{\flat}  |- |\ell^{\flat}  ]
 [q_\ell  | }{ [\ell^{\flat}\, q_\ell  ]} \nn
&+ (m+i\mu  ) \frac{|\ell^{\flat}  \rangle  \langle q_\ell  |- |q_\ell  \rangle  \langle \ell^{\flat}  |   }{ \langle q_\ell \, \ell^{\flat}  \rangle }  \nn
\stackrel{\mbox{\small Eq.}\, (\ref{Eq:Prope})}{=}{}& \slashed \ell^{\flat}+\frac{m^{2}+\mu^{2}}{2\ell^{\flat}\cdot q_\ell}\slashed q_\ell+ (m-i\mu  ) \omega_-  +  (m+i\mu  )\omega_+  \nn
\stackrel{\mbox{\small Eq.} \, (\ref{Eq:Dec})}{=}{}& \slashed \ell + i\mu\gamma^{5}  + m\, .
\end{align}

\section{Color-ordered Feynman rules} 
\label{App:COFR}

 In the FDF, the $d$-dimensional  color-ordered Feynman rules collected in Ref.~\cite{Dixon:1996wi}  become:
\vspace{-0.4cm}
\begin{subequations}
\begin{align}
\parbox{20mm}{\unitlength=0.20bp%
\begin{feynartspicture}(300,300)(1,1)
\FADiagram{}
\FAProp(4.,10.)(16.,10.)(0.,){/Cycles}{0}
\FALabel(5.5,8.93)[t]{\tiny $\alpha$}
\FALabel(14.5,8.93)[t]{\tiny $\beta$}
\FALabel(10.,12.5)[]{\tiny $k$}
\FAVert(4.,10.){0}
\FAVert(16.,10.){0}
\end{feynartspicture}} &= -i  \,\frac{ g^{\alpha \beta} }{k^2 -\mu^2+i0} \, , \quad (\mbox{gluon}), \label{Eq:FRgluCO} \\[-3.ex]
\parbox{20mm}{\unitlength=0.20bp%
\begin{feynartspicture}(300,300)(1,1)
\FADiagram{}
\FAProp(4.,10.)(16.,10.)(0.,){/ScalarDash}{0}
\FALabel(5.5,8.93)[t]{\tiny $A$}
\FALabel(14.5,8.93)[t]{\tiny $B$}
\FALabel(10.,12.5)[]{\tiny $k$}
\FAVert(4.,10.){0}
\FAVert(16.,10.){0}
\end{feynartspicture}} &= -i \,  \frac{ \GG^{AB}}{k^2 -\mu^2+i0} \, ,   \quad (\mbox{scalar}),   \\[-3.ex]
\parbox{20mm}{\unitlength=0.20bp%
\begin{feynartspicture}(300,300)(1,1)
\FADiagram{}
\FAProp(4.,10.)(16.,10.)(0.,){/Straight}{1}
\FALabel(10.,12.5)[]{\tiny $k$}
\FAVert(4.,10.){0}
\FAVert(16.,10.){0}
\end{feynartspicture}} &= i  \,\frac{ \slashed k + i \mu \gamma^5 +m }{k^2 -m^2 -\mu^2 +i0}  \,  , \; (\mbox{fermion}),  \label{Eq:FRferCO} \\[-3.ex]
 \parbox{20mm}{\unitlength=0.20bp%
\begin{feynartspicture}(300,300)(1,1)
\FADiagram{}
\FAProp(3.,10.)(10.,10.)(0.,){/Cycles}{0}
\FALabel(5.3,8.93)[t]{\tiny $1, \alpha$}
\FAProp(16.,15.)(10.,10.)(0.,){/Cycles}{0}
\FALabel(12.2273,13.5749)[br]{\tiny $2,  \beta$}
\FAProp(16.,5.)(10.,10.)(0.,){/Cycles}{0}
\FALabel(12.8873,5.81315)[tr]{\tiny $3, \gamma$}
\FAVert(10.,10.){0}
\end{feynartspicture}} &= \frac{i}{\sqrt{2}}  \big [
 g_{\alpha \beta} (k_1-k_2)_\gamma \nonumber \\[-4.5ex]
 &\qquad  + g_{\beta\gamma} (k_2-k_3)_\alpha \nonumber \\
 &\qquad + g_{\gamma \alpha} (k_3-k_1)_\beta 
 \big ] \, ,   \\[-0.ex]
 \parbox{20mm}{\unitlength=0.20bp%
\begin{feynartspicture}(300,300)(1,1)
\FADiagram{}
\FAProp(3.,10.)(10.,10.)(0.,){/Cycles}{0}
\FALabel(5.3,8.93)[t]{\tiny $1, \alpha$}
\FAProp(16.,15.)(10.,10.)(0.,){/ScalarDash}{0}
\FALabel(12.2273,13.5749)[br]{\tiny $2,  B$}
\FAProp(16.,5.)(10.,10.)(0.,){/ScalarDash}{0}
\FALabel(12.8873,5.81315)[tr]{\tiny $3, C$}
\FAVert(10.,10.){0}
\end{feynartspicture}} &=  \frac{i}{\sqrt{2}}    (k_2-k_3)_\alpha  \GG^{BC}  \, , \\[-3.ex]
 \parbox{20mm}{\unitlength=0.20bp%
\begin{feynartspicture}(300,300)(1,1)
\FADiagram{}
\FAProp(3.,10.)(10.,10.)(0.,){/Cycles}{0}
\FALabel(5.3,8.93)[t]{\tiny $1, \alpha$}
\FAProp(16.,15.)(10.,10.)(0.,){/ScalarDash}{0}
\FALabel(12.2273,13.5749)[br]{\tiny $2, B$}
\FAProp(16.,5.)(10.,10.)(0.,){/Cycles}{0}
\FALabel(12.8873,5.81315)[tr]{\tiny $3, \gamma$}
\FAVert(10.,10.){0}
\end{feynartspicture}} &=  \pm \frac{i}{\sqrt{2}}    g_{\alpha \gamma} (i \mu) \QQ^B \, ,  \nn[-4.5ex]
 & \qquad    (\tilde k_1 = 0, \quad  \tilde k_3 =\pm \tilde \ell)  \label{Eq:FRggsCO1} \\[-0.0ex]
  \parbox{20mm}{\unitlength=0.20bp%
\begin{feynartspicture}(300,300)(1,1)
\FADiagram{}
\FAProp(3.,10.)(10.,10.)(0.,){/Cycles}{0}
\FALabel(5.3,8.93)[t]{\tiny $1,  \alpha$}
\FAProp(16.,15.)(10.,10.)(0.,){/Cycles}{0}
\FALabel(12.2273,13.5749)[br]{\tiny $2,  \beta$}
\FAProp(16.,5.)(10.,10.)(0.,){/ScalarDash}{0}
\FALabel(12.8873,5.81315)[tr]{\tiny $3,  C$}
\FAVert(10.,10.){0}
\end{feynartspicture}} &=   \mp \frac{i}{\sqrt{2}}  g_{\alpha \beta} (i \mu)  \QQ^C  \, ,  \nn[-4.5ex] 
   & \qquad    (\tilde k_1 = 0, \quad  \tilde k_3 =\pm \tilde \ell)   \label{Eq:FRggsCO2} \\[-0.0ex]
\parbox{20mm}{\unitlength=0.20bp%
\begin{feynartspicture}(300,300)(1,1)
\FADiagram{}
\FAProp(4.,15.)(10.,10.)(0.,){/Cycles}{0}
\FALabel(3.0,12.)[]{\tiny $1,\alpha$}
\FAProp(10.,10.)(4.,5.)(0.,){/Cycles}{0}
\FALabel(3.0,8.)[]{\tiny $4,\delta$}
\FAProp(16.,15.)(10.,10.)(0.,){/Cycles}{0}
\FALabel(11.,15.)[]{\tiny $2,\beta$}
\FAProp(10.,10.)(16.,5.)(0.,){/Cycles}{0}
\FALabel(11,5.)[]{\tiny $3,\gamma$}
\FAVert(10.,10.){0}
\end{feynartspicture}} &=   i g_{\alpha \gamma} g_{\beta \delta}  - \frac{i}{2} \big (  g_{\alpha \beta} g_{\gamma \delta}  +  g_{\alpha \delta} g_{\beta \gamma} \big )\, ,  \\[-3.ex]
\parbox{20mm}{\unitlength=0.20bp%
\begin{feynartspicture}(300,300)(1,1)
\FADiagram{}
\FAProp(4.,15.)(10.,10.)(0.,){/Cycles}{0}
\FALabel(3.0,12.)[]{\tiny $1,\alpha$}
\FAProp(10.,10.)(4.,5.)(0.,){/Cycles}{0}
\FALabel(3.0,8.)[]{\tiny $4,\delta$}
\FAProp(16.,15.)(10.,10.)(0.,){/ScalarDash}{0}
\FALabel(11.,15.)[]{\tiny $2,B$}
\FAProp(10.,10.)(16.,5.)(0.,){/ScalarDash}{0}
\FALabel(11,5.)[]{\tiny $3, C$}
\FAVert(10.,10.){0}
\end{feynartspicture}} &=  - \frac{i}{2}   g_{\alpha \delta}  \GG^{BC}\, ,  \\[-3.ex]
 \parbox{20mm}{\unitlength=0.20bp%
\begin{feynartspicture}(300,300)(1,1)
\FADiagram{}
\FAProp(3.,10.)(10.,10.)(0.,){/Straight}{1}
\FALabel(5.3,8.93)[t]{\tiny $1$}
\FAProp(16.,15.)(10.,10.)(0.,){/Cycles}{0}
\FALabel(12.2273,13.5749)[br]{\tiny $2,  \beta$}
\FAProp(16.,5.)(10.,10.)(0.,){/Straight}{-1}
\FALabel(12.8873,5.81315)[tr]{\tiny $3$}
\FAVert(10.,10.){0}
\end{feynartspicture}} &=-\frac{i}{\sqrt{2}} \gamma^\beta \, ,  \\[-3.ex]
 \parbox{20mm}{\unitlength=0.20bp%
\begin{feynartspicture}(300,300)(1,1)
\FADiagram{}
\FAProp(3.,10.)(10.,10.)(0.,){/Straight}{-1}
\FALabel(5.3,8.93)[t]{\tiny $1$}
\FAProp(16.,15.)(10.,10.)(0.,){/Cycles}{0}
\FALabel(12.2273,13.5749)[br]{\tiny $2, \beta$}
\FAProp(16.,5.)(10.,10.)(0.,){/Straight}{1}
\FALabel(12.8873,5.81315)[tr]{\tiny $3$}
\FAVert(10.,10.){0}
\end{feynartspicture}} &=\frac{i}{\sqrt{2}} \gamma^\beta \, ,  \\[-3.ex]
\parbox{20mm}{\unitlength=0.20bp%
\begin{feynartspicture}(300,300)(1,1)
\FADiagram{}
\FAProp(3.,10.)(10.,10.)(0.,){/Straight}{1}
\FALabel(5.3,8.93)[t]{\tiny $1$}
\FAProp(16.,15.)(10.,10.)(0.,){/ScalarDash}{0}
\FALabel(12.2273,13.5749)[br]{\tiny $2, B$}
\FAProp(16.,5.)(10.,10.)(0.,){/Straight}{-1}
\FALabel(12.8873,5.81315)[tr]{\tiny $3$}
\FAVert(10.,10.){0}
\end{feynartspicture}} &=  -\frac{i}{\sqrt{2}} \gamma^5  \GA^B\, , \\[-3.ex]
\parbox{20mm}{\unitlength=0.20bp%
\begin{feynartspicture}(300,300)(1,1)
\FADiagram{}
\FAProp(3.,10.)(10.,10.)(0.,){/Straight}{-1}
\FALabel(5.3,8.93)[t]{\tiny $1$}
\FAProp(16.,15.)(10.,10.)(0.,){/ScalarDash}{0}
\FALabel(12.2273,13.5749)[br]{\tiny $2, B$}
\FAProp(16.,5.)(10.,10.)(0.,){/Straight}{1}
\FALabel(12.8873,5.81315)[tr]{\tiny $3$}
\FAVert(10.,10.){0}
\end{feynartspicture}} &= \frac{i}{\sqrt{2}} \gamma^5  \GA^B \, ,
\end{align}
\label{Eq:Rules}
\end{subequations}
The color-ordered Feynman rules describing the interaction  among an external Higgs boson and gluons in the infinite top-mass limit are given by
\vspace{-0.4cm}
\begin{subequations}
\begin{align}
\parbox{20mm}{\unitlength=0.20bp%
\begin{feynartspicture}(300,300)(1,1)
\FADiagram{}
\FAProp(3.,10.)(10.,10.)(0.,){/ScalarDash}{0}
\FALabel(5.3,8.93)[t]{\tiny $H$}
\FAProp(16.,15.)(10.,10.)(0.,){/Cycles}{0}
\FALabel(12.2273,13.5749)[br]{\tiny $2,  \beta$}
\FAProp(16.,5.)(10.,10.)(0.,){/Cycles}{0}
\FALabel(12.8873,5.81315)[tr]{\tiny $3, \gamma$}
\FAVert(10.,10.){0}
\end{feynartspicture}} &= - 2 i \left [ k_3^\beta k_2^\gamma - g^{\beta \gamma} (k_2\cdot k_3 + \mu^2)  \right ] \, ,\\[-3.ex]
\parbox{20mm}{\unitlength=0.20bp%
\begin{feynartspicture}(300,300)(1,1)
\FADiagram{}
\FAProp(3.,10.)(10.,10.)(0.,){/ScalarDash}{0}
\FALabel(5.3,8.93)[t]{\tiny $1H$}
\FAProp(16.,15.)(10.,10.)(0.,){/ScalarDash}{0}
\FALabel(12.2273,13.5749)[br]{\tiny $2,  B$}
\FAProp(16.,5.)(10.,10.)(0.,){/Cycles}{0}
\FALabel(12.8873,5.81315)[tr]{\tiny $3, \gamma$}
\FAVert(10.,10.){0}
\end{feynartspicture}} &=   \pm 2 \, k_2^\gamma \,   \mu \, Q^B    \qquad    (\tilde k_3 =\pm \tilde \ell) \label{Eq:FRhiggs1}\, , \\[-3.ex]
\parbox{20mm}{\unitlength=0.20bp%
\begin{feynartspicture}(300,300)(1,1)
\FADiagram{}
\FAProp(3.,10.)(10.,10.)(0.,){/ScalarDash}{0}
\FALabel(5.3,8.93)[t]{\tiny $H$}
\FAProp(16.,15.)(10.,10.)(0.,){/Cycles}{0}
\FALabel(12.2273,13.5749)[br]{\tiny $2,  \beta$}
\FAProp(16.,5.)(10.,10.)(0.,){/ScalarDash}{0}
\FALabel(12.8873,5.81315)[tr]{\tiny $3, C$}
\FAVert(10.,10.){0}
\end{feynartspicture}} &=   \pm 2 \, k_3^\beta \,   \mu \, Q^C    \qquad    (\tilde k_2=\pm \tilde \ell)  \label{Eq:FRhiggs2} \, ,\\[-3.ex]
\parbox{20mm}{\unitlength=0.20bp%
\begin{feynartspicture}(300,300)(1,1)
\FADiagram{}
\FAProp(3.,10.)(10.,10.)(0.,){/ScalarDash}{0}
\FALabel(5.3,8.93)[t]{\tiny $H$}
\FAProp(16.,15.)(10.,10.)(0.,){/ScalarDash}{0}
\FALabel(12.2273,13.5749)[br]{\tiny $2,  B$}
\FAProp(16.,5.)(10.,10.)(0.,){/ScalarDash}{0}
\FALabel(12.8873,5.81315)[tr]{\tiny $3, C$}
\FAVert(10.,10.){0}
\end{feynartspicture}} &= - 2 i \, \big [\mu^2 \, Q^B Q^C \nonumber \\[-4.7ex]
&\qquad  -  G^{BC} (k_2\cdot k_3 + \mu^2) \,  \big ] \, ,   \\
\parbox{20mm}{\unitlength=0.20bp%
\begin{feynartspicture}(300,300)(1,1)
\FADiagram{}
\FAProp(4.,15.)(10.,10.)(0.,){/ScalarDash}{0}
\FALabel(3.0,12.)[]{\tiny $H$}
\FAProp(10.,10.)(4.,5.)(0.,){/Cycles}{0}
\FALabel(3.0,8.)[]{\tiny $4,\delta$}
\FAProp(16.,15.)(10.,10.)(0.,){/Cycles}{0}
\FALabel(11.,15.)[]{\tiny $2,\beta$}
\FAProp(10.,10.)(16.,5.)(0.,){/Cycles}{0}
\FALabel(11,5.)[]{\tiny $3,\gamma$}
\FAVert(10.,10.){0}
\end{feynartspicture}} &=  i \sqrt{2} \,  \big [ g_{\beta \gamma} (k_2-k_3)_\delta \nonumber \\[-4.7ex]
&\qquad + g_{\beta \delta} (k_4-k_2)_\gamma  \nonumber \\
&\qquad+ g_{\gamma \delta} (k_3-k_4)_\beta \,   \big ] \, ,\\
\parbox{20mm}{\unitlength=0.20bp%
\begin{feynartspicture}(300,300)(1,1)
\FADiagram{}
\FAProp(4.,15.)(10.,10.)(0.,){/ScalarDash}{0}
\FALabel(3.0,12.)[]{\tiny $H$}
\FAProp(10.,10.)(4.,5.)(0.,){/ScalarDash}{0}
\FALabel(3.0,8.)[]{\tiny $4,D$}
\FAProp(16.,15.)(10.,10.)(0.,){/Cycles}{0}
\FALabel(11.,15.)[]{\tiny $2,\beta$}
\FAProp(10.,10.)(16.,5.)(0.,){/ScalarDash}{0}
\FALabel(11,5.)[]{\tiny $3,C$}
\FAVert(10.,10.){0}
\end{feynartspicture}} &= i \sqrt{2}  G^{CD} (k_3-k_4)_\beta \, , \\[-3.ex]
\parbox{20mm}{\unitlength=0.20bp%
\begin{feynartspicture}(300,300)(1,1)
\FADiagram{}
\FAProp(4.,15.)(10.,10.)(0.,){/ScalarDash}{0}
\FALabel(3.0,12.)[]{\tiny $H$}
\FAProp(10.,10.)(4.,5.)(0.,){/Cycles}{0}
\FALabel(3.0,8.)[]{\tiny $4,\delta$}
\FAProp(16.,15.)(10.,10.)(0.,){/Cycles}{0}
\FALabel(11.,15.)[]{\tiny $2,\beta$}
\FAProp(10.,10.)(16.,5.)(0.,){/ScalarDash}{0}
\FALabel(11,5.)[]{\tiny $3,C$}
\FAVert(10.,10.){0}
\end{feynartspicture}} &=   \mp \sqrt{2} g_{\beta \delta} \, \mu \, Q^C  \quad    (\tilde k_4-\tilde k_2=\pm \tilde \ell) \, . \label{Eq:FRhiggs5}
\end{align}
\label{Eq:RulesH}
\end{subequations}
In the Feynman rules~(\ref{Eq:Rules}),~(\ref{Eq:RulesH})   all the momenta are outgoing. 
The terms  $\mu^2$ appearing in the the propagators~(\ref{Eq:FRgluCO})--(\ref{Eq:FRferCO}) enter only  if the corresponding momentum 
$k$ is $d$-dimensional,  {\it i.e.} only if $k$ contains the  loop momentum $\bar \ell$.  In the vertices~(\ref{Eq:FRggsCO1}),~(\ref{Eq:FRggsCO2}) 
the momentum  $k_1$ is four-dimensional  while the other two are $d$-dimensional. For these vertices the overall sign depend on which of the 
combinations~(\ref{Eq:Combo}) is present in the vertex.  Similarly the overall sign of the Feynman rules~(\ref{Eq:FRhiggs1}),~(\ref{Eq:FRhiggs2}) 
and~(\ref{Eq:FRhiggs5})
depend on the flow of the loop momentum $\bar \ell$.
As already mentioned each  cut scalar propagator carries a ($-2\epsilon$)-SRs factor 
of the type
\vspace{-0.4cm}
\begin{equation}
\parbox{20mm}{\unitlength=0.20bp%
\begin{feynartspicture}(300,300)(1,1)
\FADiagram{}
\FAProp(4.,10.)(16.,10.)(0.,){/ScalarDash}{0}
\FALabel(5.5,8.93)[t]{\tiny $A$}
\FALabel(14.5,8.93)[t]{\tiny $B$}
\FAVert(4.,10.){0}
\FAVert(16.,10.){0}
\FAProp(10.,6.)(10.,14.)(0.,){/GhostDash}{0}
\end{feynartspicture}} =  \hat G^{AB}   \, ,
\label{Eq:Rules2}
\end{equation}
where $ \hat G^{AB}$ is defined in Eq.~(\ref{Eq:Pref}).

\section{Master integrals} 
\label{App:MI}

In this appendix we present  the MIs
entering in the decomposition of the four-point 
amplitudes computed in Sections~\ref{sec:gggg},~\ref{sec:ggqq} and~\ref{sec:gggH}.
\smallskip

The MIs in the decomposition~(\ref{Eq:Decomposition}) of the one-loop amplitude of the process~(\ref{Eq:ProM}) are given by
\begin{align}
I_{1|2|3|4} & =\frac{r_{\Gamma}}{s_{12}s_{14}}\bigg [\frac{2}{\epsilon^{2}}\left((-s_{12})^{-\epsilon}+(-s_{14})^{-\epsilon}\right)\nn 
&-\log^{2}\left(\frac{-s_{12}}{-s_{14}}\right)-\pi^{2}\bigg ]\, , \nn
I_{ij | k | m} & =I_{k | m | ij} =   I_{m | ij | k} = -\frac{r_{\Gamma}}{\epsilon^{2}s_{ij}}(-s_{ij})^{-\epsilon}\, , \nn
I_{ij | km} & =\frac{r_{\Gamma}}{\epsilon(1-2\epsilon)}(-s_{ij})^{-\epsilon}\, , \nn
I_{1|2|3|4}\left[\mu^{4}\right]   
&=   \frac{4 \epsilon \,  (\epsilon -1) \,  I_{1 | 2 | 3 | 4} }{b_0^2 (2 \epsilon -3) (2 \epsilon -1)}    \nn
&+  \frac{b_{1} (\epsilon-1)}{2 \epsilon -3} \left [ I_{1|2|34}[\mu^2] -   \frac{2 \epsilon \,  I_{1|2|34} }{b_0 (2\epsilon -1)}  \right ] \nn
&+  \frac{ b_{2} (\epsilon-1)}{2 \epsilon -3} \left [ I_{2|3|41}[\mu^2] -   \frac{2 \epsilon \,  I_{2|3|41} }{b_0 (2\epsilon -1)}  \right ] \nn
&= - \frac{1}{6} + \mathcal{O}(\epsilon) \, ,  \nn 
I_{ij | k | m}\left[\mu^{2}\right]&=I_{k | m | ij}\left[\mu^{2}\right] =I_{m | ij | k}\left[\mu^{2}\right]   \nn
&=-\frac{r_{\Gamma}\, (-s_{ij})^{-\epsilon}}{2\left(1-\epsilon\right)\left(1-2\epsilon\right)} =  -\frac{1}{2} + \mathcal{O}(\epsilon)\, , \nn
I_{ij | km}\left[\mu^{2}\right] 
&= \frac{r_{\Gamma}\, (-s_{ij})^{1-\epsilon}}{2\left(1-2\epsilon\right)\left(3-2\epsilon\right)}\nn
 &= - \frac{1}{6} s_{ij}  + \mathcal{O}(\epsilon)\, .
\label{Eq:MIproM}
\end{align}
The factor $r_\Gamma$  is defined as
\begin{align}
r_{\Gamma}\equiv\frac{\Gamma^{2}(1-\epsilon)\Gamma(1+\epsilon)}{\Gamma(1-2\epsilon)}\, ,
\label{Eq:Rgamma}
\end{align}
while  the coefficients $b$ read as~\cite{Bern:1995db,Bern:1995ix} 
\begin{align}
b_0 =  - \frac{4 \, s_{13}}{s_{12}s_{23}} \, , \qquad 
b_1 = \frac{s_{12}}{s_{13}} \, , \qquad
b_2 = \frac{s_{23}}{s_{13}}  \, .
\end{align}
\smallskip

The MIs entering the decomposition~(\ref{Eq:DecompositionH})  for  the process~(\ref{Eq:ProH}) are given by
\begin{align}
I_{i | j | k | H}&=  I_{j | i | H | k} = I_{i | H | k | j} \nn
&=\frac{2r_{\Gamma}}{s_{ij}s_{jk}}\frac{1}{\epsilon^{2}}\left[\left(-s_{ij}\right)^{-\epsilon}+\left(-s_{jk}\right)^{-\epsilon}-\left(m_{H}^{2}\right)^{-\epsilon}\right]\nn 
&-\frac{2r_{\Gamma}}{s_{ij}s_{jk}}\bigg[\text{Li}_{2}\left(1-\frac{m_{H}^{2}}{s_{ij}}\right)+\text{Li}_{2}\left(1-\frac{m_{H}^{2}}{s_{jk}}\right)\nn 
& + \frac{1}{2} \, \log^2\frac{  s_{ij}  }{ s_{jk}  }  + \frac{\pi^2}{6} \bigg ]  \, ,\nn
I_{ij | k | H}&=  I_{ij | H | k} = I_{ k | H | ij}  =  I_{ H | k | ij}    =  I_{ k  | ij  | H}   \nn 
& = \frac{r_{\Gamma}}{\epsilon^{2}}\frac{\left(-s_{ij}\right)^{-\epsilon}-\left(-m_{H}^{2}\right)^{-\epsilon}}{\left(-s_{ij}\right)-\left(-m_{H}^{2}\right)} \, , \nn 
I_{i | j | kH}&= I_{kH  | ij} = I_{i  | kH | j} = \frac{r_{\Gamma}}{\epsilon^{2}}\left(-s_{ij}\right)^{-1-\epsilon} \, , \nn
I_{ij |  Hk}&= I_{H k | i j}  = I_{ij | k H} =
\frac{r_{\Gamma}}{\epsilon\left(1-2\epsilon\right)}\left(-s_{ij}\right)^{-\epsilon} \, , \nn
I_{123 |  H}&=
\frac{r_{\Gamma}}{\epsilon\left(1-2\epsilon\right)}\left(-m^2_H\right)^{-\epsilon} \, , \nn
I_{i | j | k | m} \left [ \mu^4 \right ]   &= \frac{   4 \epsilon (\epsilon -1)  }{a_0^2 (2 \epsilon -3)  (2\epsilon -1)  } I_{i | j | k | m}   \nn
& + \frac{a_1 (\epsilon -1)    }{  a_0 (2 \epsilon -3)  }  \left [   I_{j | k | m i}  \left [ \mu^2 \right ]     - \frac{  2 \epsilon   I_{j | k | m i}     }{ a_0  (2 \epsilon -1)  }                \right ] \nn
& + \frac{a_2 (\epsilon -1)    }{  a_0 (2 \epsilon -3)  }  \left [   I_{i j | k | m }  \left [ \mu^2 \right ]     - \frac{  2 \epsilon   I_{i j | k | m}     }{ a_0  (2 \epsilon -1)  }                \right ] \nn
& + \frac{a_3 (\epsilon -1)    }{  a_0 (2 \epsilon -3)  }  \left [   I_{i | j  k | m }  \left [ \mu^2 \right ]     - \frac{  2 \epsilon   I_{i  | j  k | m}     }{ a_0  (2 \epsilon -1)  }                \right ] \nn
& + \frac{a_4 (\epsilon -1)    }{  a_0 (2 \epsilon -3)  }  \left [   I_{i | j | k  m }  \left [ \mu^2 \right ]     - \frac{  2 \epsilon   I_{i  | j | k  m}     }{ a_0  (2 \epsilon -1)  }                \right ] \nn
&=  - \frac{1}{6}   + \mathcal{O}( \epsilon ) \, ,    \nn
I_{ij | k | H}\left [ \mu^2 \right ]  &=  I_{ij | H | k}\left [ \mu^2 \right ]  = I_{ k | H | ij} \left [ \mu^2 \right ]  \nn
&=  I_{ H | k | ij}  \left [ \mu^2 \right ]   =  I_{ k  | ij  | H} \left [ \mu^2 \right ]   \nn 
& = \frac{ - r_{\Gamma}}{   2 (1-\epsilon)  (1 - 2\epsilon)  }\frac{\left(-s_{ij}\right)^{1-\epsilon}-\left(-m_{H}^{2}\right)^{1-\epsilon}}{\left(-s_{ij}\right)-\left(-m_{H}^{2}\right)}  \nn
& =  -\frac{1}{2}   + \mathcal{O}( \epsilon )  \, , \nn 
I_{i | j | kH}\left [ \mu^2 \right ] &= I_{kH  | ij}\left [ \mu^2 \right ]  = I_{i  | kH | j} \left [ \mu^2 \right ]  \nn
&= \frac{ - r_{\Gamma}   \, \left(-s_{ij}\right)^{-\epsilon}  }{ 2 (1-\epsilon)  (1 -2 \epsilon)    } =   -\frac{1}{2}   + \mathcal{O}( \epsilon ) \, , \nn
I_{ij |  Hk}\left [ \mu^2 \right ] &= I_{H k | i j}\left [ \mu^2 \right ]   = I_{ij | k H}\left [ \mu^2 \right ] \nn
&= \frac{ r_{\Gamma}\, \left(-s_{ij}\right)^{1-\epsilon}  }{2 (3 - 2\epsilon) \left(1-2\epsilon\right)} =  - \frac{1}{6}  \, s_{ij} + \mathcal{O}( \epsilon ) \, , \nn
I_{123 |  H}\left [ \mu^2 \right ] 
&= \frac{ r_{\Gamma}\, \left(-  m_H^2 \right)^{1-\epsilon}  }{2 (3 - 2\epsilon) \left(1-2\epsilon\right)} =  - \frac{1}{6}  \, m_H^2 + \mathcal{O}( \epsilon ) \, ,
\end{align}
The coefficients $a$  read as follows, 
\begin{align} 
a_0 \equiv \sum_{s=1}^4 a_s \, , \qquad  a_s \equiv \sum_{t=1}^4 \left (S^{-1}_{i | j | k | m} \right )_{st} \, , 
\end{align}
in terms of the cut-dependent matrix 
\begin{align}
\left ( S_{i | j | k | m}  \right )_{st} &\equiv  - \frac{1}{2}\left(  v^{(s)}_{i | j | k | m} - v^{(t)}_{i | j | k | m}\right )^2 \, ,
\end{align}
where
\begin{align}
v^{(1)}_{i | j | k | m} &= 0\, ,  &
v^{(2)}_{i | j | k | m} &= p_i \, , \nn
v^{(3)}_{i | j | k | m} &= p_i + p_j \, ,  &
v^{(4)}_{i | j | k | m} &= -p_m  \, .
\end{align}

\section{Tree-level amplitudes} 
\label{App:TREE}

In this Appendix we present the three- and four-point  tree-level amplitudes entering in the
computation described in Section~\ref{sec:ggqq}. They can be  computed
by using the Feynman rules collected in Eqs.~(\ref{Eq:Rules}). The legs with a dot are massive with mass $\mu$. 
\smallskip

The tree-level amplitudes entering in the unitarity-based decomposition of the {\it left-turning amplitudes}
amplitude can be expressed in terms of  Feynman diagrams as follows:
\vspace{-0.4cm}
 \begin{align}
A_3( g \; g^{\bullet} \; g^{\bullet}     )   ={}&  \parbox{20mm}{\unitlength=0.20bp%
\begin{feynartspicture}(300,300)(1,1)
\FADiagram{}
\FAProp(3.,10.)(10.,10.)(0.,){/Cycles}{0} 
\FAProp(16.,15.)(10.,10.)(0.,){/Cycles}{0} 
\FAVert(15.8873,12.){0}
\FAProp(16.,5.)(10.,10.)(0.,){/Cycles}{0}  
\FAVert(15.8873,8.){0}
\FAVert(10,10.){0}
\end{feynartspicture}}   \nn[-5.ex]
A_3( g \;  s^{\bullet}_g \;  s^{\bullet}_g     )   ={}&  \parbox{20mm}{\unitlength=0.20bp%
\begin{feynartspicture}(300,300)(1,1)
\FADiagram{}
\FAProp(3.,10.)(10.,10.)(0.,){/Cycles}{0} 
\FAProp(16.,15.)(10.,10.)(0.,){/ScalarDash}{0} 
\FAVert(15.8873,12.){0}
\FAProp(16.,5.)(10.,10.)(0.,){/ScalarDash}{0}  
\FAVert(15.8873,8.){0}
\FAVert(10,10.){0}
\end{feynartspicture}}   \nn[-5.ex]
A_3( g^{\bullet} \; \bar q \; q^{\bullet}     )   ={}&  \parbox{20mm}{\unitlength=0.20bp%
\begin{feynartspicture}(300,300)(1,1)
\FADiagram{}
\FAProp(3.,10.)(10.,10.)(0.,){/Cycles}{0} 
\FAVert(4.,8.){0}
\FAProp(16.,15.)(10.,10.)(0.,){/Straight}{1} 
\FAProp(16.,5.)(10.,10.)(0.,){/Straight}{-1}  
\FAVert(15.8873,8.){0}
\FAVert(10,10.){0}
\end{feynartspicture}}   \nn[-5.ex]
A_3( g^{\bullet} \; \bar q^{\, \bullet} \; q     )   ={}&  \parbox{20mm}{\unitlength=0.20bp%
\begin{feynartspicture}(300,300)(1,1)
\FADiagram{}
\FAProp(3.,10.)(10.,10.)(0.,){/Cycles}{0} 
\FAVert(4.,8.){0}
\FAProp(16.,15.)(10.,10.)(0.,){/Straight}{1} 
\FAVert(15.8873,12.){0}
\FAProp(16.,5.)(10.,10.)(0.,){/Straight}{-1}  
%
\FAVert(10,10.){0}
\end{feynartspicture}}   \nn[-5.ex]
A_3(  s^{\bullet}_g \; \bar q \; q^{\bullet}     )   ={}&  \parbox{20mm}{\unitlength=0.20bp%
\begin{feynartspicture}(300,300)(1,1)
\FADiagram{}
\FAProp(3.,10.)(10.,10.)(0.,){/ScalarDash}{0} 
\FAVert(4.,8.){0}
\FAProp(16.,15.)(10.,10.)(0.,){/Straight}{1} 
\FAProp(16.,5.)(10.,10.)(0.,){/Straight}{-1}  
\FAVert(15.8873,8.){0}
\FAVert(10,10.){0}
\end{feynartspicture}}   \nn[-5.ex]
A_3(  s^{\bullet}_g \; \bar q^{\, \bullet} \; q     )   ={}&  \parbox{20mm}{\unitlength=0.20bp%
\begin{feynartspicture}(300,300)(1,1)
\FADiagram{}
\FAProp(3.,10.)(10.,10.)(0.,){/ScalarDash}{0} 
\FAVert(4.,8.){0}
\FAProp(16.,15.)(10.,10.)(0.,){/Straight}{1} 
\FAVert(15.8873,12.){0}
\FAProp(16.,5.)(10.,10.)(0.,){/Straight}{-1}  
%
\FAVert(10,10.){0}
\end{feynartspicture}}   \nn[-5.ex]
A_4( g \;  g \;   g^{ \bullet}    \;   g^{\bullet}   )   ={}&  
\parbox{20mm}{\unitlength=0.18bp%
\begin{feynartspicture}(300,300)(1,1)
\FADiagram{}
\FAProp(0.,15.)(10.,10.)(0.,){/Cycles}{0}
\FAProp(0.,5.)(10.,10.)(0.,){/Cycles}{0}
\FAProp(20.,15.)(10.,10.)(0.,){/Cycles}{0}
\FAProp(20.,5.)(10.,10.)(0.,){/Cycles}{0}
\FAVert(10.,10.){0}
\FAVert(18.,8.){0}
\FAVert(18.,12.){0}
\end{feynartspicture}} 
+\parbox{20mm}{\unitlength=0.18bp%
\begin{feynartspicture}(300,300)(1,1)
\FADiagram{}
\FAProp(0.,15.)(6.,10.)(0.,){/Cycles}{0}
\FAProp(0.,5.)(6.,10.)(0.,){/Cycles}{0}
\FAProp(20.,15.)(14.,10.)(0.,){/Cycles}{0}
\FAProp(20.,5.)(14.,10.)(0.,){/Cycles}{0}
\FAProp(6.,10.)(14.,10.)(0.,){/Cycles}{0}
\FAVert(6.,10.){0}
\FAVert(14.,10.){0}
\FAVert(19.5,8.){0}
\FAVert(19.5,12.){0}
\end{feynartspicture}}  \nn[-5.ex]
& +\parbox{20mm}{\unitlength=0.18bp%
\begin{feynartspicture}(300,300)(1,1)
\FADiagram{}
\FAProp(0.,15.)(10.,14.)(0.,){/Cycles}{0}
\FAProp(0.,5.)(10.,6.)(0.,){/Cycles}{0}
\FAProp(20.,15.)(10.,14.)(0.,){/Cycles}{0}
\FAProp(20.,5.)(10.,6.)(0.,){/Cycles}{0}
\FAProp(10.,14.)(10.,6.)(0.,){/Cycles}{0}
\FAVert(10.,14.){0}
\FAVert(10.,6.){0}
\FAVert(17.,7.){0}
\FAVert(17.,13.){0}
\FAVert(12.,10.){0}
\end{feynartspicture}}
+\parbox{20mm}{\unitlength=0.18bp%
\begin{feynartspicture}(300,300)(1,1)
\FADiagram{}
\FAProp(0.,15.)(10.,14.)(0.,){/Cycles}{0}
\FAProp(0.,5.)(10.,6.)(0.,){/Cycles}{0}
\FAProp(20.,15.)(10.,14.)(0.,){/Cycles}{0}
\FAProp(20.,5.)(10.,6.)(0.,){/Cycles}{0}
\FAProp(10.,14.)(10.,6.)(0.,){/ScalarDash}{0}
\FAVert(10.,14.){0}
\FAVert(10.,6.){0}
\FAVert(17.,7.){0}
\FAVert(17.,13.){0}
\FAVert(12.,10.){0}
\end{feynartspicture}

}  \nn[-5.ex]
A_4( g^{\bullet}  \;  g^{\bullet}  \;  \bar q   \;   q  )   ={}&  
\parbox{20mm}{\unitlength=0.18bp%
%
\begin{feynartspicture}(300,300)(1,1)
\FADiagram{}
\FAProp(0.,15.)(6.,10.)(0.,){/Cycles}{0}
\FAProp(0.,5.)(6.,10.)(0.,){/Cycles}{0}
\FAProp(20.,15.)(14.,10.)(0.,){/Straight}{1}
\FAProp(20.,5.)(14.,10.)(0.,){/Straight}{-1}
\FAProp(6.,10.)(14.,10.)(0.,){/Cycles}{0}
\FAVert(6.,10.){0}
\FAVert(14.,10.){0}
\FAVert(.5,8.){0}
\FAVert(.5,12.){0}
\end{feynartspicture}} 
+\parbox{20mm}{\unitlength=0.18bp%
\begin{feynartspicture}(300,300)(1,1)
\FADiagram{}
\FAProp(0.,15.)(10.,14.)(0.,){/Cycles}{0}
\FAProp(0.,5.)(10.,6.)(0.,){/Cycles}{0}
\FAProp(20.,15.)(10.,14.)(0.,){/Straight}{1}
\FAProp(20.,5.)(10.,6.)(0.,){/Straight}{-1}
\FAProp(10.,14.)(10.,6.)(0.,){/Straight}{1}
\FAVert(10.,14.){0}
\FAVert(10.,6.){0}
\FAVert(3.,7.){0}
\FAVert(3.,13.){0}
\FAVert(8.,10.){0}
\end{feynartspicture}

} \nn[-5.ex]
A_4( g^{\bullet}  \;  g  \;  \bar q   \;   q^{\bullet}   )   ={}&  
\parbox{20mm}{\unitlength=0.18bp%
%
\begin{feynartspicture}(300,300)(1,1)
\FADiagram{}
\FAProp(0.,15.)(6.,10.)(0.,){/Cycles}{0}
\FAProp(0.,5.)(6.,10.)(0.,){/Cycles}{0}
\FAProp(20.,15.)(14.,10.)(0.,){/Straight}{1}
\FAProp(20.,5.)(14.,10.)(0.,){/Straight}{-1}
\FAProp(6.,10.)(14.,10.)(0.,){/Cycles}{0}
\FAVert(6.,10.){0}
\FAVert(14.,10.){0}
\FAVert(.5,8.){0}
\FAVert(19.5,8.){0}
\FAVert(10.,9.){0}
\end{feynartspicture}}
+\parbox{20mm}{\unitlength=0.18bp%
%
\begin{feynartspicture}(300,300)(1,1)
\FADiagram{}
\FAProp(0.,15.)(6.,10.)(0.,){/Cycles}{0}
\FAProp(0.,5.)(6.,10.)(0.,){/Cycles}{0}
\FAProp(20.,15.)(14.,10.)(0.,){/Straight}{1}
\FAProp(20.,5.)(14.,10.)(0.,){/Straight}{-1}
\FAProp(6.,10.)(14.,10.)(0.,){/ScalarDash}{0}
\FAVert(6.,10.){0}
\FAVert(14.,10.){0}
\FAVert(.5,8.){0}
\FAVert(19.5,8.){0}
\FAVert(10.,9.){0}
\end{feynartspicture}}\nn[-5.ex]
&+\parbox{20mm}{\unitlength=0.18bp%
\begin{feynartspicture}(300,300)(1,1)
\FADiagram{}
\FAProp(0.,15.)(10.,14.)(0.,){/Cycles}{0}
\FAProp(0.,5.)(10.,6.)(0.,){/Cycles}{0}
\FAProp(20.,15.)(10.,14.)(0.,){/Straight}{1}
\FAProp(20.,5.)(10.,6.)(0.,){/Straight}{-1}
\FAProp(10.,14.)(10.,6.)(0.,){/Straight}{1}
\FAVert(10.,14.){0}
\FAVert(10.,6.){0}
\FAVert(3.,7.){0}
\FAVert(17.,7.){0}
\end{feynartspicture}

}  \nn[-5.ex]
A_4( g  \;  g^\bullet  \;  \bar q^{\, \bullet}    \;   q  )   ={}&  
\parbox{20mm}{\unitlength=0.18bp%
%
\begin{feynartspicture}(300,300)(1,1)
\FADiagram{}
\FAProp(0.,15.)(6.,10.)(0.,){/Cycles}{0}
\FAProp(0.,5.)(6.,10.)(0.,){/Cycles}{0}
\FAProp(20.,15.)(14.,10.)(0.,){/Straight}{1}
\FAProp(20.,5.)(14.,10.)(0.,){/Straight}{-1}
\FAProp(6.,10.)(14.,10.)(0.,){/Cycles}{0}
\FAVert(6.,10.){0}
\FAVert(14.,10.){0}
\FAVert(.5,12.){0}
\FAVert(19.5,12.){0}
\FAVert(10.,12.5){0}
\end{feynartspicture}
}
+\parbox{20mm}{\unitlength=0.18bp%
%
\begin{feynartspicture}(300,300)(1,1)
\FADiagram{}
\FAProp(0.,15.)(6.,10.)(0.,){/Cycles}{0}
\FAProp(0.,5.)(6.,10.)(0.,){/Cycles}{0}
\FAProp(20.,15.)(14.,10.)(0.,){/Straight}{1}
\FAProp(20.,5.)(14.,10.)(0.,){/Straight}{-1}
\FAProp(6.,10.)(14.,10.)(0.,){/ScalarDash}{0}
\FAVert(6.,10.){0}
\FAVert(14.,10.){0}
\FAVert(.5,12.){0}
\FAVert(19.5,12.){0}
\FAVert(10.,11.){0}
\end{feynartspicture}}\nn[-5.ex]
&+\parbox{20mm}{\unitlength=0.18bp%
\begin{feynartspicture}(300,300)(1,1)
\FADiagram{}
\FAProp(0.,15.)(10.,14.)(0.,){/Cycles}{0}
\FAProp(0.,5.)(10.,6.)(0.,){/Cycles}{0}
\FAProp(20.,15.)(10.,14.)(0.,){/Straight}{1}
\FAProp(20.,5.)(10.,6.)(0.,){/Straight}{-1}
\FAProp(10.,14.)(10.,6.)(0.,){/Straight}{1}
\FAVert(10.,14.){0}
\FAVert(10.,6.){0}
\FAVert(3.,13.){0}
\FAVert(17.,13.){0}
\end{feynartspicture}

} \nn[-5.ex]
A_4( g \;  g \;  s^{ \bullet}_g    \;    s^{\bullet}_g   )   ={}&  
\parbox{20mm}{\unitlength=0.18bp%
\begin{feynartspicture}(300,300)(1,1)
\FADiagram{}
\FAProp(0.,15.)(10.,10.)(0.,){/Cycles}{0}
\FAProp(0.,5.)(10.,10.)(0.,){/Cycles}{0}
\FAProp(20.,15.)(10.,10.)(0.,){/ScalarDash}{0}
\FAProp(20.,5.)(10.,10.)(0.,){/ScalarDash}{0}
\FAVert(10.,10.){0}
\FAVert(18.,8.){0}
\FAVert(18.,12.){0}
\end{feynartspicture}
}
+\parbox{20mm}{\unitlength=0.18bp%
\begin{feynartspicture}(300,300)(1,1)
\FADiagram{}
\FAProp(0.,15.)(6.,10.)(0.,){/Cycles}{0}
\FAProp(0.,5.)(6.,10.)(0.,){/Cycles}{0}
\FAProp(20.,15.)(14.,10.)(0.,){/ScalarDash}{0}
\FAProp(20.,5.)(14.,10.)(0.,){/ScalarDash}{0}
\FAProp(6.,10.)(14.,10.)(0.,){/Cycles}{0}
\FAVert(6.,10.){0}
\FAVert(14.,10.){0}
\FAVert(19.5,8.){0}
\FAVert(19.5,12.){0}
\end{feynartspicture}

}  \nn[-5.ex]
& +\parbox{20mm}{\unitlength=0.18bp%
%
\begin{feynartspicture}(300,300)(1,1)
\FADiagram{}
\FAProp(0.,15.)(10.,14.)(0.,){/Cycles}{0}
\FAProp(0.,5.)(10.,6.)(0.,){/Cycles}{0}
\FAProp(20.,15.)(10.,14.)(0.,){/ScalarDash}{0}
\FAProp(20.,5.)(10.,6.)(0.,){/ScalarDash}{0}
\FAProp(10.,6.)(10.,14.)(0.,){/Cycles}{0}
\FAVert(10.,14.){0}
\FAVert(10.,6.){0}
\FAVert(17.,7.){0}
\FAVert(17.,13.){0}
\FAVert(12.,10.){0}
\end{feynartspicture}}
+\parbox{20mm}{\unitlength=0.18bp%
%

\begin{feynartspicture}(300,300)(1,1)
\FADiagram{}
\FAProp(0.,15.)(10.,14.)(0.,){/Cycles}{0}
\FAProp(0.,5.)(10.,6.)(0.,){/Cycles}{0}
\FAProp(20.,15.)(10.,14.)(0.,){/ScalarDash}{0}
\FAProp(20.,5.)(10.,6.)(0.,){/ScalarDash}{0}
\FAProp(10.,14.)(10.,6.)(0.,){/ScalarDash}{0}
\FAVert(10.,14.){0}
\FAVert(10.,6.){0}
\FAVert(17.,7.){0}
\FAVert(17.,13.){0}
\FAVert(12.,10.){0}
\end{feynartspicture}

}  \nn[-5.ex]
A_4( g  \;   s^{\bullet}_g  \;  \bar q^{\, \bullet}    \;   q  )   ={}&  
\parbox{20mm}{\unitlength=0.18bp%
%
%
\begin{feynartspicture}(300,300)(1,1)
\FADiagram{}
\FAProp(0.,15.)(6.,10.)(0.,){/ScalarDash}{0}
\FAProp(0.,5.)(6.,10.)(0.,){/Cycles}{0}
\FAProp(20.,15.)(14.,10.)(0.,){/Straight}{1}
\FAProp(20.,5.)(14.,10.)(0.,){/Straight}{-1}
\FAProp(14.,10.)(6.,10.)(0.,){/Cycles}{0}
\FAVert(6.,10.){0}
\FAVert(14.,10.){0}
\FAVert(.5,12.){0}
\FAVert(19.5,12.){0}
\FAVert(10.,11.){0}
\end{feynartspicture}
}
+\parbox{20mm}{\unitlength=0.18bp%
\begin{feynartspicture}(300,300)(1,1)
\FADiagram{}
\FAProp(0.,15.)(6.,10.)(0.,){/ScalarDash}{0}
\FAProp(0.,5.)(6.,10.)(0.,){/Cycles}{0}
\FAProp(20.,15.)(14.,10.)(0.,){/Straight}{1}
\FAProp(20.,5.)(14.,10.)(0.,){/Straight}{-1}
\FAProp(6.,10.)(14.,10.)(0.,){/ScalarDash}{0}
\FAVert(6.,10.){0}
\FAVert(14.,10.){0}
\FAVert(.5,12.){0}
\FAVert(19.5,12.){0}
\FAVert(10.,11.){0}
\end{feynartspicture}
}\nn[-5.ex]
&+\parbox{20mm}{\unitlength=0.18bp%
\begin{feynartspicture}(300,300)(1,1)
\FADiagram{}
\FAProp(0.,15.)(10.,14.)(0.,){/ScalarDash}{0}
\FAProp(0.,5.)(10.,6.)(0.,){/Cycles}{0}
\FAProp(20.,15.)(10.,14.)(0.,){/Straight}{1}
\FAProp(20.,5.)(10.,6.)(0.,){/Straight}{-1}
\FAProp(10.,14.)(10.,6.)(0.,){/Straight}{1}
\FAVert(10.,14.){0}
\FAVert(10.,6.){0}
\FAVert(3.,13.){0}
\FAVert(17.,13.){0}
\end{feynartspicture}

} \nn[-5.ex]
A_4(  s^{\bullet}_g  \;  g  \;  \bar q   \;   q^{\bullet}   )   ={}&  
\parbox{20mm}{\unitlength=0.18bp%
%
\begin{feynartspicture}(300,300)(1,1)
\FADiagram{}
\FAProp(0.,15.)(6.,10.)(0.,){/Cycles}{0}
\FAProp(0.,5.)(6.,10.)(0.,){/ScalarDash}{0}
\FAProp(20.,15.)(14.,10.)(0.,){/Straight}{1}
\FAProp(20.,5.)(14.,10.)(0.,){/Straight}{-1}
\FAProp(6.,10.)(14.,10.)(0.,){/ScalarDash}{0}
\FAVert(6.,10.){0}
\FAVert(14.,10.){0}
\FAVert(.5,8.){0}
\FAVert(19.5,8.){0}
\FAVert(10.,9.){0}
\end{feynartspicture}
} 
+\parbox{20mm}{\unitlength=0.18bp%
%
\begin{feynartspicture}(300,300)(1,1)
\FADiagram{}
\FAProp(0.,15.)(6.,10.)(0.,){/Cycles}{0}
\FAProp(0.,5.)(6.,10.)(0.,){/ScalarDash}{0}
\FAProp(20.,15.)(14.,10.)(0.,){/Straight}{1}
\FAProp(20.,5.)(14.,10.)(0.,){/Straight}{-1}
\FAProp(6.,10.)(14.,10.)(0.,){/Cycles}{0}
\FAVert(6.,10.){0}
\FAVert(14.,10.){0}
\FAVert(.5,8.){0}
\FAVert(19.5,8.){0}
\FAVert(10.,9.){0}
\end{feynartspicture}} \nn[-5.ex]
&+\parbox{20mm}{\unitlength=0.18bp%
\begin{feynartspicture}(300,300)(1,1)
\FADiagram{}
\FAProp(0.,15.)(10.,14.)(0.,){/Cycles}{0}
\FAProp(0.,5.)(10.,6.)(0.,){/ScalarDash}{0}
\FAProp(20.,15.)(10.,14.)(0.,){/Straight}{1}
\FAProp(20.,5.)(10.,6.)(0.,){/Straight}{-1}
\FAProp(10.,14.)(10.,6.)(0.,){/Straight}{1}
\FAVert(10.,14.){0}
\FAVert(10.,6.){0}
\FAVert(3.,7.){0}
\FAVert(17.,7.){0}
\end{feynartspicture}

} \nn[-5.ex]
A_4(  s^{\bullet}_g  \;   s^{\bullet}_g  \;  \bar q   \;   q  )   ={}&  
\parbox{20mm}{\unitlength=0.18bp%
%
\begin{feynartspicture}(300,300)(1,1)
\FADiagram{}
\FAProp(0.,15.)(6.,10.)(0.,){/ScalarDash}{0}
\FAProp(0.,5.)(6.,10.)(0.,){/ScalarDash}{0}
\FAProp(20.,15.)(14.,10.)(0.,){/Straight}{1}
\FAProp(20.,5.)(14.,10.)(0.,){/Straight}{-1}
\FAProp(6.,10.)(14.,10.)(0.,){/Cycles}{0}
\FAVert(6.,10.){0}
\FAVert(14.,10.){0}
\FAVert(.5,8.){0}
\FAVert(.5,12.){0}
\end{feynartspicture}
} 
+ \parbox{20mm}{\unitlength=0.18bp%
\begin{feynartspicture}(300,300)(1,1)
\FADiagram{}
\FAProp(0.,15.)(10.,14.)(0.,){/ScalarDash}{0}
\FAProp(0.,5.)(10.,6.)(0.,){/ScalarDash}{0}
\FAProp(20.,15.)(10.,14.)(0.,){/Straight}{1}
\FAProp(20.,5.)(10.,6.)(0.,){/Straight}{-1}
\FAProp(10.,14.)(10.,6.)(0.,){/Straight}{1}
\FAVert(10.,14.){0}
\FAVert(10.,6.){0}
\FAVert(3.,7.){0}
\FAVert(3.,13.){0}
\FAVert(8.,10.){0}
\end{feynartspicture}

} 
\end{align}
The  three- and the four-point tree-level amplitudes  entering in the computation of  the {\it right-turning amplitude} are
\vspace{-0.4cm}
 \begin{align}
A_3( g \; \bar q^{\, \bullet} \; q^{\bullet}     )   ={}& \parbox{20mm}{\unitlength=0.20bp%
\begin{feynartspicture}(300,300)(1,1)
\FADiagram{}
\FAProp(3.,10.)(10.,10.)(0.,){/Cycles}{0} 
\FAProp(16.,15.)(10.,10.)(0.,){/Straight}{1} 
\FAVert(15.8873,12.){0}
\FAProp(16.,5.)(10.,10.)(0.,){/Straight}{-1}  
\FAVert(15.8873,8.){0}
\FAVert(10,10.){0}
\end{feynartspicture}}   \nn[-5.ex]
A_3( g^\bullet \;  q^{\bullet} \; \bar q     )   ={}&  \parbox{20mm}{\unitlength=0.20bp%
\begin{feynartspicture}(300,300)(1,1)
\FADiagram{}
\FAProp(3.,10.)(10.,10.)(0.,){/Cycles}{0} 
\FAVert(4.,8.){0}
\FAProp(16.,15.)(10.,10.)(0.,){/Straight}{-1} 
\FAVert(15.8873,12.){0}
\FAProp(16.,5.)(10.,10.)(0.,){/Straight}{1}  
\FAVert(10,10.){0}
\end{feynartspicture}}    \nn[-5.ex]
A_3( g^\bullet \;  q \; \bar q^{\, \bullet}     )   ={}&  \parbox{20mm}{\unitlength=0.20bp%
\begin{feynartspicture}(300,300)(1,1)
\FADiagram{}
\FAProp(3.,10.)(10.,10.)(0.,){/Cycles}{0} 
\FAVert(4.,8.){0}
\FAProp(16.,15.)(10.,10.)(0.,){/Straight}{-1} 
\FAProp(16.,5.)(10.,10.)(0.,){/Straight}{1}  
\FAVert(15.8873,8.){0}
\FAVert(10,10.){0}
\end{feynartspicture}} \nn[-5.ex]  
A_3(  s^{\bullet}_g \;  q^{\bullet} \; \bar q     )   ={}&   \parbox{20mm}{\unitlength=0.20bp%
\begin{feynartspicture}(300,300)(1,1)
\FADiagram{}
\FAProp(3.,10.)(10.,10.)(0.,){/ScalarDash}{0} 
\FAVert(4.,8.){0}
\FAProp(16.,15.)(10.,10.)(0.,){/Straight}{-1} 
\FAVert(15.8873,12.){0}
\FAProp(16.,5.)(10.,10.)(0.,){/Straight}{1}  
\FAVert(10,10.){0}
\end{feynartspicture}}  \nn[-5.ex]
A_3(  s^{\bullet}_g \;  q \; \bar q^{\, \bullet}     )   ={}&   \parbox{20mm}{\unitlength=0.20bp%
\begin{feynartspicture}(300,300)(1,1)
\FADiagram{}
\FAProp(3.,10.)(10.,10.)(0.,){/ScalarDash}{0} 
\FAVert(4.,8.){0}
\FAProp(16.,15.)(10.,10.)(0.,){/Straight}{-1} 
\FAProp(16.,5.)(10.,10.)(0.,){/Straight}{1}  
\FAVert(15.8873,8.){0}
\FAVert(10,10.){0}
\end{feynartspicture}} \nn[-5.ex]  
A_4( g \; g \;  \bar q^{\, \bullet} \; q^{\bullet}     )    ={}& 
\parbox{20mm}{\unitlength=0.18bp%
%
\begin{feynartspicture}(300,300)(1,1)
\FADiagram{}
\FAProp(0.,15.)(6.,10.)(0.,){/Cycles}{0}
\FAProp(0.,5.)(6.,10.)(0.,){/Cycles}{0}
\FAProp(20.,15.)(14.,10.)(0.,){/Straight}{1}
\FAProp(20.,5.)(14.,10.)(0.,){/Straight}{-1}
\FAProp(6.,10.)(14.,10.)(0.,){/Cycles}{0}
\FAVert(6.,10.){0}
\FAVert(14.,10.){0}
\FAVert(19.5,8.){0}
\FAVert(19.5,12.){0}
\end{feynartspicture}
}
+\parbox{20mm}{\unitlength=0.18bp%
\begin{feynartspicture}(300,300)(1,1)
\FADiagram{}
\FAProp(0.,15.)(10.,14.)(0.,){/Cycles}{0}
\FAProp(0.,5.)(10.,6.)(0.,){/Cycles}{0}
\FAProp(20.,15.)(10.,14.)(0.,){/Straight}{1}
\FAProp(20.,5.)(10.,6.)(0.,){/Straight}{-1}
\FAProp(10.,14.)(10.,6.)(0.,){/Straight}{1}
\FAVert(10.,14.){0}
\FAVert(10.,6.){0}
\FAVert(17.,7.){0}
\FAVert(17.,13.){0}
\FAVert(12.,10.){0}
\end{feynartspicture}

}  \nn[-5.ex]   
A_4(\bar q^{\, \bullet} \; q^{\bullet}  \; \bar q \; q  )    ={}& 
\parbox{20mm}{\unitlength=0.18bp%
%
\begin{feynartspicture}(300,300)(1,1)
\FADiagram{}
\FAProp(0.,15.)(10.,14.)(0.,){/Straight}{-1}
\FAProp(0.,5.)(10.,6.)(0.,){/Straight}{1}
\FAProp(20.,15.)(10.,14.)(0.,){/Straight}{1}
\FAProp(20.,5.)(10.,6.)(0.,){/Straight}{-1}
\FAProp(10.,14.)(10.,6.)(0.,){/Cycles}{0}
\FAVert(10.,14.){0}
\FAVert(10.,6.){0}
\FAVert(3.,7.){0}
\FAVert(3.,13.){0}
\FAVert(8.,10.){0}
\end{feynartspicture}
} 
+\parbox{20mm}{\unitlength=0.18bp%
\begin{feynartspicture}(300,300)(1,1)
\FADiagram{}
\FAProp(0.,15.)(10.,14.)(0.,){/Straight}{-1}
\FAProp(0.,5.)(10.,6.)(0.,){/Straight}{1}
\FAProp(20.,15.)(10.,14.)(0.,){/Straight}{1}
\FAProp(20.,5.)(10.,6.)(0.,){/Straight}{-1}
\FAProp(10.,14.)(10.,6.)(0.,){/ScalarDash}{0}
\FAVert(10.,14.){0}
\FAVert(10.,6.){0}
\FAVert(3.,7.){0}
\FAVert(3.,13.){0}
\FAVert(8.,10.){0}
\end{feynartspicture}

} \nn[-5.ex]   
A_4( q^{ \bullet} \; g \; \bar q  \; g^\bullet )    ={}& 
\parbox{20mm}{\unitlength=0.18bp%
%
\begin{feynartspicture}(300,300)(1,1)
\FADiagram{}
\FAProp(0.,15.)(6.,10.)(0.,){/Cycles}{0}
\FAProp(0.,5.)(6.,10.)(0.,){/Straight}{-1}
\FAProp(20.,15.)(14.,10.)(0.,){/Straight}{1}
\FAProp(20.,5.)(14.,10.)(0.,){/Cycles}{0}
\FAProp(6.,10.)(14.,10.)(0.,){/Straight}{-1}
\FAVert(6.,10.){0}
\FAVert(14.,10.){0}
\FAVert(.5,8.){0}
\FAVert(19.5,8.){0}
\FAVert(10.,8.5){0}
\end{feynartspicture}}
+\parbox{20mm}{\unitlength=0.18bp%
\begin{feynartspicture}(300,300)(1,1)
\FADiagram{}
\FAProp(0.,15.)(10.,14.)(0.,){/Cycles}{0}
\FAProp(0.,5.)(10.,6.)(0.,){/Straight}{-1}
\FAProp(20.,15.)(10.,14.)(0.,){/Straight}{1}
\FAProp(20.,5.)(10.,6.)(0.,){/Cycles}{0}
\FAProp(10.,14.)(10.,6.)(0.,){/Straight}{1}
\FAVert(10.,14.){0}
\FAVert(10.,6.){0}
\FAVert(3.,7.){0}
\FAVert(17.,7.){0}
\end{feynartspicture}

}  \nn[-5.ex]
A_4( g \; \bar q^{\, \bullet} \; g^\bullet \;  q  )    ={}& 
\parbox{20mm}{\unitlength=0.18bp%
%
\begin{feynartspicture}(300,300)(1,1)
\FADiagram{}
\FAProp(0.,15.)(6.,10.)(0.,){/Straight}{1}
\FAProp(0.,5.)(6.,10.)(0.,){/Cycles}{0}
\FAProp(20.,15.)(14.,10.)(0.,){/Cycles}{0}
\FAProp(20.,5.)(14.,10.)(0.,){/Straight}{-1}
\FAProp(6.,10.)(14.,10.)(0.,){/Straight}{1}
\FAVert(6.,10.){0}
\FAVert(14.,10.){0}
\FAVert(.5,12.){0}
\FAVert(19.5,12.){0}
\FAVert(10.,11.5){0}
\end{feynartspicture}}
+\parbox{20mm}{\unitlength=0.18bp%
\begin{feynartspicture}(300,300)(1,1)
\FADiagram{}
\FAProp(0.,15.)(10.,14.)(0.,){/Straight}{1}
\FAProp(0.,5.)(10.,6.)(0.,){/Cycles}{0}
\FAProp(20.,15.)(10.,14.)(0.,){/Cycles}{0}
\FAProp(20.,5.)(10.,6.)(0.,){/Straight}{-1}
\FAProp(10.,14.)(10.,6.)(0.,){/Straight}{1}
\FAVert(10.,14.){0}
\FAVert(10.,6.){0}
\FAVert(3.,13.){0}
\FAVert(17.,13.){0}
\end{feynartspicture}

}  \nn[-5.ex]
A_4( q^{ \bullet} \; g \; \bar q  \;  s^{\bullet}_g )    ={}& 
\parbox{20mm}{\unitlength=0.18bp%
%
\begin{feynartspicture}(300,300)(1,1)
\FADiagram{}
\FAProp(0.,15.)(6.,10.)(0.,){/Cycles}{0}
\FAProp(0.,5.)(6.,10.)(0.,){/Straight}{-1}
\FAProp(20.,15.)(14.,10.)(0.,){/Straight}{1}
\FAProp(20.,5.)(14.,10.)(0.,){/ScalarDash}{0}
\FAProp(6.,10.)(14.,10.)(0.,){/Straight}{-1}
\FAVert(6.,10.){0}
\FAVert(14.,10.){0}
\FAVert(.5,8.){0}
\FAVert(19.5,8.){0}
\FAVert(10.,8.5){0}
\end{feynartspicture}}
+\parbox{20mm}{\unitlength=0.18bp%
\begin{feynartspicture}(300,300)(1,1)
\FADiagram{}
\FAProp(0.,15.)(10.,14.)(0.,){/Cycles}{0}
\FAProp(0.,5.)(10.,6.)(0.,){/Straight}{-1}
\FAProp(20.,15.)(10.,14.)(0.,){/Straight}{1}
\FAProp(20.,5.)(10.,6.)(0.,){/ScalarDash}{0}
\FAProp(10.,14.)(10.,6.)(0.,){/Straight}{1}
\FAVert(10.,14.){0}
\FAVert(10.,6.){0}
\FAVert(3.,7.){0}
\FAVert(17.,7.){0}
\end{feynartspicture}

}  \nn[-5.ex]
A_4( g \; \bar q^{\, \bullet} \;  s^{\bullet}_g \;  q  )    ={}& 
\parbox{20mm}{\unitlength=0.18bp%
%
\begin{feynartspicture}(300,300)(1,1)
\FADiagram{}
\FAProp(0.,15.)(6.,10.)(0.,){/Straight}{1}
\FAProp(0.,5.)(6.,10.)(0.,){/Cycles}{0}
\FAProp(20.,15.)(14.,10.)(0.,){/ScalarDash}{0}
\FAProp(20.,5.)(14.,10.)(0.,){/Straight}{-1}
\FAProp(6.,10.)(14.,10.)(0.,){/Straight}{1}
\FAVert(6.,10.){0}
\FAVert(14.,10.){0}
\FAVert(.5,12.){0}
\FAVert(19.5,12.){0}
\FAVert(10.,11.5){0}
\end{feynartspicture}} 
+\parbox{20mm}{\unitlength=0.18bp%
%
\begin{feynartspicture}(300,300)(1,1)
\FADiagram{}
\FAProp(0.,15.)(10.,14.)(0.,){/Straight}{1}
\FAProp(0.,5.)(10.,6.)(0.,){/Cycles}{0}
\FAProp(20.,15.)(10.,14.)(0.,){/ScalarDash}{0}
\FAProp(20.,5.)(10.,6.)(0.,){/Straight}{-1}
\FAProp(10.,14.)(10.,6.)(0.,){/Straight}{1}
\FAVert(10.,14.){0}
\FAVert(10.,6.){0}
\FAVert(3.,13.){0}
\FAVert(17.,13.){0}
\end{feynartspicture}

} 
 \end{align}
 \vspace{2.5cm}

\bibliography{references}

\end{document}